\documentclass[prx,twocolumn,english,longbibliography,superscriptaddress]{revtex4-2}

\usepackage{ifthen}
\usepackage{xcolor}
\usepackage{graphics,graphicx,epsfig}
\usepackage{epsf,epstopdf,wrapfig}
\usepackage{amssymb,amsfonts,amsmath}
\usepackage[export]{adjustbox}
\usepackage[english]{babel}
\include{epsf}
\usepackage{ifthen}
\usepackage{rotating} 

\usepackage{booktabs}
\usepackage{multirow}

\definecolor{ocre}{RGB}{10,100,185}
\usepackage[colorlinks = true,
            linkcolor = ocre,
            urlcolor  = ocre,
            citecolor = ocre,
            anchorcolor = ocre]{hyperref}

\newcommand{\AN}[1]{{\color{black}#1}}

\usepackage[normalem]{ulem}

\usepackage{hyperref}

\newcommand{\EQ}{\begin{equation}}
\newcommand{\EE}{\end{equation}}
\newcommand{\EQA}{\begin{eqnarray}}
\newcommand{\EEA}{\end{eqnarray}}

\newcommand{\RNum}[1]{\uppercase\expandafter{\romannumeral #1\relax}}

\newcommand{\AffActInfoTable}{S1}
\newcommand{\SpearmanrAffinityTable}{S2}
\newcommand{\SpearmanrActivityTable}{S3}
\newcommand{\AfMetricsTable}{S4}
\newcommand{\DesignInfoTable}{S5}
\newcommand{\AllPdbsTable}{S6}
\newcommand{\pMHCPWMsTInfoable}{S7}

\newcommand{\ProteinMPNNScoringSchemesFigure}{S1} 
\newcommand{\HsiueHisFigure}{S2}
\newcommand{\ActivityHeatmapsFigure}{S3}  
\newcommand{\AfScoringResultsFigure}{S4} 
\newcommand{\AfExamplesFigure}{S5} 
\newcommand{\HermesDesignsEnergiesFigure}{S6} 
\newcommand{\FlowMicroscopyFigure}{S7} 
\newcommand{\DesignTCRDockAFScatterFigure}{S8} 
\newcommand{\DesignExpNyesoFigure}{S9} 
\newcommand{\DesignExpEbvFigure}{S10} 
\newcommand{\DesignExpMageFigure}{S11}  
\newcommand{\DesignPWMsWithAf}{S12} 
\newcommand{\DesignTcrdockNyesoFigure}{S13} 
\newcommand{\DesignTcrdockEbvFigure}{S14} 
\newcommand{\DesignTcrdockMageFigure}{S15} 
\newcommand{\DesignAUCFigure}{S16} 
\newcommand{\pMHCPWMsFigure}{S17}

\newcommand{\AfInfoDataset}{S1}
\newcommand{\experimentalMeasurementsDataset}{S2}

\newcommand{\corres}{Correspondence should be addressed to:
Gian Marco Visani: {gvisan01@cs.washington.edu}, and Armita Nourmohammad: {armita@uw.com}.}

\begin{document}

\title{ T-cell receptor specificity landscape revealed through de novo peptide design}
\author{Gian Marco Visani}
\thanks{\corres}
\affiliation{Paul G. Allen School of Computer Science and Engineering, University of Washington, 85 E Stevens Way NE, Seattle, WA 98195, USA}
\author{Michael N. Pun}
\affiliation{Department of Physics, University of Washington, 3910 15th Avenue Northeast, Seattle, WA 98195, USA}
\author{{Anastasia A. Minervina}}
\affiliation{St. Jude Children’s Research Hospital, Memphis, TN, USA}
\author{Philip Bradley}
 \affiliation{Fred Hutchinson Cancer Center, 1241 Eastlake Ave E, Seattle, WA 98102, USA}
\author{Paul Thomas}
\affiliation{St. Jude Children’s Research Hospital, Memphis, TN, USA}
\author{Armita Nourmohammad}
\thanks{\corres}
\affiliation{Paul G. Allen School of Computer Science and Engineering, University of Washington, 85 E Stevens Way NE, Seattle, WA 98195, USA}
 \affiliation{Department of Physics, University of Washington, 3910 15th Avenue Northeast, Seattle, WA 98195, USA}
 \affiliation{Fred Hutchinson Cancer Center, 1241 Eastlake Ave E, Seattle, WA 98102, USA}
\affiliation{Department of Applied Mathematics, University of Washington, 4182 W Stevens Way NE, Seattle, WA 98105, USA}

\begin{abstract}
\noindent T-cells play a key role in  adaptive immunity by mounting specific responses against diverse pathogens. An effective binding between  T-cell receptors (TCRs) and pathogen-derived peptides presented on Major Histocompatibility Complexes (MHCs) mediate an immune response. However, predicting these interactions remains challenging due to limited functional data on  T-cell reactivities. Here, we introduce a computational approach to predict TCR interactions with peptides presented on {MHC} class I alleles, and to design novel immunogenic peptides for specified TCR-MHC complexes. Our method leverages HERMES, a structure-based, physics-guided machine learning model trained on the protein universe to predict amino acid preferences based on local structural environments. Despite no direct training on TCR-pMHC data, the implicit physical reasoning in HERMES enables us to make accurate predictions of both TCR-pMHC binding affinities and T-cell activities across diverse viral epitopes and cancer {neoantigens}, achieving up to {0.72} correlation with experimental data. Leveraging our TCR recognition model, we develop a computational protocol for {de novo} design of immunogenic peptides. Through experimental validation in three TCR-MHC systems targeting viral and cancer {peptides}, we demonstrate that our {designs}---with up to five substitutions from the native sequence---activate T-cells at success rates of up to 50\%. Lastly, we use our generative framework to quantify the diversity of the peptide recognition landscape for various TCR-MHC complexes, offering key insights into T-cell specificity in both humans and mice. \AN{Our approach provides a platform for immunogenic peptide and neoantigen design, as well as for evaluating TCR specificity, offering a computational framework to inform design of engineered T-cell therapies and vaccines.}
\end{abstract}

\maketitle

\section{Introduction} 

T-cells play a key role in the adaptive immune system, contributing to immune surveillance and response to pathogens. T-cells recognize pathogen-derived epitopes in the form of short protein fragments (peptides) displayed on specific molecules known as Major Histocompatibility Complexes (MHCs). The binding between a T-cell receptor (TCR) and a peptide-MHC (pMHC) is essential to mediate a protective T-cell response to an infection. To defend the host against a multitude of pathogens, a highly diverse TCR repertoire is generated through a somatic VDJ recombination process, and selected to target infected cells with remarkable sensitivity and specificity.

Predicting the TCR-pMHC binding specificity is an important step in characterizing the efficacy of the immune system in countering different pathogens. Such understanding would aid with disease diagnoses and development of diagnostic tests for early detection of autoimmune diseases~\cite{Attaf2015-tw}. A predictive model for TCR-pMHC specificity can be used to design engineered TCRs that recognize cancer antigens~\cite{Liu2025-hl}, enhancing the effectiveness of adoptive cell transfer and {Chimeric Antigen Receptor (CAR) T-cell} therapies~\cite{Leidner2022-ca,Mirazee2024-bo}, as well as the design of soluble TCRs and bispecific T-cell engager therapeutics~\cite{Hsiue2021-ag}. Moreover, it can enable the design of optimized antigens for vaccines to elicit robust T-cell responses against specific pathogens or tumors~\cite{Sahin2018-fv,Blass2021-jd,Rojas2023-nx}.

One bottleneck in deciphering the TCR-pMHC code is the lack of well-curated datasets. High-throughput immune repertoire sequencing, combined with comparative analyses of hosts facing similar immune challenges, and the development of functional assays (e.g.,  {pMHC-tetramer staining}~\cite{Zhang2018-pe} or MIRA~\cite{Nolan2025-fu}), have enabled the identification of diverse TCR pools with potential reactivity to specific antigens~\cite{Dash2017-if}, collected in repositories like Immune Epitope Database (IEDB)~\cite{Vita2019-qn} and VDJdb~\cite{Shugay2018-nt}. However, the current databases are unbalanced, with only a handful of dominant peptides (e.g. derived from SARS-CoV-2, CMV, EBV, Influenza, etc.) and  {MHC alleles} present, each linked to hundreds of TCRs. 

As a result, statistical inference and machine learning methods trained on such data to predict TCR-pMHC  reactivities lack generalizability beyond their training data~\cite{Isacchini2021-fs,Weber2021-ik,Meynard-Piganeau2024-po}. On the structural side, only a few hundred  TCR-pMHC crystal structures have been experimentally generated to date. This is in part because TCR-pMHC interactions are typically of low affinity and are difficult to stabilize to the level needed for CryoEM or X-ray crystallography~\cite{Piepenbrink2009-jy}. Importantly, this scarcity  hinders computational prediction methods like AlphaFold in modeling reliable TCR-pMHC  structures~\cite{Bradley2023-rl,Yin2023-zm}. Machine learning models, such as large language models, pre-trained on vast unannotated protein datasets can generate meaningful data representations~\cite{Rives2021-sr,Lin2023-ub}. Fine-tuning these models on smaller, labeled datasets for specific tasks (e.g., predicting mutation impacts on stability or function) has proven highly effective~\cite{Madani2023-mg,Schmirler2024-wp,Wang2024-ep,Hie2024-vb}. Pre-trained sequence-based protein language models have been employed to predict reactivities of antibodies to antigens~\cite{Wang2024-ep,Hie2024-vb}, or TCRs to pMHC complexes~\cite{Bravi2023-aj,Meynard-Piganeau2024-po}. However, lack of generalizability due to sparse and skewed training data has remained a challenge in these models.

\begin{figure*}[th!]
\includegraphics[width=0.95\textwidth]{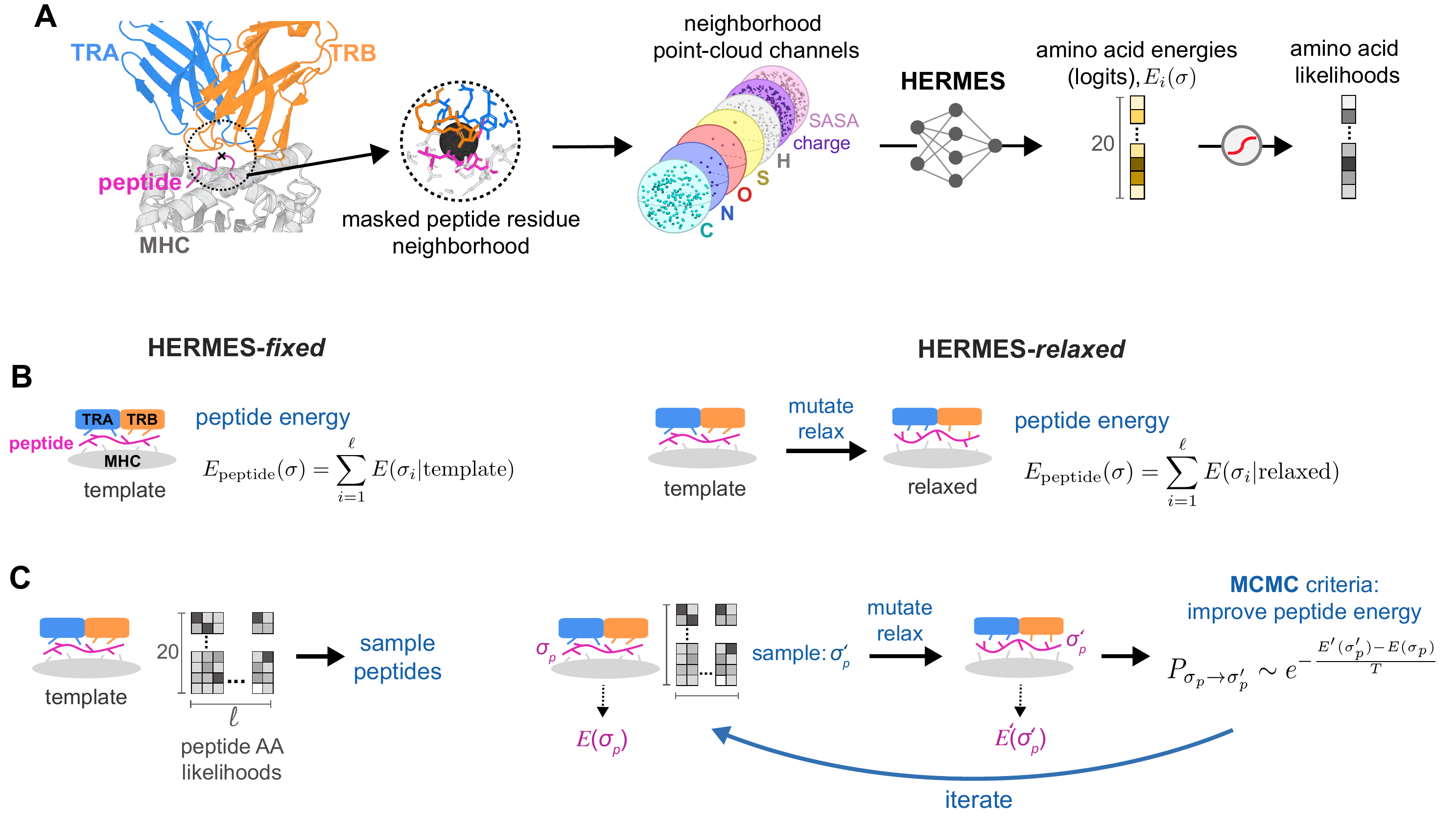}
\caption{{\bf HERMES model for predicting and designing TCR-pMHC interactions.} {\bf (A)} HERMES is a structure-based, rotationally equivariant neural network pre-trained on the protein universe to predict amino acid preferences at a residue from its 3D structural environment~\cite{Visani2024-sh,Pun2024-qz}. It takes as input all atoms within 10\AA\ of a focal residue (with that residue’s atoms masked), featurizing them by atom type, charge, and SASA. HERMES then outputs energies $E_i(\sigma)$ for the 20 possible amino acids $\sigma$ at residue $i$, which serve as logits in a softmax to estimate amino acid likelihoods.
{\bf (B)} To predict TCR-pMHC affinity, we use two protocols: HERMES-{\em fixed} (left) and HERMES-{\em relaxed} (right). In HERMES-{\em fixed} we use a fixed structural template closest to the TCR-pMHC of interest as input to HERMES, and evaluate the peptide energy as the sum of  HERMES-computed per-residue energies for the peptide. In HERMES-{\em relaxed} mutations are introduced to the template's peptide to match the peptide of interest, followed by local repacking in PyRosetta~\cite{Chaudhury2010-mo}. Peptide energy is then evaluated similar to  HERMES-{\em fixed}, but using the relaxed structure. {\bf (C)} We also use two protocols for peptide design. Similar to (B) we fix the closest structural template in HERMES-{\em fixed}, and use HERMES' likelihood matrix for possible amino acids at each peptide position, which we sample to generate new peptide sequences.  In  HERMES-{\em relaxed} (right), we begin with a random peptide $\sigma$, and locally pack it with PyRosetta, and also generate an amino acid likelihood matrix  (similar to HERMES-{\em fixed}) for this packed structure. We sample a new sequence $\sigma'$ from this likelihood matrix, and pack it with PyRosetta. We evaluate peptide energies $E_{\text{peptide}}(\sigma)$, and $E_{\text{peptide}}(\sigma')$ for both peptides, using their respective PyRosetta-relaxed structures, and we use these energies as the criterion to sample  improved sequences  with MCMC (SI). This procedure is then iterated until convergence. Peptides designed by both approaches are then filtered by TCRdock~\cite{Bradley2023-rl} or  Alpha-Fold 3~\cite{Abramson2024-jp} to assure the stability of the complex.
\label{Fig:1}}
\end{figure*}
 
Protein-protein interactions in general, and immune-pathogen interaction in particular, rely on the complementarity in the 3D structures and amino acid compositions of the interacting protein pairs. Even though the interacting amino acids can be far apart in a protein sequence, they are close to each other in structure, resulting in local interactions that can determine immune-pathogen recognition. Machine learning models trained on protein structures instead of sequences can learn local structural representations for amino acid statistics, which tend to be more generalizable across different protein families and beyond their training sets~\cite{Dauparas2022-fu,Blaabjerg2023-oq,Michalewicz2024-hw,Visani2024-sh,Pun2024-qz,Diaz2024-vv}.

Here, we present a structure-based approach to predict  interaction affinities between TCRs and peptides presented on  {MHC class I},  and to  design reactive peptides for specific TCR-MHC-I complexes (TCR-MHC for short). Due to the limited availability of structural data for TCRs, we use HERMES-- a physics-guided, 3D-equivariant machine learning model trained on the entire protein universe to predict amino acid preferences based on their local structural environments~\cite{Pun2024-qz,Visani2024-sh}.
 
{We demonstrate that HERMES accurately predicts both binding affinity of TCR-pMHC Class I complexes and the functional responses and T-cell activities induced by a range of peptides presented on MHC molecules. Affinity of TCR-pMHC complexes, measured as the dissociation constant $K_d$ by surface plasmon resonance, is labor intensive to obtain. Functional readout used to measure T-cell activities, including measurements of T-cell proliferation,  labeled cellular activation markers, and cytokine (e.g., IFN-$\gamma$) release, are easier to obtain than affinity measurements. Yet,  functional readouts are shaped not only by the TCR-pMHC affinity, but also by other cellular factors such as the T-cell's differentiation state (e.g., naive or memory), self-reactivity of the TCR~\cite{Altan-Bonnet2005-lm,Achar2022-tb}, and the avidity effects driven by TCR/co-receptor clustering and pMHC density on antigen-presenting cells~\cite{Stone2009-ab}. Although HERMES is more suited to predict molecular affinities, we show that its scores nonetheless can well track T-cell activities, making it a reliable tool for predicting T-cell  response to antigens.}

Lastly, we show that HERMES can be used to design novel peptides  {binding specific} TCR-MHC complexes, achieving up to 50\% experimental validation accuracy across diverse TCR-MHC systems. By leveraging our design algorithm, we characterize the specificity of a diverse range of {TCRs}, providing a quantitative measure for the diversity of the peptide-MHC antigens that a TCR can recognize in humans and mice. 

\begin{figure*}[ht!]
	\centering
	\includegraphics[width=0.8\textwidth]{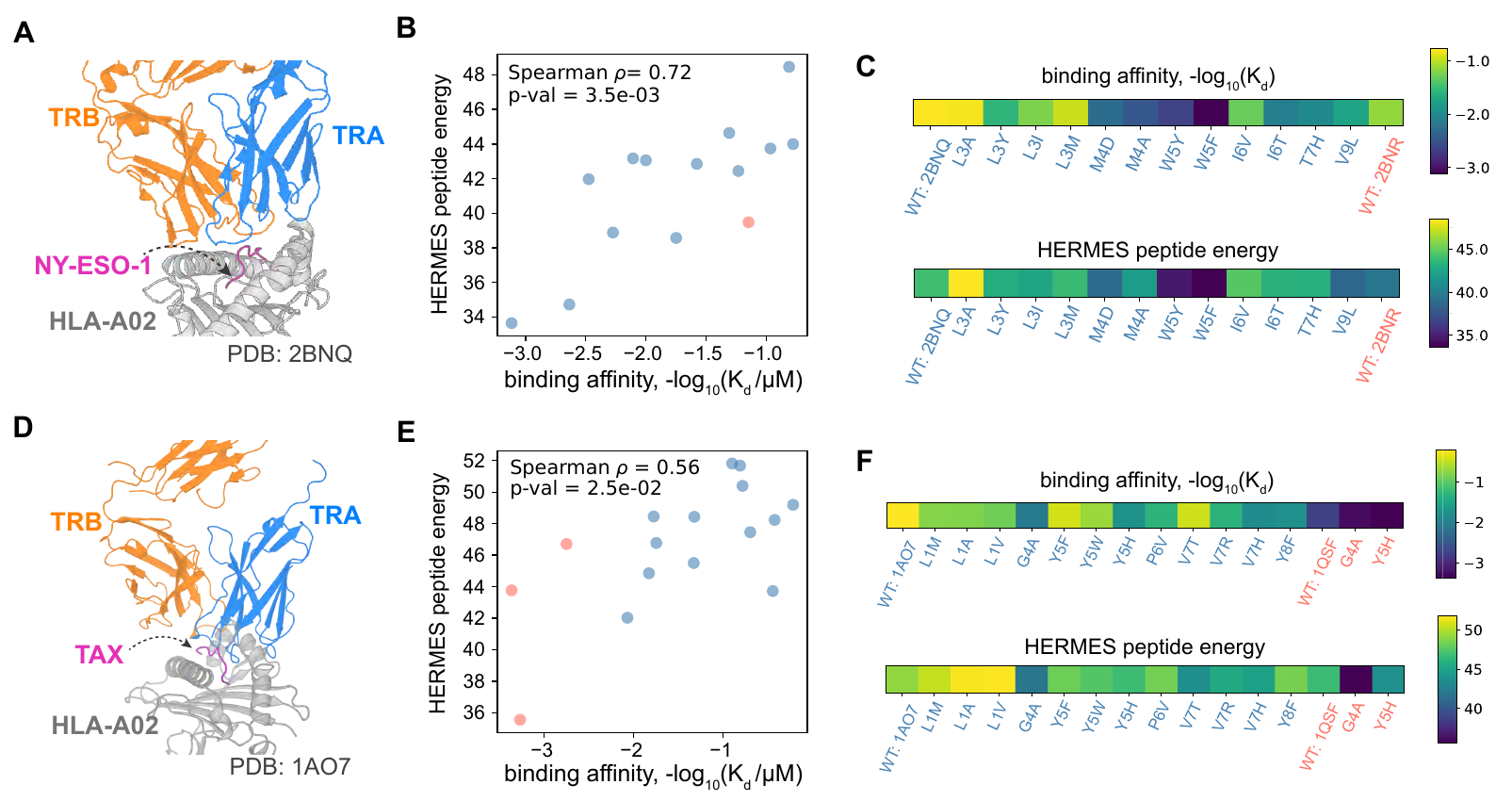}
	\caption{
	{\bf Predicting binding affinity of TCR-pMHC complexes.}  {\bf(A)} The structure of the NY-ESO-1 peptide (purple), in complex with HLA-A*02:01 and a purified specific 1G4 TCR is shown (PDB ID: 2BNQ~\cite{Chen2005-jk}). This structure together with that of a peptide variant with C9V mutation (PDB ID: 2BNR~\cite{Chen2005-jk}) are used as templates to predict the binding affinity of the TCR-MHC complex to other peptide variants. {\bf (B)} The binding affinities ($-\log_{10} (K_d/\mu\text{M})$)  of TCR-MHC complexes with 14 distinct peptides, measured using surface plasmon resonance, are shown against the predicted HERMES-{\em relaxed} peptide energies, using the closest structural template for each variant; 2BNQ is used for  blue points and  2BNR for the orange point. The Spearman $\rho$ between predictions and measurements is reported in the panel. {\bf(C)} The heatmaps show  the experimental affinity values (top) and predicted peptide energies (bottom) for all the peptide variants, with the identity of the mutations color coded based on the closest structural template. {\bf(D-F)} Similar to (A-C) but for the Tax viral peptide in complex with HLA-A*02:01 and a specific human TCR. Two structural templates are used~\cite{Garboczi1996-fr} for evaluating peptide energies with PDB IDs: 1AO7 (shown in D, and used for blue points), and 1QSF (used for orange points).
	\label{Fig:2}
	}
\end{figure*}

\section*{Model}

 T-cell response is mediated by the interactions between TCRs and pMHC complexes. To model the TCR-pMHC interaction, we use HERMES~\cite{Visani2024-sh},  a 3D rotationally equivariant structure-based neural network model, trained on the protein universe; see Fig.~\ref{Fig:1}A. 
HERMES predicts the amino acid propensity from its surrounding 3D protein environment, a measure that we have previously shown to serve as a reliable proxy for assessing the effect of mutations on protein stability or binding~\cite{Pun2024-qz,Visani2024-sh}.

For TCR-pMHC complexes, we seek to determine how changes in a peptide sequence impact the  binding affinity of a TCR-pMHC complex, and ultimately, the T-cell response to the antigen. To do so, we characterize a score, which we term peptide energy, for a peptide $\sigma$, given its surrounding TCR and MHC structural environment,
{
\EQ
E_{\text{peptide}}(\sigma ; \text{TCR, MHC}) = \sum_{i=1}^\ell  E(\sigma_i ; \text{TCR, MHC},\sigma_{/i})
\label{eq:pepEn}
\EE
We assume that each peptide residue contributes linearly to the total energy by the amount  $E(\sigma_i ; \text{TCR, MHC}, \sigma_{/i})$, where $\sigma_i$ is the amino acid at position $i$ of the peptide. A residue's energy contribution $E$ is evaluated by HERMES (logits of the network), taking in as input the atomic composition of the surrounding TCR, MHC, and the rest of the peptide $\sigma_{/i}$ (excluding the $i^{th}$ residue) within a 10~\AA~of $\sigma_i$'s $\alpha$-Carbon}; see Fig.~\ref{Fig:1} and SI for details.

 To compute peptide energy, we input to HERMES  an experimentally or computationally determined structure of a specified TCR-MHC complex bound to at least one peptide variant. Experimental data on the impact of peptide substitutions on the TCR-pMHC structure is  limited, and computational models often fail to capture subtle conformational changes in the structure due to amino acid substitutions in a peptide. Given these constraints, we adopt two approaches to estimate  peptide energies across diverse  peptide variants for a given TCR-MHC complex:
\begin{itemize}
	\item HERMES-{\em fixed}: In the simplest approach, we choose a TCR-pMHC structure with the same TCR and MHC as our query and a peptide sequence closest to the peptide of interest. This structure is  then used as our  template in HERMES to  compute the peptide energy as described in eq.~\ref{eq:pepEn} (Fig.~\ref{Fig:1}B). This method does not alter the underlying structure and assumes that peptide amino acid substitutions do not significantly change the conformation of the TCR-pMHC complex.
	
	\item HERMES-{\em relaxed}: Since amino acid substitutions can locally reshape a protein complex, we introduce a more involved protocol to account for these structural changes. We begin with the closest available  structure for the TCR-pMHC complex. We then mutate the original peptide  to the desired state  and apply a relaxation procedure using PyRosetta to  pack  the substituted amino acid~\cite{Chaudhury2010-mo} (SI). During relaxation, side-chain and backbone atoms of the peptide are allowed to move, while only the side-chain atoms of the MHC and TCR chains within a 10~\AA\,radius of the peptide are flexible. We then calculate the peptide energy with HERMES (eq.\ref{eq:pepEn}), using the relaxed structure as input (Fig.~\ref{Fig:1}B). Since PyRosetta's relaxation procedure is stochastic, we average the peptide energy across 100 realizations of these relaxations. {For ablation purposes, we also include  results obtained by selecting the relaxed conformation with the  lowest (most favorable) Rosetta energy, and using its HERMES peptide energy.}
\end{itemize}

\section*{Results}
\subsection*{Predicting binding affinity of TCR-pMHC complexes} 
 Binding between TCRs and pMHC complexes is necessary to mediate a T-cell response. The binding affinity of a natural TCR-pMHC complex is relatively low, with a dissociation constant $K_d$ in the range of $1-100\mu{\rm M}$~\cite{Davis1998-bv,Piepenbrink2009-jy,Zhong2013-ja}, in contrast to nanomolar affinities for antibody-antigen complexes. Engineered TCRs  can achieve much higher affinities in a nanomolar~\cite{Soto2013-df} to picomolar~\cite{Li2005-md, Dunn2006-dh} range.

Surface plasmon resonance (SPR) spectroscopy provides accurate measurements of the dissociation constant ($K_d$) for specific TCR-pMHC systems. In these experiments, one protein--either the TCR or the MHC--is immobilized on a conducting plate, and the other is introduced in solution to bind to it. This binding alters the local refractive index near the plate, affecting the resonance signal. We  {used a published} dataset of SPR-measured affinities  for two TCR-MHC complexes binding to an ensemble of peptides, for which crystal structures of the complexes with at least  one peptide are available~\cite{Aleksic2010-wz,Pettmann2021-kl}; see Table~\AffActInfoTable~for details.

\begin{table*}[ht!]
\centering
\begin{tabular}{ll|cc|cccccccc}
\toprule
 & & \multicolumn{2}{c|}{TCR-pMHC affinity} & \multicolumn{8}{c}{T-cell activity} \\
\cmidrule(lr){3-4}\cmidrule(lr){5-12}
Model type & Model & 1G4 TCR & A6 TCR & H2-scDb & TCR1 & TCR2 & TCR3 & TCR4 & TCR5 & TCR6 & TCR7 \\
\midrule
\multirow{1}{*}{Substitution Matrix} & BLOSUM62~\cite{Henikoff1992-jq} & \textbf{\underline{0.83}} & {0.56}*  & -0.05  & -0.07 & 0.14  & 0.38 & \textbf{\underline{0.33}} & 0.42* & \textbf{\underline{0.55}} & -0.01 \\
\midrule
\multirow{2}{*}{Structure Prediction} & TCRdock~\cite{Bradley2023-rl} & {0.62*} & \textbf{\underline{0.88}} & N/A & 0.02  & 0.13 & 0.21 & 0.18* & -0.02 & 0.19 & 0.06  \\
& TCRdock-{\em no template}~\cite{Bradley2023-rl} & 0.25 & -0.14 & N/A & -0.03  & 0.44 & 0.11  & -0.22  & 0.28*  & 0.14 & 0.06  \\
\midrule
\multirow{1}{*}{Sequence based ML} & TAPIR~\cite{Fast2023-tk} &  0.29 & 0.35 & N/A & 0.36 & 0.36 & -0.00 & 0.12 & 0.39* & 0.08 & -0.01 \\
\midrule
\multirow{2}{*}{Structure based ML} 
& ESM-IF1~\cite{Hsu2022-km} & 0.38* & 0.17 & 0.12 & 0.43* & 0.41 & 0.19 & -0.15 & 0.28* & 0.25 & 0.08 \\
& ProteinMPNN~\cite{Dauparas2022-fu} & -0.01 & 0.49 & \textbf{\underline{0.55}} & 0.46* & 0.52 & 0.49* & -0.28 & \textbf{\underline{0.45}} & 0.41* & \textbf{\underline{0.31}} \\
\midrule
\multirow{2}{*}{Structure based ML} & HERMES-\textit{fixed} & 0.29 & 0.63* & 0.46* & \textbf{\underline{0.56}} & \textbf{\underline{0.71}} & \textbf{\underline{0.59}} & -0.06 & 0.37* & 0.42* & -0.01 \\
& HERMES-\textit{relaxed} & 0.72* & 0.56* & 0.29 & 0.31 & 0.57 & 0.55* & -0.20 & 0.37* & 0.42* & 0.14* \\
\bottomrule
\end{tabular}
\caption{{{\bf Benchmarking of models for predicting TCR-pMHC binding affinities and T-cell activities.}
\normalfont{The table lists the Spearman~$\rho$ of models' predictions against experimentally measured binding affinities (Fig.~\ref{Fig:2}) and T-cell activities (Fig.~\ref{Fig:3}) in response to peptide variants. The HERMES-{\em fixed}  and HERMES-{\em relaxed}  performance are from  models with no noise (0.00) in training data,  and the ProteinMPNN performance is from the model with a  small noise amplitude of 0.02~\AA; see SI for details. We show in \textbf{bold} the best model for each dataset, and any model whose performance is \textit{not} significantly different  from the best model (p-value$>0.05$  associated with the difference in the Fisher's z-transformed Spearman correlations) is indicated with  an asterisk (*). See Tables~\SpearmanrAffinityTable~and~\SpearmanrActivityTable~in the SI for $p$-values and additional models.}}}
\label{table:spearmanr_affinity_and_activity}
\end{table*}

In our first example, we examined TCR 1G4 binding to NY-ESO-1 peptide variants,  which is a cancer-testis antigen commonly expressed in many cancers and is targeted by immunotherapies~\cite{Esfandiary2015-jf}. We used two structural templates, one with the wild-type  peptide SLLMWITQC (NY-ESO-1), and the other with a more immunogenic peptide, in which the Cysteine at position 9 is substituted  by Valine (C9V)~\cite{Chen2000-vq,Chen2005-jk} (Fig.~\ref{Fig:2}A). For variants differing by a single amino acid, our predictions with HERMES-{\em relaxed} correlated well (Spearman {$\rho=0.72$}) with experimentally measured affinities from refs.~\cite{Aleksic2010-wz,Pettmann2021-kl}; see Fig.~\ref{Fig:2}B,~C.

Next, we tested the A6 TCR binding to the HTLV-1 Tax peptide LLFGYPVYV. We employed two structural templates: one with the wild-type, the other with the mutant peptide with substitution Y8A~\cite{Garboczi1996-fr} (Fig.~\ref{Fig:2}D and {Table~\AffActInfoTable}). Again, the HERMES-{\em relaxed}  peptide energies  correlated  well (Spearman {$\rho=0.56$}) with the  affinities, measured  in ref.~\cite{Pettmann2021-kl} (Fig.~\ref{Fig:2}E,~F).

{Overall, HERMES predictions aligned closely with these limited experimental affinity measurements, with  HERMES-{\em relaxed} showing a better agreement with affinity measurements than HERMES-{\em fixed}. 
We compared our approach for scoring peptides to alternative methods, including the baseline substitution matrices (BLOSUM62~\cite{Henikoff1992-jq} and the TCR-specific substitution matrices from ref.~\cite{Luksza2022-gx}), the  score from the TCR-pMHC structure prediction algorithm TCRdock~\cite{Bradley2023-rl}, the sequence-based machine learning algorithms for TCRs (TAPIR~\cite{Fast2023-tk}, NetTCR~\cite{Jensen2024-xs}, and TULIP~\cite{Meynard-Piganeau2024-po}), and the structure-based inverse folding models (ESM-IF1~\cite{Hsu2022-km}  and ProteinMPNN~\cite{Dauparas2022-fu}); see Table~\ref{table:spearmanr_affinity_and_activity} for key benchmarking results and Table~\SpearmanrAffinityTable~and Fig.~\ProteinMPNNScoringSchemesFigure~for a more extensive comparison of different methods, including the different HERMES models, and different scoring schemes by ProteinMPNN.

For the 1G4 TCR, BLOSUM62~\cite{Henikoff1992-jq} achieves the highest correlation with experimental affinity data, whereas template-guided TCRdock~\cite{Bradley2023-rl} leads for the A6 system; in contrast, TCRdock without using a homologous structure as a template underperforms.  In both systems, HERMES achieves performance that is statistically indistinguishable from the top performers (Fisher's z-test p-value $>0.05$). Overall, BLOSUM62, template-guided TCRdock, and HERMES rank strongly as predictors of TCR-pMHC affinities across the tested datasets. In addition, for the  1G4 system,  ESM-IF1~\cite{Hsu2022-km} and the TCR-specific substitution matrices from ref.~\cite{Luksza2022-gx} also perform competitively. However, due to the limited number of affinity measurements, we lack statistical power  to resolve finer differences among the models. 

It should be noted that although BLOSUM62 is one of the best contenders for affinity prediction, its sequence‑averaged construction ignores local structural context and therefore assigns identical substitution scores regardless of a residue's environment (Fig.~\HsiueHisFigure); similar limitations apply to the TCR-specific substitution matrices from ref.~\cite{Luksza2022-gx}. HERMES, by contrast, is  structure‑aware and can attribute distinct scores to the same mutation when it occurs in different conformational settings--an ability that is critical for nuanced mutational forecasts (Fig.~\HsiueHisFigure). Determining the biophysical circumstances under which such context‑agnostic scoring schemes suffice, and when structure‑aware approaches become necessary, will require systematic evaluation on broader, more diverse datasets.

Lastly, we probed HERMES's robustness to potential inaccuracies in structures generated by the widely used AlphaFold3 (AF3) algorithm~\cite{Abramson2024-jp}, replacing the crystal structure inputs with AF3 models; see SI for details.
Both the template-guided and the de novo AF3 structures remained within  $1.1$~\AA\,  RMSD of the respective crystal structures (Table~\AfMetricsTable). 
For both HERMES and ProteinMPNN, replacing experimental structures with AF3 predictions altered predictions for the 1G4 and A6 TCRs binding to pMHC variants, but the changes were not statistically significant (Fisher's z-test p-value $>0.05$); see Fig.~\AfScoringResultsFigure. It is nonetheless expected that the predictions should suffer in the absence of reliable structural templates, as seen also in other methods such as TCRdock~\cite{Bradley2023-rl} (Tables~\ref{table:spearmanr_affinity_and_activity},~\SpearmanrAffinityTable). }

\begin{figure*}[ht!]
	\centering
	\includegraphics[width=0.9\textwidth]{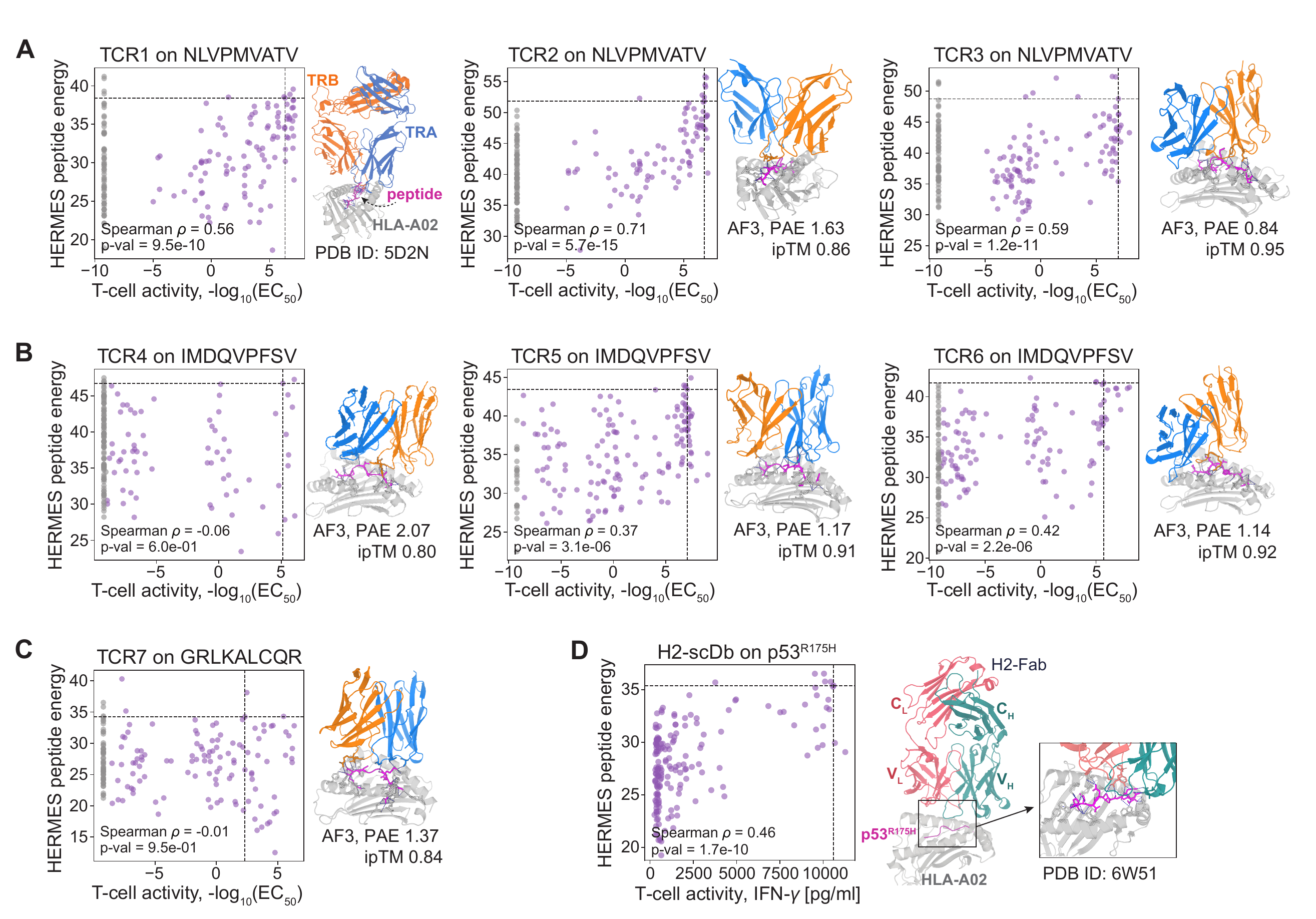}
	\caption{{\bf Predicting T-cell activity in response to peptide variants.}
	{\bf (A-C)}  The predicted  HERMES-{\em fixed} peptide energies versus experimentally measured T-cell activity are shown for (A) HLA-A*02:01-restricted CMV peptide variants interacting with TCR1-TCR3 (panels), (B)  HLA-A*02:01-restricted  melanoma gp100 peptide variants interacting with TCR4-TCR6, and (C) HLA-B*27:05-restricted pancreatic cancer peptide variants interacting with TCR7. In each case, T-cell activity in response to all single amino acid mutations from the wild-type is measured. T-cell activity is based on fitted EC$_{50}$ values from dose-response curves of 4-1BB$^+$ CD8$^+$ T-cells (SI; data from \cite{Luksza2022-gx}). Gray points indicate peptides with T-cell responses below detection limit (no EC$_{50}$ obtained) and are  {excluded from the reported Spearman correlations.}    Except for TCR1 (resolved structure, PDB ID: 5D2N~\cite{Yang2015-cj}), the other TCR-pMHC complexes were modeled with AF3, and the model's confidence for each wild-type peptide is denoted. These structures are used  as templates to estimate peptide energies with HERMES-{\em fixed}; see Table~\AffActInfoTable~and Dataset~\AfInfoDataset~for details on the templates. In all panels, TCR-$\alpha$ (TRA) is shown in blue, TCR-$\beta$ (TRB) in orange, the peptide in magenta, and HLA in gray.    {\bf (D)} The predicted HERMES-{\em fixed} peptide energies are shown against  experimentally measured T-cell activity for single point mutants  of the HLA-A*02:01-restricted p53$^{\rm R175H}$ {neoepitope} from ref.~\cite{Hsiue2021-ag}. T-cell response (IFN-$\gamma$ production) is mediated through  a bispecific antibody binding to the pMHC complex.  The co-crystallized structure of the bispecific antibody with  the p53$^{\rm R175H}$-HLA-A*02:01 complex (PDB ID: 6W51~\cite{Hsiue2021-ag}) is used as the template for HERMES-{\em fixed} predictions.  For all systems (A-D), the heatmaps for the predicted and the experimentally measured  mutational effects are shown in Fig.~\ActivityHeatmapsFigure.
		\label{Fig:3}
	}
\end{figure*}

\subsection*{\bf Predicting T-cell response to peptide antigens}
{T-cell activation is shaped by factors beyond TCR-pMHC affinities, including the T-cell's differentiation state, its cross-reactivity to self-antigens~\cite{Altan-Bonnet2005-lm,Achar2022-tb}, and avidity effects from TCR/co-receptor cross-linking~\cite{Stone2009-ab}. Nonetheless, because  readouts of T-cell activiy (e.g., T-cell proliferation, and production of activation markers) are easier to obtain than SPR-based affinity measurements, we next asked whether HERMES scores, which are more attuned for predicting TCR-pMHC binding affinity, can serve as predictors of T-cell activities.}

As the first system, we examined T-cell responses to all single-point mutants of a native peptide, as measured in ref.~\cite{Luksza2022-gx} for different TCR-pMHC systems. Specifically, we make predictions for (i) three TCRs (TCRs~1-3) recognizing variants of a highly immunogenic  HLA-A*02:01-restricted  human cytomegalovirus (CMV) epitope NLVPMVATV (NLV), (ii) three TCRs (TCRs~4-6) recognizing variants of the  {weaker} HLA-A02:01-restricted melanoma gp100 self-antigen IMDQVPFSV,  and (iii) a TCR (TCR~7) recognizing variants of weakly immunogenic HLA-B*27:05-restricted pancreatic cancer neo-peptide GRLKALCQR, believed to have elicited an effective immune response in long-term cancer survivors~\cite{Luksza2022-gx}. T-cell activity was quantified by determining the half-maximal effective concentration (EC$_{50}$) from the dose-response curves of 4-1BB$^+$ CD8$^+$ T-cells across varying peptide concentrations; see Methods on details of inferring EC$_{50}$.

With the exception of TCR1 (where a TCR-pMHC template with HLA-A*02:01 and one peptide variant was already available), we relied on AlphaFold3 (AF3)~\cite{Abramson2024-jp} to model the remaining TCR-pMHC complexes; see Table~\AffActInfoTable~and Dataset~\AfInfoDataset~for template information and AF3 sequence inputs for TCRs~2-7.  {As shown in Figs.~\ref{Fig:3}A, and~\ActivityHeatmapsFigure,~HERMES-{\em fixed} peptide energy predictions (eq.~\ref{eq:pepEn}) correlate well with experimental EC$_{50}$ measurements for  the CMV peptide variants, reaching up to  Spearman correlation $\rho= 0.71$ for TCR2, and outperforming the alternative models. For TCR1 and TCR3, the performance of ProteinMPNN~\cite{Dauparas2022-fu} is worse but statistically indistinguishable from HERMES (Tables~\ref{table:spearmanr_affinity_and_activity},~\SpearmanrActivityTable, and Fig.~\ProteinMPNNScoringSchemesFigure).  Moreover, replacing the crystal structure for TCR1 with an AF3 model  yields comparable performance, underscoring HERMES's robustness to structural inputs so long as the structural models are of  high quality (Figs.~\AfExamplesFigure,~\AfScoringResultsFigure). Correlations are  lower for TCRs~4-6 targeting variants of the melanoma gp100 self-antigen, with HERMES achieving Spearman correlation  {$\rho=0.42$} for TCR6 as its best prediction in this class, with an accuracy comparable to ProteinMPNN~\cite{Dauparas2022-fu} (Fig.~\ref{Fig:3}B {and Tables~\ref{table:spearmanr_affinity_and_activity},~\SpearmanrActivityTable}).}
HERMES underperforms for TCR7 responding to the pancreatic cancer neo-peptide GRLKALCQR (Fig.~\ref{Fig:3}C){, for which only the ProteinMPNN's predictions show significant albeit weak correlation with data (Table~\ref{table:spearmanr_affinity_and_activity})}. The reduced accuracy of HERMES for cancer epitopes compared to the variants of the viral CMV epitope could reflect the impact of antagonism from self-antigens similar to tumor epitopes, hindering T-cell responses~\cite{Altan-Bonnet2005-lm,Achar2022-tb}---though additional experiments are needed to confirm this. 

{Sequence-based models (TAPIR~\cite{Fast2023-tk}, NetTCR~\cite{Jensen2024-xs}, and TULIP~\cite{Meynard-Piganeau2024-po}) are outperformed by structure-based models tested here in all cases except for TCR4, which has the lowest AF3 confidence at the TCR-pMHC interface (i.e., a lower confidence in its fold); see    Table~\AfMetricsTable~and \AN{Fig.~\AfScoringResultsFigure~for the AF3 confidence metrics for all systems}. Here, even the basic substitution models (BLOSUM62~\cite{Henikoff1992-jq} and the TCR-specific  matrices from ref.~\cite{Luksza2022-gx}) outperform the structure-based predictors \AN{(Tables~\ref{table:spearmanr_affinity_and_activity},~\SpearmanrActivityTable). This suggests that structural approaches hinge on a reliable structure as input, though more data is needed to draw definitive conclusions (Fig.~\AfScoringResultsFigure).}
}

Next, we examined the  {therapeutic T-cell responses}  to all single-point mutants of the HLA-A*02:01-restricted p53-derived {neoepitope} p53$^{\rm R175H}$, HMTEVVRHC, measured in ref.~\cite{Hsiue2021-ag}. This T-cell therapy approach used a designed bi-specific Fab antibody that binds both to the TCR-CD3 complex on T-cells and to the variants of the p53$^{\rm R175H}$-MHC complex on tumor cells (Table~\AffActInfoTable). A strong binding to an {epitope} activates the engaged T-cell, and the extent of such activity was measured by the production of IFN-$\gamma$.  The resulting HERMES-{\em fixed} predicted peptide energies (eq.~\ref{eq:pepEn}) show a {Spearman $\rho=0.46$}  with the experimentally measured IFN-$\gamma$ production (Figs.~\ref{Fig:3}D,~\ActivityHeatmapsFigure).  {ProteinMPNN achieves a slightly  higher correlation (Spearman $\rho=0.55$), but this improvement over   HERMES is not statistically significant  (Fisher's z-test p-value $>0.05$); see Table~\ref{table:spearmanr_affinity_and_activity}. Lastly, we found that replacing the crystal structure with the AF3 model  substantially degrades HERMES and ProteinMPNN performances (Fig.~\AfScoringResultsFigure, Tables~\ref{table:spearmanr_affinity_and_activity},~\SpearmanrActivityTable), as the AF3 model consistently mis-docks the bispecific antibody on the pMHC complex (Table~\AfMetricsTable~and Fig.~\AfExamplesFigure).
}

\AN{Across the limited number of systems examined, our analyses indicate that--when a suitable structural template is available--HERMES can estimate epitope mutational effects at the pMHC-TCR interface and, in the H2-scDb case, at a Fab-pMHC interface. In this benchmark, ProteinMPNN performs comparably to HERMES, indicating that inverse-folding algorithms not explicitly trained for this task may nonetheless capture the impact of peptide mutations on T-cell responses. However, definitive conclusions about the generalizability of these methods to  neoepitopes will require larger datasets with broader mutational coverage across peptide variants. Importantly, predictive performance for both approaches depends on access to a high-quality structure (experimental or computational), which limits applicability in systems without suitable structural models (Fig.~\AfScoringResultsFigure); }
see SI and Tables~\ref{table:spearmanr_affinity_and_activity}~and~\SpearmanrActivityTable~for a detailed benchmark of different approaches, including different HERMES models.

\begin{figure*}[th!]
\centering
\includegraphics[width=0.63\textwidth]{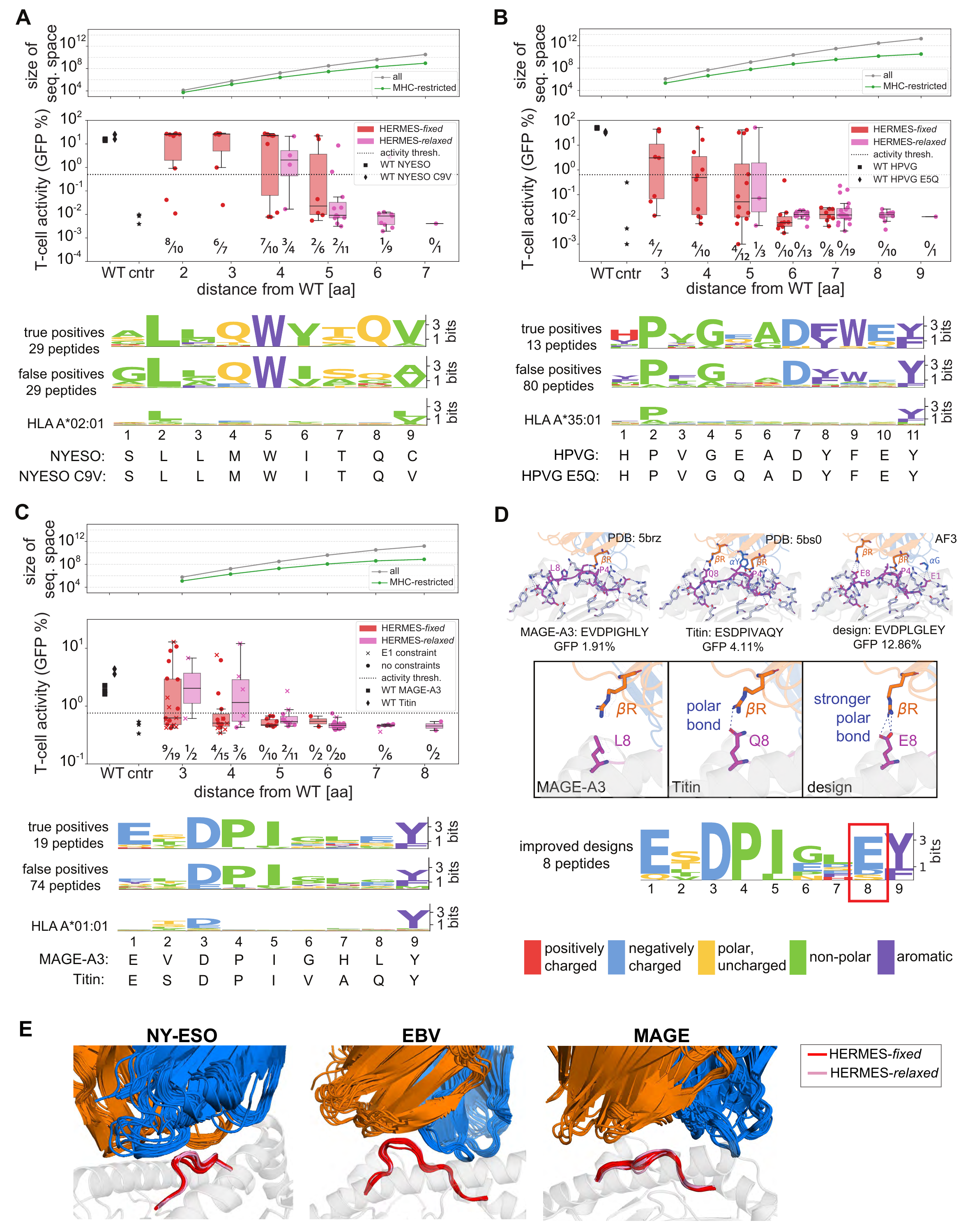}
\caption{
\textbf{{De novo} peptide design to elicit a T-cell response.} {\bf (A-C)} Validation results are shown for designed peptides built upon the wild-type templates from  (A)  NY-ESO peptide against  TCR 1G4 and HLA-A*02:01 (PDB IDs: 2BNR, 2BNQ~\cite{Chen2005-jk}), (B) EBV epitopes against  the TK3 TCR and the HLA-B*35:01 (PDB IDs: 3MV7~\cite{Gras2010-xp}, 4PRP~\cite{Liu2014-cz}), and (C) MAGE-A3 and Titin epitopes against an engineered TCR and HLA-A*01:01 (PDB IDs: 5BRZ, 5BS0~\cite{Raman2016-ew}).  Designs were validated using Jurkat cells  {with endogenous NFAT-eGFP reporter} expressing the paired TCR, interacting with peptide presented by the MHC in each system.  The reported T-cell activities (percentage GFP-positive) are averaged over three replicates. The box plots show T-cell activity for designed peptides at different Hamming distances from the wild-type; the black line is the median, the box spans the 25-75\% quantiles, and dots indicate individual data points. Red and pink boxes/markers  correspond to designs with HERMES-{\em fixed} and HERMES-{\em relaxed} models, respectively; square markers denote wild-type peptides, and stars are negative controls (no peptide). The dashed line marks the activation threshold, and numbers below each bar show the relative number of successful designs out of total designs. The top plot in each panel shows the size of the  peptide sequence space at different distances from the wild-types, with the gray line indicating the  complete sequence space, and the green indicating the typical number of sequences likely to be presented by the MHC of each system; see SI for details.  The bottom shows sequence motifs as PWMs for (i) true-positive designs that induced T-cell activation, (ii) false positives that did not, and (iii) the MHC presentation motif from the MHC Motif Atlas~\cite{Tadros2023-ij}. {\bf (D)} 3D structures of TCR–pMHC complexes focused on the peptide are shown for MAGE-A3 (PDB ID: 5BRZ), Titin (PDB ID: 5BS0), and the highest-activity design in (C) modeled by AF3~\cite{Abramson2024-jp}. The enhanced  activity for Titin versus MAGE-A3, and for the designed peptide versus both wild-types follows from the formation of  polar bonds between position 8 of the peptide (magenta) and the CDR3 loop of TCR$\beta$ (orange). The PWM highlights the prominence of E$_8$ among the eight designs with improved T-cell activity in (C). {{\bf (E)} Structures predicted with AF3 for the activity-inducing designed peptides (above the dashed lines in A-C)  are displayed in complex with their cognate TCR-MHC's. For each system, designs generated by {\em HERMES-fixed} (red) and {\em HERMES-relaxed} (pink) are overlaid after aligning the MHC molecule (in gray). Slight shifts are seen in the backbone conformations of the peptides and the TRA (blue) and TRB (orange) loops.}
\label{Fig:4}}
\end{figure*}

\subsection*{\bf Design of novel peptides reactive to a TCR-MHC complex} 
\AN{TCR-engineered T-cell therapies are designed to target defined neoantigens, but specificity remains a central challenge. Engineered high-affinity TCRs can cross-react with untested self-peptides, causing severe off-target toxicity; in a MAGE-A3 directed trial, recognition of a Titin-derived peptide led to fatal cardiac events~\cite{Linette2013-tc}. Therefore, developing approaches to predict the distribution of peptides recognized by engineered TCRs could enable more stringent preclinical screening and help mitigate off-target risk in patients.}

\AN{To map the peptide-recognition landscape of TCRs,} we present a structure-guided pipeline for designing immunogenic peptides predicted to elicit strong TCR recognition when presented by MHC class I. Leveraging HERMES's ability to predict T-cell responses to peptides presented by MHC-I, our approach begins with an existing template structure--experimentally or computationally resolved---of a TCR-MHC in complex with at least one  peptide variant, which we refer to as the wild-type peptide. We then generate candidate peptides using two strategies, with different degrees of complexity; see Fig.~\ref{Fig:1}C for a schematic description of the two pipelines.

In the basic pipeline (HERMES-{\em fixed}), we sequentially mask the atoms of each amino acid along the peptide, one at a time. 
Using the HERMES model, we predict the probability of each amino acid type for a given residue, based on the local structural environment within 10~\AA\, of the masked residue. Repeating this procedure for every amino acid  along the peptide yields a position weight matrix (PWM) that represents the probabilities of different amino acids at each peptide position. Peptides are then sampled by drawing from this PWM. It should be noted that HERMES-{\em fixed} retains the structural fingerprints of the original peptide, as it does not account for local structural changes induced by amino acid substitutions;  see Fig.~\ref{Fig:1}C and~\HermesDesignsEnergiesFigure A, and SI for more details.

To incorporate structural flexibility and relaxation following amino acid substitutions, we introduce the design pipeline of HERMES-{\em relaxed}, using simulated annealing and Markov chain Monte Carlo (MCMC) sampling. We begin with the template TCR-MHC structure and a completely random peptide sequence, which is packed {\em  in silico} within the structure using PyRosetta's packing functionality, followed by the Relax protocol~\cite{Chaudhury2010-mo}. During relaxation, both side-chain and backbone atoms of the peptide are allowed to move, while only the side-chain atoms of the MHC and TCR chains within 10~\AA~of the peptide are flexible. With this relaxed structure, we apply the HERMES-{\em fixed} method to generate a PWM and sample a new peptide, which is then  packed and relaxed to form a new structure. We use MCMC to  determine whether to accept the newly sampled peptide by comparing its HERMES peptide energy (favorability), evaluated in its relaxed conformation, to that of the original peptide (eq.~\ref{eq:pepEn}). We then iteratively perform this procedure, incrementally reducing the MCMC temperature at each step, until the results converge; see Fig.s~\ref{Fig:1}C,~\HermesDesignsEnergiesFigure B-C,~and SI for more details.

The peptides designed by both approaches are then filtered by TCRdock~\cite{Bradley2023-rl} or  {AlphaFold3}~\cite{Abramson2024-jp}, using the Predicted Alignment Error (PAE) between TCR and pMHC interfaces, to assure folding and binding of the resulting TCR-pMHC structures. For each system, we define a specific PAE threshold close to the TCRDock PAE of the system's TCR-MHC in complex with the native (wild-type) peptide,  which is known a priori to activate T-cells. As a result, the PAE thresholds vary slightly across different systems; see SI for details.

We tested the accuracy of our peptide design pipeline in three systems, with the structural  templates  taken from (i) the cancer-testis antigen NY-ESO-1 in complex with the TCR 1G4 and HLA-A*02:01, (ii) a peptide derived from the Epstein-Barr virus (EBV) in complex with the TK3 TCR and the HLA-B*35:01, and (iii) the immuno-therapeutic target MAGE-A3 in complex with an engineered TCR and HLA-A*01:01. For brevity, we refer to these three systems as NY-ESO, EBV and MAGE, respectively; See Table~\DesignInfoTable~for information on the structure templates used in each case. {These systems were selected because each one has high-resolution crystal structures of the same TCR-MHC scaffold bound to at least two distinct peptide variants, and that they span a therapeutically relevant spectrum that includes both engineered (for MAGE and NY-ESO) and natural (for EBV) TCRs directed against viral as well as tumor antigens.} It should be noted that as we progressed from NY-ESO to EBV and MAGE, we chose to explore the sequence space farther from the wild-type peptides, leading to more challenging designs.

We validated our designs using Jurkat cells expressing  {a specific} TCR {and an endogenous} NFAT-eGFP reporter to indicate T-cell activation in response to different peptides. We tested the TCR-MHC specificity in all systems--NY-ESO, MAGE, EBV--under 96 different conditions, including the  {de novo} designed peptides,  positive controls (wild-type peptides from the template structures), and an unstimulated ``no peptide" control.  Peptides were presented on artificial APCs (aAPCs), expressing the specified HLA in each system. For NY-ESO designs, we determined the  {the percentage of GFP-positive cells} through both  flow cytometry  and fluorescence microscopy. Given the consistency between the two  experimental approaches (Fig.~\FlowMicroscopyFigure) we relied on fluorescence microscopy only  to measure GFP levels induced by different peptides for MAGE and EBV designs; see SI for experimental details and Dataset~\experimentalMeasurementsDataset~for data on these measurements.

We used two structural templates for the NY-ESO system: one containing the original 9-amino acid peptide SLLMWITQC, and another with  the more immunogenic  C9V substitution~\cite{Chen2000-vq,Chen2005-jk}. Our {de novo} designs differed from the closest wild-type peptide by 2 to 7 amino acids, with a TCRDock PAE~$\leq 5.5$. For this system only, we selected 35 negative designs (designed by HERMES, but with PAE $> 5.5$) for experimental testing: only one of these was a false negative, which interestingly, had a desirable AF3 PAE score (as low as the wild-type peptides); see Figs.~\DesignTCRDockAFScatterFigure,~\DesignExpNyesoFigure. For the 58 positive designed peptides (PAE~$\leq~5.5$), we achieved an overall design accuracy of 50\%, with GFP levels significantly higher than the negative control; see SI for details. The accuracy of predictions decreases as the sequence divergence of the  {de novo} designs from the wild-type peptides increases (Fig.~\ref{Fig:4}A,~\DesignExpNyesoFigure). HERMES-{\em fixed} achieved higher design accuracy of 70\% within its smaller designed sequence subspace of up to 5 substitutions from the wild-types, whereas HERMES-{\em relaxed} explored sequences further from the wild-type peptides (4-7 substitutions)  but with a reduced accuracy of 24\% (Fig.~\ref{Fig:4}A). The main difference between the true and false positive sequences was at position 8--from strongly preferring glutamine within the true positives, to having similar preferences among glutamine, glycine, arginine, or methionine within the false positives (Fig.~\ref{Fig:4}A). 

For the EBV system, we used two structural templates: one with the 11-amino acid peptide HPVGEADYFEY---commonly referred to as HPVG---derived from the viral latent nuclear protein EBNA-1, and the other with an epitope from a  wide-spread viral {variant}, with a glutamine at position 5~\cite{Liu2014-cz}. Unlike NY-ESO, the HPVG peptide has a more flexible conformation with a helical turn within the {MHC} binding pocket.  Our {de novo} designs differed from the closest wild-type peptide by 3 to 9 amino acids, with a TCRDock PAE~$\leq~5.1$. Across the 93 {de novo} designs, we achieved an overall accuracy of 14\%, associated with the peptides that  induced significant GFP levels  in T-cells.  The accuracy of predictions decreases with increasing the sequence divergence from the wild-types (Fig.~\ref{Fig:4}B,~\DesignExpEbvFigure). Notably, we achieved 40\% accuracy among the designs within 3-5  amino acid distance of one of the wild-type templates, while none of  the designs with larger than 5 amino acid distance were successful.  Lastly, we observe a strong preference for  {g}lutamic acid at position 10 of  the successful designs relative to the false positives (Fig.~\ref{Fig:4}B).

\begin{figure*}[th!]
\includegraphics[width=0.9\textwidth]{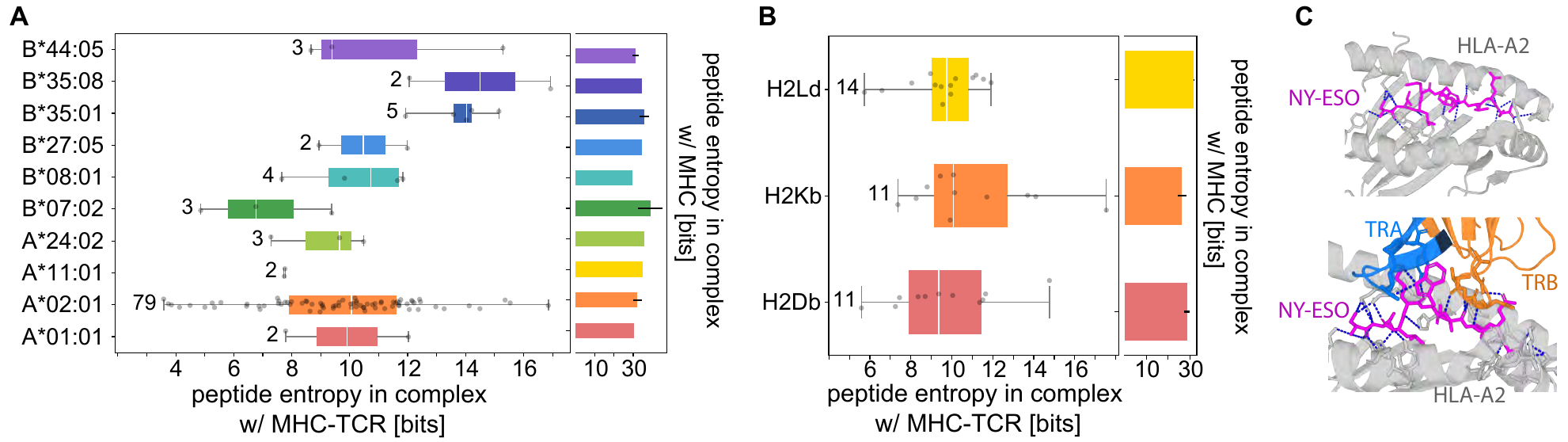}
\caption{
{\bf Evaluating T-cell specificity from {de novo} designed peptides.} {\bf (A)} 
T-cell specificity is measured by peptide entropy, which is computed from the HERMES-{\em fixed} predicted amino acid likelihood matrix, starting from a TCR-pMHC structural template. Each point on the plot corresponds to the entropy estimate from a distinct TCR-pMHC template, grouped by their MHC allele. The box plots display the distribution of peptide entropies for TCR-pMHC templates sharing the same MHC allele, with a white line marking the median and the box spanning the 25\% to 75\% quantiles. The number on the side of each box indicates the number of distinct structural templates we used for that MHC class; see Table~\AllPdbsTable~for details  on these structures. The side plot shows peptide degeneracy for MHC presentation alone, evaluated as the entropy of the peptide position weight matrices gathered from the MHC Motif Atlas~\cite{Tadros2023-ij}. The error bars indicate the standard deviation of peptide degeneracies across  motifs of different lengths, associated with a given MHC allele. {\bf (B)} Similar to (A) but for  degeneracy  of peptides interacting with TCRs and MHCs in mice. {\bf (C)} Polar bonds (blue dashed lines) between the NY-ESO peptide (magenta) and HLA-A*02:01 (gray) is shown at the top (PDB ID: 1S9W~\cite{Webb2004-am}). The additional polar bonds formed between the peptide and TRA (blue) and TRB (orange) of an interacting TCR are shown at the bottom (PDB ID: 2BNR~\cite{Chen2005-jk}). These additional polar bonds limit the degeneracy of the  peptides that can interact with a TCR-MHC complex compared to those  presented by the same MHC.
\label{Fig:5}
}
\end{figure*}

For the MAGE system, we used two structural templates: one with the 9-amino acid MAGE-A3 peptide EVDPIGHLY  and another with the Titin-derived self-peptide ESDPIVAQY, which is expressed in cardiomyocytes~\cite{Raman2016-ew}. Our {de novo} designs differed from the closest wild-type peptide by 3 to 8 amino acids, with a TCRDock PAE~$\leq~5.2$ (Fig.~\ref{Fig:4}C). From the resulting 93 {de novo} designs,  20\% activated T-cells with significant levels of GFP expression. Similar to the other systems, the accuracy of predictions decreases with increasing  the sequence divergence  from the wild-types (Fig.~\ref{Fig:4}C). We achieved 30\% accuracy among the designs with 3-5 to amino acid distance, while  none of the designs with larger than 5 amino acid distance from the wild-types were successful.

Fig.~\ref{Fig:4}C shows differences in amino acid  composition between  successful and failed designs. Alanine and glycine scanning experiments have previously suggested that MAGE-A3 peptide variants with an E-DPI-\,-\,-Y motif  activate  T-cells in this system~\cite{Cameron2013-vw}. Consistent with these findings, the fraction of T-cell activating variants in our designs increases as their sequences more closely resemble this motif (Fig.~\DesignExpMageFigure).

In a subset of designs using the MAGE-A3 template, we fixed the glutamic acid at position 1 (E$_1$), which is part of the E-DPI---Y pattern. In both MAGE-A3 and Titin E$_1$ is within 4~\AA~of the MHC and forms strong polar interactions with it~(Fig. 4D). Fixing  E$_1$ improved the design accuracy to $1/3$, compared to $1/9$ when this amino acid was not constrained. Even without this constraint in the design protocol, E$_1$ appeared in $2/3$ of successful designs, and in none of the unsuccessful ones (Fig.~\DesignExpMageFigure). Notably, among the more distant designs made with HERMES-{\em relaxed}, only those containing a fixed E$_1$ were able to activate T-cells (Fig.~\ref{Fig:4}C). This suggests that constraining essential amino acids can enable broader exploration of sequence space at other positions, possibly due to epistatic interactions. However, further investigation is needed to confirm this. 

{Several} of our designs outperformed the wild-type peptides in activating T-cells: eight surpassed Titin’s GFP levels, and eleven exceeded MAGE-A3’s (Fig.~\ref{Fig:4}C). These improved designs strongly favor glutamic acid (E) at position 8 of the peptide, where MAGE-A3 has leucine (L), and Titin has glutamine (Q) (Fig.~\ref{Fig:4}D). In Titin, the glutamine forms a polar bond with an arginine (R) in the interacting TCR$\beta$, and it is likely responsible for increasing Titin's activity relative to MAGE-A3's. The glutamic acid substitution in our designs further strengthens this polar bond (Fig.~\ref{Fig:4}D), resulting in a stronger T-cell response. {For NY-ESO and EBV, the wild-type peptides were strongly immunogenic and induced high levels of GFP, and many of our designs induced comparable T-cell responses, but did not improve upon those (Fig.~\ref{Fig:4})}.

{To gauge the impact of using AF3 structures as design templates, we reran the HERMES-{\em fixed}  protocol with AF3 predicted structures (with and without template guidance), and obtained peptide PWMs that were almost identical to those derived from using crystal structures  (SI and Fig.~\DesignPWMsWithAf). An interesting exception is the MAGE system, where using the AF3 template introduced a strong  preference for glutamic acid at positing 1 of the peptide (Fig.~\DesignPWMsWithAf C), mirroring the pattern observed among the designed peptides that induced strong T-cell response (Fig.~\ref{Fig:4}D).}

Although no direct experimental comparisons were performed, computational evaluations using TCRdock PAE and  presentation score by MHC-I alleles indicate that HERMES designs outperform or match those generated by ProteinMPNN~\cite{Dauparas2022-fu}, the corresponding MHC position weight matrix~\cite{Lundegaard2008-ao, Lundegaard2008-wz, Andreatta2016-uh,Tadros2023-ij}, or the BLOSUM substitution matrix~\cite{Henikoff1992-jq} (SI and Figs.~\DesignTcrdockNyesoFigure,~\DesignTcrdockEbvFigure,~\DesignTcrdockMageFigure). A notable exception is in the MAGE system, where ProteinMPNN designs generally have more favorable PAE scores, albeit with a lower proportion of designs passing the antigen presentation filter by the MHC molecule (Fig.~\DesignTcrdockMageFigure).

In all the three systems we used the PAE values from TCRdock and AF3 as filtering criteria to select designs for experimental validation (see SI). The PAE values reported by the structure prediction algorithms have been shown to be noisy indicators of protein folding fidelity and design quality~\cite{Roney2022-am}. While larger experimental libraries are needed to draw definitive conclusions, it appears that PAEs from TCRdock and AF3 have some complementary predictive strengths ({Figs.~\DesignTCRDockAFScatterFigure,~\DesignAUCFigure}), and therefore, combining their information could lead to a more robust  design decisions (SI). {Both of these scores   correlate only weakly
with HERMES  (Figs.~\DesignTCRDockAFScatterFigure,~\DesignTcrdockNyesoFigure,~\DesignTcrdockEbvFigure,~\DesignTcrdockMageFigure), so they can provide complementary design filters. Furthermore, designed peptides with higher (more favorable) HERMES energies  more often activated T-cells (Fig.~\DesignAUCFigure), suggesting that a more restrictive sampling  (e.g., with a lower MCMC temperature for  HERMES-{\em relaxed} protocol) could  yield a higher success rate, albeit at the expense of peptide diversity.}

Overall, HERMES  demonstrates a potential for designing highly immunogenic  peptides up to five amino acid substitutions away from the wild-type--a task that would otherwise require exploring a vast sequence space of roughly $10^8$ to  $10^{10}$ possibilities for a 9-residue to an 11-residue peptide (Fig.~\ref{Fig:4}A-C). {This capability raises an important question: to what extent do  these substantial  changes in sequence alter the binding geometry  of the peptide inside the MHC-TCR cleft? To  this end, we used AF3 to model the TCR-pMHC interfaces for the experimentally  validated immunogenic designs from the three systems. Although we see slight variation in both peptide and TCR backbone conformations,  overall, we found that our designed immunogenic peptides adopt conformations very close to their wild‑type templates, even when those designs were generated with HERMES‑{\em relaxed} (Fig.~\ref{Fig:4}E). We attribute this structural convergence to two potential factors: (i) residual limitations in HERMES peptide‑backbone sampling strategy, and (ii) an inherent AF3 bias toward conformations it encountered during training. Given that TCRs can cross-react with peptides of markedly different  conformational poses~\cite{Riley2018-zi}, developing methods that can better explore peptide conformational space within the TCR-MHC groove remains a compelling direction for future research.}

\subsection*{Structural basis for TCR specificity}
TCRs exhibit substantial degeneracy in their recognition, with some autoimmune T-cells experimentally shown to recognize over one million distinct peptides~\cite{Wooldridge2012-zq}. However, the full extent of TCR degeneracy for typical receptors remains unclear, largely due to the limitations in high-throughput experimental assays for TCR recognition of many peptides presented on different MHCs.

Our computational peptide design framework helps address this limitation, at least for the TCR-MHC complexes with known structural templates. Specifically, for a given TCR-MHC pair, we can use HERMES-{\em fixed} protocol to generate a peptide position weight matrix (PWM), representing the ensemble of peptides presented by the MHC and recognized by the TCR. We use the entropy of this PWM as a proxy for TCR degeneracy.

Fig.~\ref{Fig:5}A shows the entropy of these peptide distributions for 105 TCR-MHC pairs across 10 human MHC-I alleles; see Table~\AllPdbsTable~for details  on these structures. The median peptide entropy across these TCR-MHC pairs is approximately 10~bits, indicating that a typical TCR-MHC pair can recognize on the order of  $ \simeq 10^3$   peptides.  In contrast, examining the degeneracy of MHC-I presentation alone in humans--using the  peptide PWMs gathered from the MHC Motif Atlas~\cite{Tadros2023-ij}--yields a median entropy of 31 bits, implying that MHC-I molecules can  typically recognize $\simeq 2\times 10^{9}$ distinct peptides (Fig.~\ref{Fig:5}A). A similar analysis on 36 TCR-MHC pairs spanning three mouse MHC-I alleles shows comparable trends, with a MHC-I degeneracy on the order of  $\simeq5 \times 10^8$ peptides (29 bits), and the TCR-MHC degeneracy of  $\simeq 10^3$ {peptides} (10 bits); see Fig.~\ref{Fig:5}B, and Table~\AllPdbsTable~for details  on these structures. This pronounced reduction in entropy---observed by comparing the distribution of peptides recognized by MHC-I molecules alone to that recognized by TCR-MHC-I complexes---quantifies how TCRs constrain the accessible antigenic shape space, as illustrated by the polar interactions that they form with  peptides in Fig.~\ref{Fig:5}C. 

{As an alternative to MHC Motif Atlas~\cite{Tadros2023-ij}, we explored the ability of HERMES-{\em fixed} to model MHC degeneracy. When structural coverage is extensive (roughly $>30$ structures of distinct peptides for a given MHC allele), the motifs produced by HERMES align closely with those in the MHC Motif Atlas (SI, Fig.~\pMHCPWMsFigure, and Table~\pMHCPWMsTInfoable). With sparser structural coverage, HERMES underestimates the peptide diversity, likely because backbone conformations are insufficiently sampled. Given  the imbalanced distribution of structures  across MHC alleles,  we therefore continue to use MHC Motif Atlas~\cite{Tadros2023-ij} as our reference for computing MHC motif specificity.}

Because our analyses focus on interactions involving typical TCR and MHC molecules, our entropy-based degeneracy estimates are smaller than those from previous theoretical studies, which often relied on experimental observations from  autoimmune T-cells or limited synthetic peptide libraries~\cite{Sewell2012-du}. Moreover, the HERMES-{\em fixed} model generates PWMs based on peptides whose backbone conformations match those of the wild-type template, likely underestimating the true degeneracy of TCR-MHC complexes (Fig.~\ref{Fig:5}). Although HERMES-{\em relaxed} can partially address this issue, its designs are only reliable within roughly five amino acids of the native peptides {with similar backbone conformations (Fig.~\ref{Fig:4})}, preventing a full exploration of the possible peptide poses for a given TCR-MHC complex. Nonetheless, our estimates can serve as informative lower bounds on typical MHC-I and TCR degeneracies.

\section*{Discussion}
 In this work, we introduced a structure-based machine learning model to predict TCR-pMHC interactions and to design novel reactive peptides for TCR-MHC complexes. Our approach builds on HERMES, a physically motivated equivariant neural network model that we previously developed~\cite{Visani2024-sh,Pun2024-qz}. Trained on the protein universe, HERMES predicts amino acid propensities based on their atomic environments in protein structures, and has been shown to implicitly learn biophysical interatomic interactions in proteins~\cite{Pun2024-qz}, enabling it to make highly generalizable predictions. This generalizability is particularly important for TCR-pMHC complexes, where only a modest number of experimentally resolved structures are available, and computational algorithms often struggle to accurately predict new structures~\cite{Bradley2023-rl,Yin2023-zm}.

We found that HERMES-predicted peptide energies closely tracked both TCR-pMHC binding affinities and peptide-induced T-cell activities across diverse systems, including variants from a CMV-derived viral epitope~\cite{Luksza2022-gx}, the NY-ESO cancer-testis antigen~\cite{Aleksic2010-wz,Pettmann2021-kl}, and a melanoma self-antigen~\cite{Luksza2022-gx}.  Notably, HERMES also accurately predicted the activity of p53-derived neoepitope variants in the context of a bispecific antibody-mediated T-cell therapy~\cite{Hsiue2021-ag}. \AN{This result suggests the algorithm's potential applicability:} given a reliable structural template close to the system of interest, HERMES can  predict peptide-induced T-cell responses, whether through direct TCR interactions or antibody-mediated mechanisms. \AN{However, drawing more definitive conclusions about the generalizability of HERMES would require examining a larger datasets of TCRs interacting with pMHC variants.}

Based on the HERMES predictions for TCR-pMHC reactivities, we presented a computational protocol  for  {de novo} design of immunogenic peptides that can elicit robust T-cell responses. Starting from the structural template of a TCR-MHC with a native (wild-type) peptide, we developed an MCMC search guided by HERMES peptide energies to explore  the  vast sequence space of  roughly $10^8$ ($10^{10}$) possibilities for 9-residue (11-residue) peptides, to  identify immunogenic  candidates. Experimental validation across three systems---with native peptides from NY-ESO and MAGE cancer epitopes, plus a viral epitope from EBV---revealed that our designs reliably activated T-cells, even with  up to five substitutions from the native peptide used in the structural template. Notably, several MAGE-derived variants outperformed or matched the native peptide in T-cell activation, demonstrating the method’s potential to enhance peptide potency. 

This capability can be  relevant to personalized peptide-based cancer vaccines, which aim to  induce durable T-cell memory against a patient's  tumor antigens to reduce the risk of rebound~\cite{Sahin2018-fv,Vormehr2019-aw,Rojas2023-nx}. Despite their promise, the  success of cancer vaccines has been limited by low peptide immunogenicity  and poor uptake by antigen-presenting cells~\cite{Song2024-di}. Recent approaches have employed deep mutational scanning experiments to identify more immunogenic cancer  {neoantigens} against specific TCR candidates~\cite{Luksza2022-gx}. However, these experiments are often limited to scanning single amino acid mutations from the wild-type peptide due to the combinatorial explosion when exploring multiple mutations. Moreover, the precise TCRs responding to peptide vaccines in a patient are often unknown, and many immune responses are induced de novo~\cite{Vormehr2019-aw}. With the caveat of access to suitable structural templates,  HERMES could inform T-cell vaccine design by computationally generating diverse peptide libraries predicted to engage a broad set of candidate TCRs (identified independently), nominating vaccine epitopes for subsequent safety and efficacy testing.

Accurately modeling T-cell specificity remains challenging, in part due to the limited functional data on  {TCR-epitope} pairs, and the highly cross-reactive nature of {TCR-epitope} interactions. A design protocol like the one presented here, coupled with experimental validation, offers a promising path forward for future {studies}. By iteratively generating candidate peptide libraries, testing them experimentally, and refining the model with each iteration, it should be possible to   more efficiently explore the  high dimensional antigenic sequence space. This active learning strategy has a potential for revealing the ``shape space" of {TCR-epitope} interactions, with broad applicabilities for early disease diagnosis, and for design of targeted immunotherapies. Moreover, this approach can help mitigate immunotherapy-related toxicity by identifying and excluding potential self-reactivities--beyond single-point mutations in the target epitope--such as the off-target Titin reactivity observed in an engineered T-cell therapy against MAGE-A3  that led to a severe cardiac toxicity~\cite{Linette2013-tc}.

Although our findings highlight the promise of structure-based models for immune recognition, a key limitation remains the scarcity of high-quality TCR-pMHC structural templates, despite advances from tools like AlphaFold3~\cite{Abramson2024-jp} and TCRDock~\cite{Bradley2023-rl}.  On the other hand, while TCR sequences are more widely available,  sequence-based models often lack generalizability because existing paired TCR-pMHC data are skewed toward just a few peptides~\cite{Shugay2018-nt,Vita2019-qn}. A promising direction for future work lies in developing multi-modal flexible methods that integrate structural and sequence-based models, leveraging both the depth and generalizability offered by structural information and the scalability of sequence data. 

{TCRs can cross-react with  diverse peptides in multiple orientations and conformation~\cite{Riley2018-zi}. To reflect this flexibility, we  sampled peptide conformational space (i) implicitly, by  applying noise to the coordinates during training, and (ii) explicitly, by locally relaxing the atomic environment after each amino acid substitution. Even so,  AlphaFold3 predicts that our experimentally  validated immunogenic designed peptides retain similar structural poses as their wild-type templates. This structural convergence may partly stem from  AlphaFold3's bias  toward structures represented  in its training set, but it also exposes limitations in our current sampling strategy. Future work should therefore aim to co-design  peptide sequence, peptide backbone geometry, together with the CDR loop conformations of the interacting TCR, enabling a more comprehensive exploration of the peptide-TCR conformational landscape.}

Although our study primarily examines  {MHC class I}-restricted peptides, the broad applicability of structure-based modeling suggests that similar methods could be adapted for the more complex TCR-peptide-MHCII interactions, thereby offering new insights into CD4+ T-cell responses. Moreover, incorporating additional immunological factors---such as TCR clustering, co-receptor interactions, cross-reactivity with self-antigens, and functional immune profiling---can further enhance the accuracy of T-cell response predictions and improve the resulting {de novo} peptide desig pipeline.

\section{Data and code availability}
 Data and code for all the analyses can be found in the  Github repository:~\url{https://github.com/StatPhysBio/tcr_antigen_design}. Code for the original HERMES~\cite{Visani2024-sh} model can be found in the Github repository:~\url{https://github.com/StatPhysBio/hermes}.

\section*{Acknowledgment}
We thank Mikhail V. Pogorelyy for valuable comments and help with CellCyte. This work has been supported by the National Institutes of Health MIRA awards R35~GM142795 (AN, GV, MNP) and R35~GM141457 (PB), the CAREER award from the National Science Foundation grant 2045054  (AN), the National Institutes of Health R01 award AI136514 (PT and PB), the Royalty Research Fund from the University of Washington no. A153352 (AN, MNP),  the Allen School Computer Science \& Engineering Research Fellowship from the Paul G. Allen School of Computer Science \& Engineering at the University of Washington (GV), the Microsoft Azure award from the eScience institute at the University of Washington (AN, GV, MNP), the TIRTL Bluesky Initiative at St. Jude Children’s Research Hospital  (PT), and the  ALSAC at St. Jude Children’s Research Hospital (PT and AAM). This work is also supported, in part, through the Departments of Physics and Computer Science and Engineering, and the College of Arts and Sciences at the University of Washington (AN, GV, MNP). This work was performed in part during the  2023 summer workshop ``Statistical Physics and Adaptive Immunity" at the Aspen Center for Physics, which is supported by National Science Foundation grant PHY-2210452. This work also benefited from discussions  during  the  2024 program ``Interactions and Co-evolution between Viruses and Immune Systems" at the Kavli Institute for theoretical physics (KITP), which is supported by  National Science Foundation grants PHY-2309135, and PHY-2309135, and the Gordon and Betty Moore Foundation Grant No. 2919.02.
 
\section*{Competing Interest Statement}
PT is on the Scientific Advisory Board of Immunoscape and Shennon Bio, has received research support and personal fees from Elevate Bio, and consulted for 10X Genomics, Illumina, Pfizer, Cytoagents, Sanofi, Merck, and JNJ. \\\\


%

\end{document}


\counterwithin{table}{section}
\setcounter{table}{0}
\renewcommand{\thetable}{S\arabic{table}}

\counterwithin{figure}{section}
\setcounter{figure}{0}
\renewcommand{\thefigure}{S\arabic{figure}}

\counterwithin{equation}{section}
\setcounter{equation}{0}
\renewcommand{\theequation}{S\arabic{equation}}

\counterwithin{algocf}{section}
\setcounter{algocf}{0}
\renewcommand{\thealgocf}{S\arabic{algocf}}

\noindent{\bf \Large 
Supplementary Information\\}
\vspace{-1ex}

\noindent {\bf \large T-cell receptor specificity landscape revealed through de-novo peptide design}\\
\vspace{-2ex}

\noindent Gian Marco Visani, Michael N. Pun,  {Anastasia A. Minervina}, Philip Bradley, Paul Thomas, and Armita Nourmohammad\\\\

 \vspace{0.5cm}

\noindent {\bf Data and code availability.} Data and code for all the analyses can be found in the  Github repository:~\url{https://github.com/StatPhysBio/tcr_antigen_design}. Code for the original HERMES~\cite{Visani2024-sh} model can be found in the Github repository:~\url{https://github.com/StatPhysBio/hermes}.

\section{HERMES neural network model for amino acid propensities}
 HERMES is a self-supervised structure-based machine learning model for proteins~\cite{Visani2024-sh,Pun2024-qz}. HERMES has a 3D rotationally equivariant architecture and is trained to predict a residue's amino acid identity from its surrounding atomic environment within a 10 \AA~ of the central residue's C-$\alpha$ in the 3D structure.  The input to the neural network is a point cloud of atoms from a given structural neighborhood, in which all atoms associated with the central residue are masked. These point clouds are featurized by atom types, including computationally-added hydrogens, partial  charge, and Solvent Accessible Surface Area (SASA). The point clouds  are first projected onto the orthonormal Zernike Fourier basis, centered at the position of the (masked) central residue's C-$\alpha$ (Fig.~1A). The resulting \textit{holographic} projections are fed to a stack of SO(3)-equivariant layers, the output of which is SO(3)-invariant, and is then passed through a multi-layer perceptron (MLP) to generate the desired predictions. The MLP's outputs are termed pseudo-energies, associated with the 20 different possible amino acids that can be placed at the masked residue, and act as logits within a softmax function to compute probabilities of each amino acid.

Each HERMES model is an ensemble of 10 individually-trained architectures. In this work, we used HERMES models trained on structural data with different levels of added noise:  0.00 noise (the original PDB structure), and 0.50 noise, which is obtained by adding Gaussian noise to the 3D coordinates of the original structure, with standard deviation of 0.50~\AA. We only use models that were trained on the wild-type amino acid classification task~\cite{Visani2024-sh}. Models were pre-trained on structures from 30\% similarity split of ProteinNet's CASP12 set~\cite{alquraishi_proteinnet_2019}, resulting in around $\sim$10k training structures.

We refer the reader to refs.~\cite{Pun2024-qz, Visani2024-sh} for further details on the architecture, training procedure, training data preparation, and the mathematical introduction to SO(3)-equivariant models in Fourier space.

\section{TCR-pMHC binding affinity and T-cell activity prediction}
\label{sec:aff_act}

\noindent \textbf{Peptide energy calculation with HERMES.} To predict the affinity of TCR-pMHC complexes or the activity of TCRs induced by their interactions with a pMHC complex, we introduce the predicted energy for a peptide $\sigma$, given its surrounding TCR and MHC,

\EQ
E_{\text{peptide}}(\sigma; \text{TCR, MHC}) = \sum_{i=1}^\ell  E(\sigma_i ; \text{TCR, MHC},\sigma_{/i}).
\label{eq:pepEn}\EE

\noindent We assume that each peptide residue contributes linearly to the total energy by the amount  $E(\sigma_i ; \text{TCR, MHC}, \sigma_{/i})$, where $\sigma_i$ is the amino acid at position $i$ of the peptide. A residue's energy contribution $E$ is evaluated as its associated logit value from HERMES, taking in as input the atomic composition of the surrounding TCR, MHC and the rest of the peptide $\sigma_{/i}$ (Fig.~1A).

To compute peptide energies, HERMES operates on the structural composition of the protein complex. However, data on the impact of substitutions on the TCR-pMHC structure is  limited, and computational models often fail to capture subtle conformational changes in the structure due to amino acid substitutions in a peptide.  To address this issue, we adopt two protocols to estimate  energy for a diverse array of peptides for a given TCR-MHC complex:

\begin{itemize}
\item HERMES-{\em fixed}: In the simplest approach, we select the TCR-pMHC structure with matching TCR and MHC to our query and with the peptide sequence that most closely matches the peptide of interest and  directly input it into HERMES. The peptide energy is subsequently calculated as described in eq.~\ref{eq:pepEn} (Fig.~1B). This method does not modify the underlying structure, effectively assuming that any amino acid substitutions do not significantly alter the protein conformation.

\item HERMES-{\em relaxed}: Since amino acid substitutions can locally reshape the protein complex, we introduce a more involved protocol to account for these structural changes. We begin with the available crystal structure of the TCR-pMHC complex that closely matches our query as template. We then mutate the original peptide to the desired state and relax the structure using PyRosetta's cartesian\_ddg fastrelax protocol~\cite{Chaudhury2010-mo}. During relaxation, side chain and backbone atoms of the peptide are allowed to move, while only the side chain atoms of the MHC and TCR chains within a 10\AA\, of the peptide are flexible. We then calculate the peptide energy with HERMES (eq.\ref{eq:pepEn}), using the  relaxed structure  of the mutated variant as input (Fig.~1B). \GM{Since PyRosetta's relaxation procedure is stochastic in nature, we run 100 realizations of the relaxation procedure and average the peptide energies across them. 
 For ablation purposes, we also use an alternative protocol, whereby we use the HERMES peptide energy from the  conformation with the minimum (most favorable) Rosetta energy computed on the relaxed atoms; we call this protocol {\em relaxed-min\_energy}.}
\end{itemize}

\noindent \GM{Overall, we use four protocols for scoring peptides that differ in their restrained protocol (HERMES-{\em fixed} or HERMES-{\em relaxed}), and the amount of noise injected during training of the underlying HERMES model. Specifically, we train HERMES models with no added noise ($\sigma =0.00~\AA$) and another with moderate Gaussian noise with standard deviation $\sigma= 0.50~\AA$~applied to the atomic coordinates of the  protein structure data during training. We refer to the resulting peptide scoring protocols as HERMES-{\em fixed} 0.00, HERMES-{\em fixed} 0.50, HERMES-{\em relaxed} 0.00, HERMES-{\em relaxed} 0.50, specifying both the model and the noise amplitude.
}

\noindent{\bf Evaluation of EC$_{50}$ from dose response measurements against peptide libraries.} Ref.~\cite{Luksza2022-gx} presents dose response measurements for fraction of  4-1BB$^+$ CD8$^+$ T-cells across varying peptide concentrations.  These measurements are done for peptide libraries of single point mutants from a given wild-type peptide, across 7 different TCR-MHC systems: three TCRs recognizing variants of the highly immunogenic HLA-A02:01-restricted  human cytomegalovirus (CMV) epitope NLVPMVATV (NLV); three TCRs recognizing variants of the {weaker} HLA-A02:01-restricted melanoma gp100 self-antigen IMDQVPFSV; and one TCR recognizing variants of the {weakly immunogenic} HLA-B*27:05-restricted pancreatic cancer neo-peptide GRLKALCQR~\cite{Luksza2022-gx}.

The half-maximal effective concentration (EC$_{50}$) from the dose response curves of 4-1BB$^+$ CD8$^+$ T-cell fractions can quantify the level of T-cell activity in response to each peptide. We used our own fits to estimate the  EC$_{50}$ associated with T-cell responses. Specifically, we fitted a Hill function with background to the dose response measurements  for each peptide from ref.~\cite{Luksza2022-gx},
\EQ
f(x) = \frac{(A-A_0) x^n}{x^n + K^n}+ A_0
\label{eq:EC50}\EE
where $f(x)$ is the fraction of 4-1BB$^+$ CD8$^+$ T-cells  pulsed by a peptide at concentration $x$, $A$ is the response amplitude,  $A_0$ is the background activation, $n$ is the Hill coefficient, and $K$ is the half-maximal effective concentration EC$_{50}$. Similar to ref.~\cite{Luksza2022-gx}, we  regularized our fitted Hill functions, whereby the deviation of the amplitude $A$ and the Hill coefficient $n$ from 1, and the background activation $A_0$ from 0 are penalized. The code to fit these curves is available in our github repository.\\

\section{De-novo peptide design}
\label{sec:peptide_design}

\noindent {\bf Peptide design algorithm.}
Given a TCR and an MHC, we use HERMES to design de novo peptides that can form a stable TCR-pMHC complex, and potentially, elicit a T-cell response. Our approach starts with an existing template TCR-pMHC structure. We refer to the peptide in the template as wild-type; see Table~\ref{table:design_info} for the templates we use in our analyses. Similar to evaluating the peptide energy, we use two pipelines to sample novel peptides:

\begin{itemize}
\item HERMES-{\em fixed}: This is our basic pipeline, in which we  generate novel peptide sequences by conditioning on the fixed TCR-pMHC template structure.
Specifically, we sequentially mask the atoms of each amino acid along the peptide and then  use  HERMES  to compute the probability $P(\sigma_i|{\bf x}_i)$ of different amino acid types $\sigma_i$ at the masked residue $i$, conditioned on the surrounding atomic neighborhood  ${\bf x}_i$, within a 10~\AA \, radius of the masked  residue's C-$\alpha$. The neighborhood ${\bf x}_i$ can include atoms from the TCR, the MHC, and the other amino acids in the peptide. Repeating this procedure  for every residue along the peptide allows us to construct a Position Weight Matrix (PWM) that represents the probabilities of different  amino acids at each peptide position, conditioned on the surrounding structural neighborhood. We generate peptides using multinomial sampling from the PWM; see the schematic in Fig.~1C.

\item HERMES-{\em relaxed}: This pipeline incorporates simulated annealing and Markov chain Monte Carlo (MCMC) sampling to account for structural changes resulting from peptide substitutions. We start with the template TCR-MHC structure and an initial random peptide sequence $\sigma^{(0)}$, which is {\em in-silico} packed via PyRosetta, followed by the Relax protocol, yielding a relaxed structure ${\bf x}^{(0)}$.  During relaxation, both side chain and backbone atoms of the peptide are allowed to move, while only the side chain atoms of the MHC and TCR  within 10~\AA~of the peptide are flexible.  Following the HERMES-{\em fixed} protocol, we then evaluate the PWM associated with this relaxed structure and use it to generate a new peptide sequence, $\sigma^{(1)}$. This sequence is again packed and relaxed, producing a new structure, ${\bf x}^{(1)}$. Using eq.~\ref{eq:pepEn}, we evaluate the  energies of the peptides in their respective (relaxed) structures as $E(\sigma^{(0)} ; {\bf x}^{(0)})$ and $E(\sigma^{(1)} ; {\bf x}^{(1)})$, and apply the Metropolis-Hastings criterion to  accept the transition from $\sigma^{(0)}$ to $\sigma^{(1)}$ with probability:
\EQ P_{\sigma^{(0)}\to \sigma^{(1)}} = \text{min}\left(1, e^{- \frac{ E(\sigma^{(1)} ; {\bf x}^{(1)})- E(\sigma^{(0)} ; {\bf x}^{(0)})}{T}}\right)\EE
where $T$ is the temperature. This process is iterated, incrementally reducing $T$ at each step, until convergence. The peptide obtained at the final iteration represents the protocol’s de novo design. Repeating this procedure many times produces a diverse set of peptides spanning various distances from the wild-type sequence; see the schematic in Fig.~1C and Algorithm~\ref{alg:hermes_relaxed_design} for more details, and the the convergence curves in Fig.~\ref{fig:joint_design_distributions}C,F,I.\\
\vspace{1cm}

{
\begin{algorithm}[H]
\caption{HERMES-{\em relaxed} protocol for generating TCR-MHC-specific peptide sequences. For brevity, we define ${\bf x}_i \equiv (\text{TCR}, \text{MHC}, \sigma_{/i})$ and drop the arguments of $E_{\text{peptide}}$.}
\label{alg:hermes_relaxed_design}
\textbf{Input:} TCR-pMHC structure ${\bf x}$, maximum number of iterations N\;
\textbf{Output:} peptide sequence $\sigma^{(N)}$\;
$T_0 \gets 10$\;
$T_f \gets 0$\;
$a \gets 0.95$\;
$T^{(0)} \gets T_0$\;
$\sigma^{(0)} \gets \text{random sequence}$\;
${\bf x}^{(0)} \gets \text{RosettaMutate}({\bf x}^{(0)}, \sigma^{(0)})$\;
${\bf x}^{(0)} \gets \text{RosettaFastRelax}({\bf x}^{(0)})$\;
$E^{(0)}_{\text{peptide}} \gets \sum_{i \in \mathcal{R}_{\text{peptide}}} E(\sigma^{(0)}_i ; {\bf x}^{(0)}_i)$\;
\For{$k = 1, 2, \dots, N$}{
    \For{$i \in \mathcal{R}_{\text{peptide}}$}{
        $\sigma^{(\text{temp})}_i \sim \text{Multinomial}(\text{softmax}([E(\text{aa}; {\bf x}^{(k-1)}_i) / T^{(k)}]_{\text{aa} \in \text{amino-acids}}))$\;
    }
    ${\bf x}^{(\text{temp})} \gets \text{RosettaMutate}({\bf x}^{(k-1)}, \sigma^{(\text{temp})})$\;
    ${\bf x}^{(\text{temp})}  \gets \text{RosettaFastRelax}({\bf x}^{(\text{temp})})$\;
    $E^{(\text{temp})} _{\text{peptide}} \gets \sum_{i \in \mathcal{R}_{\text{peptide}}} E(\sigma^{(\text{temp})}_i ; {\bf x}^{(\text{temp})}_i)$\;
    $p \sim \text{Uniform}(0, 1)$\;
    \eIf{$p < \exp(-(E^{(\text{temp})}_{\text{peptide}} - E^{(k-1)}_{\text{peptide}}) / T^{(k)})$}{
        $E_{\text{peptide}}^{(k)} \gets E^{(\text{temp})} _{\text{peptide}}$\;
        ${\bf x}^{(k)} \gets {\bf x}^{(\text{temp})}$\;
        $\sigma^{(k)} \gets \sigma^{(\text{temp})}$\;
    }{
        $E_{\text{peptide}}^{(k)} \gets E_{\text{peptide}}^{(k-1)}$\;
        $x^{(k)} \gets {\bf x}^{(k-1)}$\;
        $\sigma^{(k)} \gets \sigma^{(k-1)}$\;
    }
    $T^{(k+1)} \gets (T_0 - T_f)a^k + T_f$\;
}
\end{algorithm}
}
\vspace{1cm}
\end{itemize}

\noindent \GM{Similar to scoring, we use four protocols for peptide design that differ in their restrained protocol (HERMES-{\em fixed} or HERMES-{\em relaxed}), and the amount of noise injected during training of the underlying HERMES model. Specifically, we train HERMES models with no added noise ($\sigma =0.00~\AA$) and another with moderate Gaussian noise with standard deviation $\sigma= 0.50~\AA$ applied to the atomic coordinates of the  protein structure data during training. We refer to the resulting peptide design protocols as HERMES-{\em fixed} 0.00, HERMES-{\em fixed} 0.50, HERMES-{\em relaxed} 0.00, HERMES-{\em relaxed} 0.50, specifying both the model and the noise amplitude.\\
}

\noindent{\bf Filtering metrics for peptide design with HERMES and benchmarking of different design approaches.} For each system, we first selected designs predicted by NetMHCPan to be at least weak binders to their target MHCs (EL\_rank $\leq 2.0$)~\cite{reynisson_netmhcpan-41_2020}. We then used TCRdock~\cite{Bradley2023-rl} to filter peptides for reactivity.  TCRdock is a structure prediction algorithm built on top of AlphaFold2~\cite{Jumper2021-hl}, and refined to model TCR-pMHC interactions. Its  PAE score for the TCR-pMHC interface can be used to distinguish reactive from decoy peptides~\cite{Bradley2023-rl}, and therefore, can provide a  metric for filtering our designed peptides; other work has similarly used the PAE of structure-prediction algorithms to filter designed protein binders~\cite{Liu2024-vi}. Using TCRdock, we retained  those peptides whose  PAE scores at the TCR-pMHC interface fell below a threshold, as illustrated in Figs.~\ref{fig:design_tcrdock_nyeso},~\ref{fig:design_tcrdock_ebv}, and~\ref{fig:design_tcrdock_mage}). The distribution of PAE scores is highly system specific~\cite{Liu2024-vi}, so we set each system's threshold near the TCRdock PAE value of its {\em wild-type} peptides in complex with the respective TCR-MHC, as the wild types are known binders. Among the filtered peptides, we selected 93 for experimental validation, ensuring that designs from both HERMES-{\em fixed} and HERMES-{\em relaxed}, as well as  from models with  0.0 and 0.5\AA\, noise levels were represented. We also enforced a minimum  difference of 2-3 amino acids (depending on the system) between the designs and the wild-type peptides. 

It should be noted that the AF3 PAE score can also discriminate between reactive and decoy peptides (Figs.~\ref{fig:design_scatterplots},~\ref{fig:design_roc_curves}), making it also suitable for filtering. However, a full-access implementation of AF3 was unavailable before our experimental validations. Therefore, we used AF3 only as a secondary filter for designs in the EBV and MAGE systems, and for benchmarking against  TCRdock's PAE score for the designed peptides (Figs.~\ref{fig:design_scatterplots}); see details below.\\

\noindent{\bf Design parameters for different TCR-pMHC systems.}  Here, we provide the details on the specifics of  peptide selection for each TCR-MHC system. It should be noted that as we progressed from NY-ESO to EBV and MAGE, we chose to explore the sequence space farther from the wild-type peptides, leading to more challenging designs.

\begin{itemize}

\item {\em NY-ESO:} The wild-type peptides (NYESO: SLLMWITQC and NYESO V9C: SLLMWITQV) have TCRdock PAEs of $5.17$ and $5.18$, respectively. 
As positive designs, we selected  58 peptides with the lowest TCRdock PAE values (all below 5.35) and a Hamming distance of at least two amino acids from their respective wild-type sequences. For negative designs, we uniformly sampled 35 peptides among those designed by HERMES whose TCRdock PAE values fell between 5.5 and 7.0. We did not enforce all HERMES models to be represented equally among these designs, nor did we enforce equal representation of the template structures. However, we note that all models are represented among positive designs (Fig.~\ref{fig:design_experimental_nyeso}).

\item {\em EBV:} The wild-type peptides (HPVG: HPVGEADYFEY and HPVG E5Q: HPVGQADYFEY)  have TCRdock PAEs of $4.90$ and $4.82$, and AF3 PAEs of $1.15$ and $1.13$, respectively. We selected 93 positive peptides using a TCRdock's PAE score cutoff of $5.1$ and AF3 PAE score cutoff of  $2.0$. Specifically, we sampled uniformly below these cutoffs with the following design protocols: HERMES-{\em fixed} at both noise levels using the HPVG structure (3MV7) as template (24 peptide), HERMES-{\em fixed} at both noise levels,  using the HPVG E5Q structure (4PRP)  as template (23 designs); HERMES-{\em relaxed} at both noise levels, using the HPVG structure (3MV7)  as template (24 peptides), and  HERMES-{\em relaxed} at both noise levels using the HPVG E5Q structure (4PRP)  as  template (23 designs);  see Fig.~\ref{fig:design_experimental_ebv}.

\item {\em MAGE:} The wild-type peptides (MAGE-A3: EVDPIGHLY and Titin: ESDPIVAQY)  have TCRdock PAEs of $5.21$ and $5.03$, and AF3 PAEs of $1.00$ and $0.98$, respectively. We selected 93 positive peptides using a TCRdock's PAE score cutoff of 5.2; the AF3 PAE scores for all of these designed peptide are comparable to that of the wild-types. For these designs, we sampled uniformly below the PAE cutoffs with the following design  protocols: HERMES-{\em fixed} at both noise levels using MAGE-A3 structure (5BRZ) as template (15 peptides); HERMES-{\em fixed} at both noise levels using the MAGE-A3 structure (5BRZ) as  template with E$_1$ fixed (15 peptides); HERMES-{\em relaxed} at both noise levels using  the MAGE-A3 structure (5BRZ)  as template (16 peptides); HERMES-{\em relaxed} at bothsee Fig.~\ref{fig:design_experimental_ebv} noise levels using the  MAGE-A3 structure (5BRZ)  as template with E$_1$ fixed (15 peptides); HERMES-{\em fixed} at both noise levels and using the Titin structure (5BS0)  as template (16 peptides); HERMES-{\em relaxed} at both noise levels using the Titin structure (5BS0) as template (16 peptides); see Fig.~\ref{fig:design_experimental_mage}. 
\end{itemize}

\noindent{\bf Characterizing the size of the sequence space for design.} The size of the sequence space to explore grows exponentially with increasing distance from the wild-type peptide. If no restriction is imposed (i.e., all amino acids are permissible with equal probabilities), the size of the sequence space follows,
\EQ
\Omega_0(d) = {\ell \choose d} 19^d
\EE
where $\ell$ is the length of the peptide, and $d$ is the distance from the wild-type sequence.

MHC presentation imposes constraints on amino acid usages in peptides, especially at the anchor residues. Let $p_i(a)$ denote the probability of amino acid $a$ at position $i$ of the peptide, e.g., set by the sequence-specific preferences of the MHC allele.  The probability of  introducing a mutation at  position $i$ depends on the prominence of the wild-type amino acid, and is equal to $q_i= 1-p_i(\text{wt})$. The typical number of non-wildtype amino acids realized at position $i$ can be estimated by $ n_i = 2^{S_i}$, where $S_i = -\sum_{a \neq \text{wt}} \hat p_i(a) \log_2 \hat p_i(a) $ is the entropy associated with the non-wildtype amino acid frequencies, and $\hat p_i(a) = \frac{p_i(a)}{1-p_i(\text{wt})}$ is the normalized probability of amino acid $a$, obtained by excluding the wild-type from the set.  Therefore, the typical size of the sequence space at Hamming distance $d$ from the wild-type peptide---subject to constraints imposed by the amino acid frequencies $p_i(a)$ (e.g., due to MHC restriction)---is given by,
\EQ
\Omega (d) =  \sum_{\{i_1,\dots,i_d\}} \left(q_{i_1}  n_{i_1} \right)\dots \left(q_{i_d}  n_{i_d} \right) \equiv  \sum_{\{i_i,\dots,i_d\}} 2^{ \hat S_{i_1}+ \dots +\hat S_{i_d} } \label{eq:Omega_d}
\EE
where the summation runs over all ${\ell \choose d}$ ways to choose $d$ distinct sites at which mutations may occur,  and $q_i n_i$ is the expected number of  substitutions associated with site $i$---evaluated as the likelihood of mutating site $i$  times the typical number of viable amino acid replacements at that site---and $\hat S_i =S_i + \log_2 q_i$ is the resulting adjusted entropy at site $i$.

In principle, evaluating $\Omega (d) $ in eq.~\ref{eq:Omega_d} involves summation over a combinatorially large number of terms, which can become computationally prohibitive.  In practice, however, many of the sites have the same adjusted entropy, which simplifies the  evaluations of $\Omega (d) $. For example,  the adjusted entropies for mutations away from the 9 amino acid NY-ESO peptide restricted by HLA-A*02:01 yield the following (rounded) values: $[4, 2, 4, 4, 4, 3, 4, 4, 2]$ in bits, indicating six sites with 4 bits of entropy (or $2^4=16$ typical substitutions), one site with 3 bits (8 typical substitutions), and two sites with 2 bits (4 typical substitutions). To demonstrate, in this case, the number of sequences at distance $d=5$ from the wild-type follows,

\EQA
\nonumber \Omega_\text{NYESO} (d=5) &=&  {6 \choose 5} 16^5  + {6 \choose 4} 16^4\times \left[  {2 \choose 1} 4^1 + 8^1   \right]+ 
{6 \choose 3} 16^3\times \left[  {2 \choose 1} 4^1\times  8^1 + {2 \choose 2}  4^2    \right]+ 
{6 \choose 2}16^2  \times {2 \choose 2}  4^2\times 8^1  \\
\nonumber&=& 29,065,216\\
\EEA
Similar calculations can be done to characterize the size of the sequence space at varying distances from the wild-type, as shown in Fig. 4.

\section{Alternative computational methods for scoring peptides and design}
\label{sec:AltModel}
 We compare our approach to alternative methods (listed below), based on which we characterize peptide scores akin to HERMES's peptide energy,  and when possible,  use them to design novel peptides. A detailed comparison between the accuracy of different methods in predicting  the affinity of TCR-pMHC interactions and the  peptide-induced activation of T-cells is presented in Tables~\ref{table:spearmanr_affinity} and~\ref{table:spearmanr_activity}.  Comparison between the fidelity of the designs--only according to computational metrics--is shown in Figures~\ref{fig:design_tcrdock_nyeso},~\ref{fig:design_tcrdock_ebv},~and~\ref{fig:design_tcrdock_mage}. All the designed  peptides  as well as the code to generate samples are available in our github repository. 

\begin{itemize}
\item {\em BLOSUM62 substitution matrix}~\cite{Henikoff1992-jq}: The BLOSUM62 substitution matrix characterizes the favorability  of substitutions between all pairs of amino acids, and has been used widely to characterize the mutational effects in proteins. As our first (and simplest) benchmark, we used BLOSUM62 to define a peptide energy (score). Specifically, we score a peptide by summing over the BLOSUM62 scores associated with the amino acid substitutions from the wild-type peptide $ \sigma^{\text{wt}}$ (the one present in the template structure) to the query peptide sequence $\sigma$ that we are interested in,
\EQ
E^{\rm (B62)}_{\text{peptide}}(\sigma; \sigma^{\text{wt}}) = \sum_{i=1}^\ell B62[\sigma^{\text{wt}}_i \to \sigma_i]
\label{eq:score_blosum}
\EE
Notably, the BLOSUM62 predicted peptide energy only depends on the substitutions in the peptide relative to the wild-type amino acid, and does not depend on the structural context of these substitutions (Fig.~\ref{fig:hsiue_h1_h8}).

For designs with BLOSUM62, we independently sample amino acids along the peptide based on their favorability with respect to the wild-type amino acid at each position. Specifically,  at a given position $i$ we interpret the substitution score $B62[\sigma^{\text{wt}}_i \to \sigma_i]$ from the wild-type to any of the other 19 amino acids as the energy difference between the two states, and by applying a softmax function with temperature $T$, we evaluate the probability of these possible substitutions. We then  use multinomial sampling to draw new amino acids from this probability distribution at each position, 
\EQ
\sigma_i \sim \text{Multinomial}\left(\text{softmax}\left[\frac{B62[\sigma_i^\text{wt} \to \sigma]}{T}\right]_{{\sigma} \in \text{amino-acids}}\right)
\EE
For our analyses, we use temperatures $T=1.0, 2.0, 3.0$ to modulate the variability of the designed peptides.\\

\GM{
\item{\em \L uksza et al. $M$ and $d \cdot M$ substitution matrices}~\cite{Luksza2022-gx}: Here, $M$ is an amino acid substitution matrix, and $d$ is a vector of position-specific scale factors for 9-mer peptides. Briefly, they were constructed to fit the EC$_{50}$ measurements on seven saturation mutagenesis experiments of three different wild-type peptides, against a total of seven TCR-MHC pairs--the very same T-cell activity data  that we used to test HERMES in Fig.~3. Specifically, $d$ and $M$ were constructed such that:
\EQ
d_i \cdot M[\sigma^{\text{wt}}_i \to \sigma^{\text{mt}}_i] \approx \log \frac{\text{EC}_{50}^{\text{mt}}}{\text{EC}_{50}^{\text{wt}}} + R_\text{B62}[\sigma^{\text{wt}}_i \to \sigma^{\text{mt}}_i]
\label{eq:luksza_matrix}
\EE
where the mutant and wild-type peptides only differ at position $i$, and $R_\text{B62}$ is a regularization term reflecting the BLOSUM62 substitution matrix; we refer the reader to the Supplementary Methods of ref.~\cite{Luksza2022-gx} for details. Since all peptides considered when constructing $d$ and $M$ are 9-mers, the position scale factors can only be used on 9-mer peptides. We use $d$ and $M$ for scoring in a manner analogous to BLOSUM62:
\EQ
E^{(d \cdot M)}_{\text{peptide}}(\sigma; \sigma^{\text{wt}}) = \sum_{i=1}^\ell d_i \cdot M[\sigma^{\text{wt}}_i \to \sigma_i]
\label{eq:score_luksza_d_m}
\EE
and
\EQ
E^{(M)}_{\text{peptide}}(\sigma; \sigma^{\text{wt}}) = \sum_{i=1}^\ell M[\sigma^{\text{wt}}_i \to \sigma_i]
\label{eq:score_luksza_m}
\EE
Similar to BLOSUM62,  the predicted peptide energies $E^{(d \cdot M)}_{\text{peptide}}$ and   $E^{(M)}_{\text{peptide}}$ only depend on the type substitutions, and not on their structural contexts.\\
}

\item {\em MHC position weight matrices}~\cite{Tadros2023-ij}:  MHC alleles exhibit distinct sequence-specific preferences for peptide presentation, most notably in the peptide's amino acid composition at anchor positions.  Computational models have characterized these preferences with position weight matrices (PWMs) that reflect the 
the probability of different amino acids at each position for peptides presented by a given MHC allele~\cite{Lundegaard2008-ao, Lundegaard2008-wz, Andreatta2016-uh,Tadros2023-ij}. As a benchmark for our designs, we use the PWMs from the MHC Motif Atlas~\cite{Tadros2023-ij} to independently sample amino acids at  each site  along the peptide. The resulting peptide sequences should reflect  biases imposed by MHC presentation but not interactions with TCRs.\\

\item {\em ProteinMPNN}~\cite{Dauparas2022-fu}: ProteinMPNN is a commonly used tool for  inverse-folding (i.e., designing protein sequences consistent with the backbone of a structure). It is primarily used to sample amino acid sequences conditioned on a protein's backbone structure, and optionally with partial sequence information (i.e.,  amino acids specified at some of the residues). As ProteinMPNN also outputs probability distributions of amino acids at different sites of the protein, it can also be used to score peptides (i.e., evaluate peptide energy) in a manner similar to HERMES, as 
\GM{
\EQA
 E^{\rm (MPNN)}_{\text{pep 1-by-1}}(\sigma; TCR, MHC) &=& \sum_{i=1}^\ell  \log P^{\rm (MPNN)}(\sigma_i ; {\rm backbone, \sigma_{\text{TCR}}, \sigma_\text{MHC}}, \sigma_{/i})
\label{eq:peplogp_proteinmpnn}
\EEA
where $P^{\rm (MPNN)}(\sigma_i ; {\rm backbone, TCR, MHC},  \sigma_{/i})$ is the ProteinMPNN estimate for the  probability of amino acid $\sigma_i$ at position $i$ of the peptide, given the backbone structure of the whole protein complex, and the amino acid sequence of the MHC $\sigma_\text{MHC}$, TCR $\sigma_\text{TCR}$, and the rest of the peptide $\sigma_{/i}$. We term this scoring scheme as peptide 1-by-1 since we mask the peptide's residues  one at a time.

We explored two additional ProteinMPNN scoring schemes that differ in the residues they mask and score. As our second scoring scheme, we simultaneously mask the entire peptide's sequence $\sigma$, instead of one amino-acid at a time, and evaluate the score for the whole peptide $E^{\rm (MPNN)}_{\text{pep}}(\sigma; TCR, MHC)$:
\EQA
 E^{\rm (MPNN)}_{\text{pep}}(\sigma; TCR, MHC) &=& \sum_{i=1}^\ell  \log P^{\rm (MPNN)}(\sigma_i ; {\rm backbone, \sigma_{\text{TCR}}, \sigma_\text{MHC}})
\label{eq:peplogp_proteinmpnn_pep}
\EEA

For the third scoring scheme, we consider masking and scoring the TCR residues that make contact with the peptide--defined as those whose C-$\alpha$ lie within  12~\AA~ of any peptide residue's  C-$\alpha$. We call this set {\bf TCR-int} and its corresponding amino acid composition as $\sigma _\text{TCR-int}$. In this scheme, both the peptide and the interacting TCR residues are masked simultaneously and scored together:
\EQA
\nonumber E^{\rm (MPNN)}_{\text{pep and tcr}}(\sigma; TCR, MHC) &=& \sum_{i=1}^\ell  \log P^{\rm (MPNN)}(\sigma_i ; {\rm backbone, \sigma_{\text{TCR / TCR-int}}, \sigma_\text{MHC}}) \\ &+& \sum_{j=1}^n \log P^{\rm (MPNN)}(\sigma_{\text{TCR-int}_j} ; {\rm backbone, \sigma_{\text{TCR / TCR-int}}, \sigma_\text{MHC}})
\label{eq:peplogp_proteinmpnn_tcr_and_pep}
\EEA

Similar to HERMES, we considered ProteinMPNN models trained with two noise levels: 0.02~\AA~(very little noise) and 0.20 \AA.}

For designs with ProteinMPNN, we use the algorithm's native functionalities, keeping the TCR and MHC structures fixed (from the design template) and prompting the model to design only the peptide. We use two ProteinMPNN models (trained with 0.02~\AA~and 0.20~\AA~noise), and use a temperature of 0.7 for all systems. \\

\GM{
\item {\em ESM-IF1}~\cite{Hsu2022-km}: ESM-IF1, like ProteinMPNN,  is an inverse-folding model, but it lacks native support  for scoring or designing proteins when only  partial sequence information is available. 

To access the ESM-IF1 model, we used the code in the esm repository at~\url{https://github.com/facebookresearch/esm/tree/main/examples/inverse_folding}. 

For scoring a peptide sequence given a TCR-MHC context, we used the script \texttt{score\_log\_likelihoods.py} with the \texttt{--multichain-backbone} flag, with the same TCR-pMHC structure inputs as for HERMES and ProteinMPNN. This script computes a score for the peptide sequence conditioned on the backbone coordinates of the entire TCR-pMHC complex, but {\em not} on the sequence of the TCR and the MHC:
\EQA
\nonumber E^{\rm (ESM-IF1)}_{\text{peptide}}(\sigma; TCR, MHC) &=& \log P^{\rm (ESM-IF1)}_{\text{peptide}}(\sigma ; {\rm backbone})
\label{eq:peplogp_esmif}
\EEA

To design peptide sequences given a TCR-MHC context, we used the script \texttt{sample\_sequences.py} with the \texttt{--multichain-backbone} flag, and a temperature of 0.7 for all systems, with the same TCR-pMHC structure inputs as for HERMES and ProteinMPNN. Similar to scoring, this script samples peptide sequences conditioned on the backbone coordinates of the entire TCR-pMHC complex, but {\em not} on the sequences of the TCR and MHC.\\
}

\item {\em TCRdock}~\cite{Bradley2023-rl}: TCRdock is a structure prediction algorithm built on top of AlphaFold2~\cite{Jumper2021-hl}, and refined to predict the structure of  TCR-pMHC complexes. The algorithm model the structure of TCR-pMHC complexes by leveraging structural templates provided by its curated database, as well as the knowledge of TCR-pMHC canonical binding poses. Crucially, TCRdock's Predicted Alignment Error (PAE) for TCR-pMHC interfaces was  found to well discriminate between true peptide binders and random decoys~\cite{Bradley2023-rl}. Thus, to benchmark against TCRdock, we interpret this PAE score as peptide energy, and use it  to predict TCR-pMHC binding affinities and peptide-induced T-cell activities. 

TCRdock  takes as input sequences of a TCR, MHC, and peptide, and identifies TCR-pMHC structures in its database with similar amino acid sequences to the query to  serve as templates to model the structure of the input TCR-pMHC complex. To examine how template-query sequence similarity influences TCRdock's predictive accuracy, we ran the algorithm under two conditions: the default mode, which uses any closest-matching templates available, and the {\em benchmark} mode, which restricts the maximum sequence similarity of templates with the input TCR-pMHC.

We use TCRdock to characterize peptide  energies, and to filter the sequences designed by HERMES. However, TCRdock does not provide a natural way to sample peptides for design, and because of its high computational cost, we did not include TCRdock PAE calculations in our design pipeline (e.g., for MCMC sampling).\\

\item {\em TULIP}~\cite{Meynard-Piganeau2024-po}: TULIP is a transformer-based model trained to generate either TCR or peptide sequences conditioned on the other binding partners in a TCR-pMHC complex. The conditional likelihood function learned by TULIP can be used to score peptides within a TCR-pMHC context. Specifically, this scoring scheme was used in ref.~\cite{Meynard-Piganeau2024-po} to predict the T-cell activity measurements from ref.~\cite{Luksza2022-gx}, and we benchmarked our predictions against TULIP's for this data.  In Table~\ref{table:spearmanr_activity}, we present the Spearman correlations between TULIP predictions and the fitted EC$_{50}$ values as reported in ref.~\cite{Luksza2022-gx}, which are the same correlations reported in the TULIP paper~\cite{Meynard-Piganeau2024-po}. Additionally, we provide correlations with the fitted EC$_{50}$ values obtained using the procedure outlined in Section~\ref{sec:aff_act} and eq.~\ref{eq:EC50}, consistent with  correlations reported in Fig.~3.\\

\GM{
\item {\em NetTCR2.2}~\cite{Jensen2024-xs}: NetTCR2.2 consists of a Convolutional Neural Network classifier to predict TCR-peptide binding from the  the amino acid sequences of the peptide and of all six CDRs from the $\alpha$ and the $\beta$ chains of a TCR. Because the training set contains many examples of different TCRs binding to the same peptide, but far fewer cases of multiple peptides tested against a given TCR, the model performs best when ranking candidate TCRs for a fixed peptide, as reported in ref.~\cite{Jensen2024-xs}. By contrast, we aim to address the complementary task for ranking peptides for a fixed TCR. 

We query NetTCR2.2 through the authors' web server (\url{https://services.healthtech.dtu.dk/services/NetTCR-2.2/}). CDR boundaries are assigned by aligning each TCR sequence to the IMGT reference sequences, using code from the TCRdock repository~\cite{Bradley2023-rl}. Of the TCR-pMHC pairs examined in this work, only one system (TCR1, Table~\ref{table:affinity_and_activity_info}) appears in the NetTCR2.2 training set, and even then with just a single annotated positive peptide. Therefore, this small overlap is unlikely to bias our predictions.\\

\item {\em TAPIR}~\cite{Fast2023-tk}: Similar to NetTCR-2.2, TAPIR consists of a Convolutional Neural Network trained to predict TCR-peptide-MHC binding. As input, TAPIR considers the CDR3's amino acid sequences, the V and J genes, the peptide's amino acid sequence, and the MHC allele. We use TAPIR for scoring via the web interface hosted by the authors~\url{https://vcreate.io/tapir/}. We extract V/J genes and CDR3 from a TCR sequence by aligning it to the IMGT reference sequences, using  code from the TCRdock repository~\cite{Bradley2023-rl}. We find that five out of nine of the TCR-pMHC systems in this study are present in the training data of TAPIR, but each only with a single annotated positive peptide (1G4 TCR, TCR1, TCR4, TCR5, TCR6, Table~\ref{table:affinity_and_activity_info}), and therefore, this small overlap is unlikely to bias our predictions. 
}
\end{itemize}

\GM{
\noindent \textbf{Computational modeling of TCR-pMHC templates with AlphaFold3.} For six out of seven systems with T-cell activity measurements from ref.~\cite{Luksza2022-gx} (TCRs~2-7), no experimentally-determined structure is available. For these, we generated structures using the AlphaFold3 (AF3) model~\cite{Abramson2024-jp}. Furthermore, to probe how sensitive HERMES is to AF3 modeled inputs, we produced AF3 predictions for every TCR-pMHC complex we studies in this work--both for scoring and design design targets--using the wild-type sequences of all the chains as inputs; see Dataset~\ref{dataset:af3_sequences} for these sequences.  Each target was folded using both the AF3 server's option with template search (with default cutoff date of September 9th 2021), and without template, the latter approximating the scenario in which no homologous structures exist. We note, however, that several of TCR-pMHC complexes we analyzed here were likely present in the AF3 training data, so the ``template-free" predictions can only serve as an approximate test of true de-novo performance.

\GM{We fold TCR-pMHC systems for which we have a structure using restricted amino-acid sequences of the TCR and MHC chains that were extracted with functionalities in the TCRdock codebase, and are equivalent to the sequences that TCRdock would use to predict the structure~\cite{Bradley2023-rl}. Specifically, we first extract the amino-acid sequences from the experimental structure. Then, we extract V/J genes, CDR3 and MHC allele from them using `parse\_tcr\_pmhc\_pdbfile.py`, and input them to `setup\_for\_alphafold.py' to get the restricted amino-acid sequences.}  For the bispecific Fab antibody-mediated T-cell therapy from ref.~\cite{Hsiue2021-ag}, we used AF3 to fold and dock the Fab and the pMHC by extracting the amino-acid sequences from the experimental structure. We found that AF3 displayed high uncertainty (as measured by plDDt) only at the Fab-pMHC interface, and not in the folds of individual chains. The  sequences we used as input to AF3 are listed in Dataset~\ref{dataset:af3_sequences}. \\
\\
}

\noindent \textbf{Benchmarking HERMES predictions for T-cell affinity and activity.}  \GM{We compared the performance of HERMES models at predicting the  binding affinity and activity of T-cells against substitution matrices (BLOSUM62~\cite{Henikoff1992-jq}, and \L uksza {\em et al}'s TCR-specific $d \cdot M$ and $M$  matrices~\cite{Luksza2022-gx}), the TCR-pMHC structure prediction algorithm TCRdock~\cite{Bradley2023-rl}, the sequence-based machine learning algorithms for TCRs (TAPIR~\cite{Fast2023-tk}, NetTCR~\cite{Jensen2024-xs}, and TULIP~\cite{Meynard-Piganeau2024-po}), and the structure-based inverse folding models (ESM-IF1~\cite{Hsu2022-km}  and ProteinMPNN~\cite{Dauparas2022-fu}). Since all of these models are unsupervised in nature (except for   \L uksza {\em et al} activity predictions for TCRs 1-7), we do not expect predictions to be in the same units as the affinity and activity values, and thus, we use Spearman correlation between data and  predictions as a metric for model performance.}

\GM{For predicting TCR-pMHC binding affinities (Table~\ref{table:spearmanr_affinity}), 
BLOSUM62 leads for the 1G4 TCR system, while template-guided TCRdock tops on the A6 TCR systems. HERMES-{\em relaxed} with no noise  (0.00)  achieves slightly lower Spearman correlation but is statistically  indistinguishable from these best performers  (p-value $>0.05$ associated with the difference in the Fisher's z-transformed Spearman correlations). For 1G4,   ESM-IF1 and the \L uksza {\em et al}'s TCR-specific $d \cdot M$ and $M$ substitution matrices are similarly competitive. HERMES-{\em relaxed} 0.00 consistently outperforms the  sequence-based models  (TAPIR, NetTCR-2.2),  ProteinMPNN, and TCRdock-{\em no template} (the TCRdock model without using structural templates similar to the query data). Lastly, choosing the minimum-Rosetta-energy conformation for HERMES-{\em relaxed} yields correlations with the data that are comparable to the baseline HERMES-{\em relaxed} model that averages over the  energies of 100 different realizations of relaxed peptides (Fisher's z-test p-value $>0.05$). Table~\ref{table:spearmanr_affinity} shows these benchmarking results in detail.}

\GM{For predicting T-cell activity  (Table~\ref{table:spearmanr_activity}), HERMES-{\em fixed} without noise (0.00) achieves the best performance for TCR1, TCR2, and TCR3, and proteinMPNN-0.02 shows competitive performances for TCR1 and TCR3.  ProteinMPNN-0.02 leads for TCR6 and H2-scDb, while HERMES-{\em fixed} 0.00  achieves slightly lower but statistically indistinguishable Spearman correlations (Fisher's z-test p-value $>0.05$). All models struggle to provide reliable predictions for TCR7, with the exception of ProteinMPNN 0.02, which attains a moderate Spearman correlation  $\rho=0.31$ when scoring one peptide amino acid at a time (eq.~\ref{eq:peplogp_proteinmpnn}), and $\rho=0.29$, when scoring the peptide and the TCR together (eq.~\ref{eq:peplogp_proteinmpnn_tcr_and_pep}). For TCR4, which has a poor structural model (AF3 PAE = 2.07), all structure-based models fail to produce reliable predictions, whereas BLOSUM62 achieves a moderate Spearman correlation $\rho=0.33$. The \L uksza {\em et al}'s $d \cdot M$  substitution matrix overall  performs well  on the TCR~1-7 benchmark; however, because it was obtained by fitting  on these same data, its predictions are not strictly zero-shot. The structure-based models HERMES and ProteinMPNN consistently outperform the sequence-based models (TAPIR, NetTCR-2.2, and TULIP), as well as the structure-based ESM-IF1. Lastly, choosing the minimum-Rosetta-energy conformation for HERMES-{\em relaxed} model yields a similar performance to the baseline  HERMES-{\em relaxed} model that averages over the  energies of 100 different realizations of relaxed peptides. Table~\ref{table:spearmanr_activity} shows these benchmarking results in detail.}

\begin{figure}[ht]
\centering
	\includegraphics[width=0.90\textwidth]{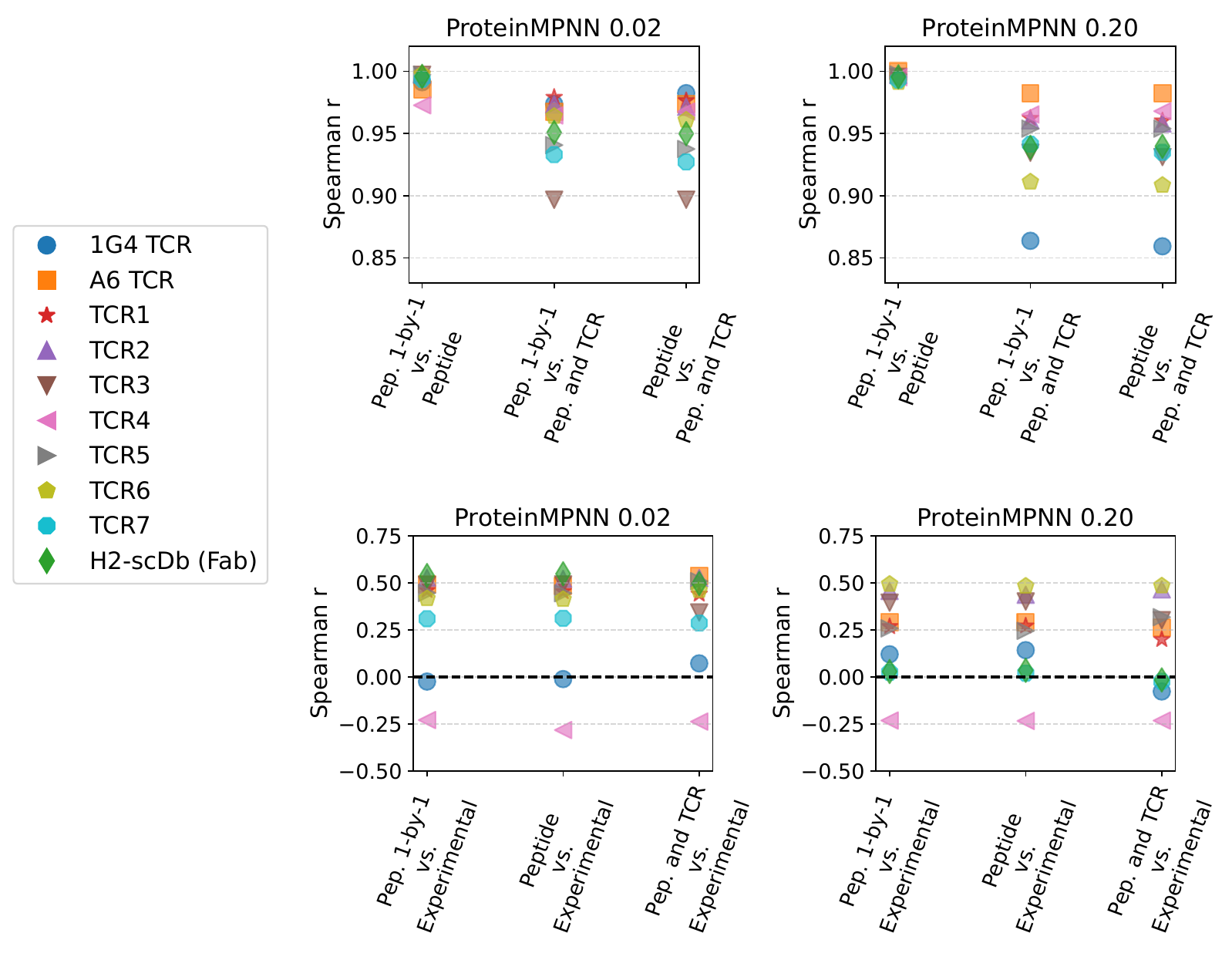}
\caption{\GM{\textbf{Robustness of   ProteinMPNN performance across scoring schemes .} The performance of ProteinMPNN across  different systems (legend) is shown for models trained with varying noise amplitudes (specified above each panel), and using  different scoring schemes. We consider three scoring schemes: (i): ``Peptide 1-by-1" in which residues within the peptide are masked and scored one at a time,  using the remainder  of the TCR-pMHC complex as structural context  (eq.~\ref{eq:peplogp_proteinmpnn}), (ii) ``Peptide" scheme, in which the entire peptide is masked  and scored simultaneously (eq.~\ref{eq:peplogp_proteinmpnn_pep}), and (iii) ``Pep. and TCR" scheme in which both the peptide and the nearby TCR residues are masked and scored simultaneously  (eq.~\ref{eq:peplogp_proteinmpnn_tcr_and_pep}). 
Top panels show the Spearman correlation between predictions from different scoring schemes, while bottom panels show  correlations between each scheme and  experimental data. Results demonstrate that ProteinMPNN predictions are robust to the choice of scoring scheme.
}
}
\label{fig:proteinmpnn_scoring_schemes_comparison}
\end{figure}

\begin{figure}[ht]
\centering
	\includegraphics[width=0.8\textwidth]{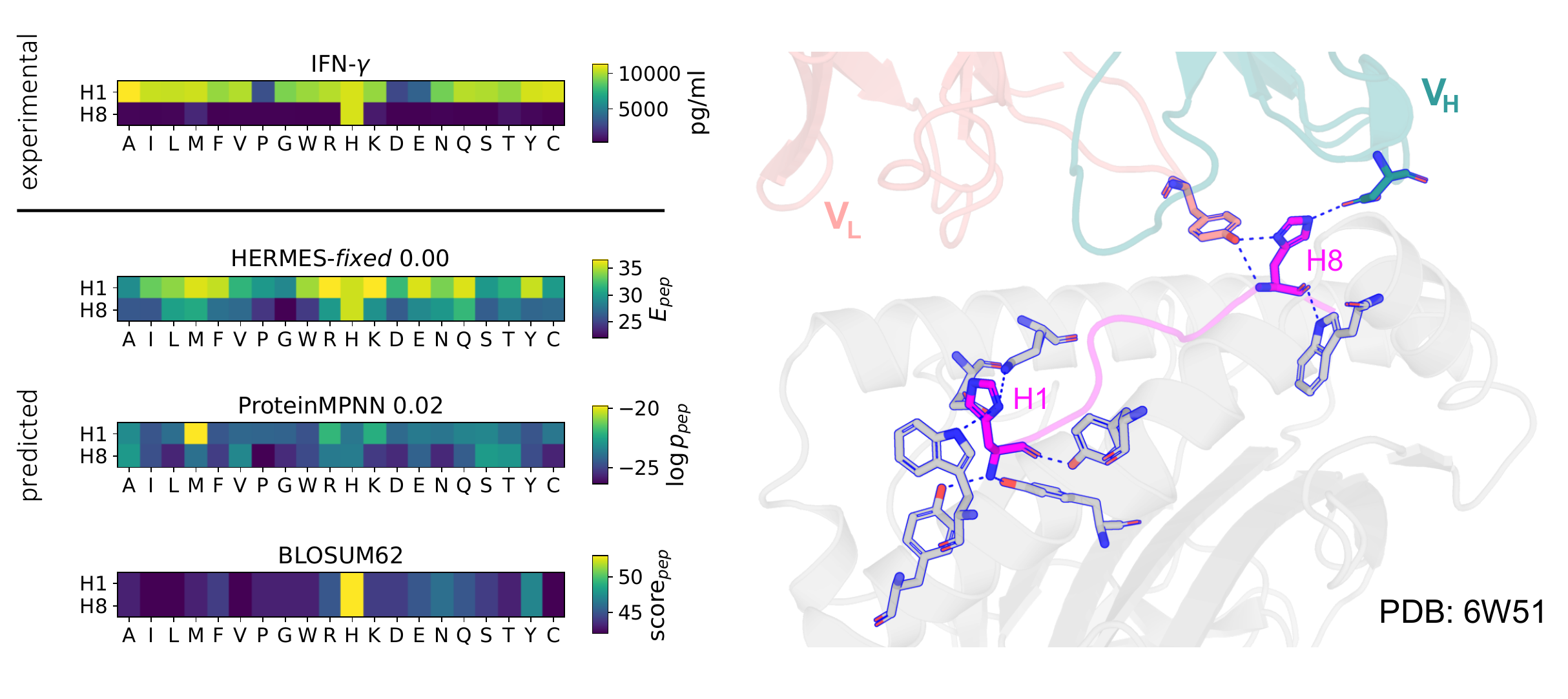}
\caption{\textbf{Context-dependent mutational effects on T-cell activity.} The heatmaps show both experimental measurements and model predictions of  T-cell activities of amino acid variants at positions 1 and 8 of the p53$^R157H$ peptide, which we study  in complex with the designed bi-specific antibody H2-scDB and HLA-A*02:01~\cite{Hsiue2021-ag}. The wild-type peptide has a histidine at both sites. A standard substitution matrix (such as BLOSUM62) would assign identical scores to mutations at these two sites, because they share the same wild-type residue. However, peptide binding and the subsequent T-cell activity depends on the structural context of each substitution. Here, H1 forms polar contacts only with the MHC (from the sides), while H8 is more tightly constrained by polar contacts with the TCR (right panel). HERMES-{\em fixed} correctly leverages structural context to predict T-cell activity as a result of substitutions away from histidines at both positions. Although ProteinMPNN also incorporates structural information, its predictions are not as accurate in this instance.}
\label{fig:hsiue_h1_h8}
\end{figure}

Our results do not indicate a clear preference between HERMES-{\em fixed} and HERMES-{\em relaxed}: HERMES-{\em fixed} predicted T-cell activity better, while HERMES-{\em relaxed} predicted binding affinity between TCR and pMHC better. However, it is important to note that in all systems tested here, the scored peptides are highly similar to the wild-type peptides, with a maximum Hamming distance of just one amino acid from a wild-type peptide found in one of the available structural templates. As a result, relaxation may not be essential for achieving high performance. Interestingly, noised HERMES models underperform compared to the un-noised counterparts. This is in contrast to prior observations, where noised models achieve higher performances in predicting the stability effect of mutations in proteins~\cite{Visani2024-sh}. \GM{
 Moreover, we observe that ProteinMPNN predictions are robust to the  choice of scoring scheme, i.e., scoring peptide amino acids one at a time $E^{\rm (MPNN)}_{\text{pep 1-by-1}}$ (eq.~\ref{eq:peplogp_proteinmpnn}),  the whole peptide simultaneously $E^{\rm (MPNN)}_{\text{pep}}$ (eq.~\ref{eq:peplogp_proteinmpnn_pep}), or peptide and the TCR interface together $E^{\rm (MPNN)}_{\text{pep and tcr}}$ (eq.~\ref{eq:peplogp_proteinmpnn_tcr_and_pep}); see Fig.~\ref{fig:proteinmpnn_scoring_schemes_comparison}. 

Lastly, we benchmarked all four HERMES models and two ProteinMPNN models (using  peptide-scoring schemes eq.~\ref{eq:peplogp_proteinmpnn_pep} with different noise levels) on AF3-predicted structures, for both activity and affinity predictions (Fig.~\ref{fig:plots_af3_struc_ablation}). We generated  the AF3 structures both with and without  template guidance, and the resulting predictions made with these AF3 structures were compared to performances using the corresponding experimental structures. AF3 produces accurate models for all complexes except the H2-scDb Fab bound to pMHC, for which the docking orientation and conformation were incorrect (Table~\ref{table:af3_folding_metrics}). Expectantly, we saw a significant drop in model performances for both HERMES and ProteinMPNN when using the inaccurate AF3 predictions for H2-scDb (Fig.~\ref{fig:plots_af3_struc_ablation}). For both HERMES and ProteinMPNN, replacing experimental structures with AF3 predictions altered predictions for the 1G4 and A6 TCRs binding to pMHC variants, but the changes were not statistically significant (p-value$>0.05$ associated with the difference in the Fisher's z-transformed Spearman correlations); see  Fig.~\ref{fig:plots_af3_struc_ablation}.  Finally, for TCR1, we observe no differences in performance at predicting T-cell activity between using the AF3 structures vs. the experimental structures. Comparisons for TCRs~2-7 show insignificant differences between predictions using AF3 models  with and without templates.

\AN{In Table~\ref{table:af3_folding_metrics} we report, where available, the C$_\alpha$ RMSD between AF3-predicted and experimentally determined structures, together with AF3 confidence metrics computed with and without template guidance. The confidence metrics include the TCR–pMHC interface predicted alignment error (PAE), the interchain predicted TM-score (ipTM), and the predicted TM-score (pTM; not interface-specific). We found ipTM and pTM to be  closely related (Table~\ref{table:af3_folding_metrics}). Because structure-based models such as HERMES and ProteinMPNN rely on accurate input structures, Fig.~\ref{fig:plots_af3_struc_ablation}B plots their best-performing variants against AF3 confidence. The AF3 model for TCR4—on which both HERMES and ProteinMPNN perform poorly—exhibits the  second-lowest ipTM confidence (after H2-scDb). However, the limited range of available AF3 confidence values prevents definitive conclusions about how these metrics impact 
the accuracy of predictions for peptide mutation-effects on T-cell responses.}
\\

}

\noindent \textbf{Benchmarking HERMES designs against alternative models.} Although we did not experimentally test the designs by the different models, we used both peptide presentation by MHC predicted by NetMHCPan and TCRdock’s PAE as benchmarking metrics across different design methods. We note that these are only computational metrics and should not be taken as ground truth; this is particularly true for results using TCRdock PAE as a proxy for binding, which are much more likely farther from the ground truth than the easier-to-predict MHC presentation.

Figs.~\ref{fig:design_tcrdock_nyeso}, \ref{fig:design_tcrdock_ebv}, and \ref{fig:design_tcrdock_mage} show the fraction of peptides meeting each criterion for designs produced by various HERMES models, ProteinMPNN, BLOSUM62, and the system's MHC peptide preferences. Consistently, across the three systems, and for all models, the fraction of designs passing the PAE threshold decreases with increasing Hamming distance  from the wild-types' sequences (Figs.~\ref{fig:design_tcrdock_nyeso},~\ref{fig:design_tcrdock_ebv}, and~\ref{fig:design_tcrdock_mage}; panels E, and G). Also consistently, but more surprisingly, models trained with Gaussian noise applied to the structure---for both HERMES and ProteinMPNN---perform worst than their noise-less counterparts  (Figs.~\ref{fig:design_tcrdock_nyeso},~\ref{fig:design_tcrdock_ebv}, and~\ref{fig:design_tcrdock_mage}; panels E, and G).

HERMES models are consistently the best at designing peptides that are scored well for presentation by the respective MHC alleles (panel C of Figs.~\ref{fig:design_tcrdock_nyeso},~\ref{fig:design_tcrdock_ebv}, and~\ref{fig:design_tcrdock_mage}). No single model stands out as being consistently best across the three systems at designing peptides above the TCRdock PAE threshold: HERMES outperforms ProteinMPNN in the NY-ESO system (Fig.~\ref{fig:design_tcrdock_nyeso}), performs similarly in the EBV system (Fig.~\ref{fig:design_tcrdock_ebv}), and performs worse in the MAGE system (Fig.~\ref{fig:design_tcrdock_mage}). \GM{BLOSUM62 makes worse designs compared to HERMES and ProteinMPNN based on TCRdock PAE scores. Surprisingly, ESM-IF1 designs have the lowest success in MHC presentation, but show better TCRdock PAE scores compared to HERMES-{\em relaxed} for the MAGE system (Figs.~\ref{fig:design_tcrdock_nyeso},~\ref{fig:design_tcrdock_ebv}, and~\ref{fig:design_tcrdock_mage}).} Designs sampled from the MHC motif, while scoring well on MHC presentation, consistently score poorly for TCRdock PAE. We expected HERMES-{\em relaxed} to design better binders than HERMES-{\em fixed} at greater sequence distances from the wild-types. However, this holds only for the NY-ESO system (Fig.~\ref{fig:design_tcrdock_nyeso}) and not for EBV or MAGE (Figs.~\ref{fig:design_tcrdock_ebv},~\ref{fig:design_tcrdock_mage}).  Overall, a broader benchmarking study, ideally with experimental validation, is needed to rigorously evaluate the effectiveness of different peptide design methods.

To assess  how using AF3 structures  as templates influence design, we ran the HERMES-{\em fixed} design protocol using experimental and AF3 structures as  templates and compared the resulting  peptide likelihood matrices (as position weight matrices or PWMs) for the three systems (Fig.~\ref{fig:joint_logoplots_for_design}). Across NY-ESO, EBV, and MAGE, the resulting PWMs were largely consistent across the two template choices: most peptide positions showed the same preferences, and when variability did appear it was typically confined to chemically similar residues. There are two noteworthy exceptions: in NY-ESO, the top choice in position 7 of the peptide was flipped from Threonine and Aspartic Acid when AF3 template was used (Fig.~\ref{fig:joint_logoplots_for_design}A). In MAGE, using the AF3 model introduced a pronounced preference for glutamic acid at position 1 (Fig.~\ref{fig:joint_logoplots_for_design}C), a residue that we have already shown to be important for eliciting T-cell activity  (Fig.~4). Aside from these cases, AF3-derived structures preserve the peptide specificity landscape learned by HERMES.

\section{Experimental protocol to measure peptide activities}
\noindent {\bf Artificial antigen-presenting cell lines.} We obtained the full-length coding sequences of HLA-A*02:01, HLA-A*01:01, and HLA-B*35:01 from the IMGT/HLA database~\cite{Robinson2003-ru}. Gene fragments were synthesized by Genscript and were cloned into the lentiviral pLVX-EF1$\alpha$-IRES-Puro (Clontech) vector. We used Lipofectamine 3000 Transfection Reagent (Invitrogen) to transduce the 293T packaging cell line (ATCC CRL-3216) with the psPAX2 packaging plasmid (Addgene plasmid \#12260), the pMD2.G envelope plasmid (Addgene plasmid \#12259), and the HLA-encoding lentiviral vector. The lentivirus-containing supernatant was collected after 24 and 48 hours post-transfection and was filtered through a 0.45um SFCA filter (Thermo Scientific). The lentivirus was concentrated using Lenti-X Concentrator (Takara Bio) according to the manufacturer’s protocol, resuspended in DPBS (Gibco), and stored at -80C for future use. To generate monoallelic HLA cell lines for HLA-A*02:01 and HLA-B*35:01, HLA-deficient K562 cells (ATCC CCL-243) were transduced with lentivirus. Since K562 cells have a high expression of both MAGE-A3 and Titin antigens, we transduced murine EL4 cell lines (ATCC TIB-39) with HLA-A*01:01 lentivirus to produce HLA-A*01:01 presenters lacking the natural expression of the tested antigens. At 48 hours post-transduction, K562 cells were transferred to Iscove’s Modified Dulbecco’s Medium containing 10\% FBS, 2mM L-glutamine, 100 U/mL penicillin/streptomycin, and 2 $\mu$g/mL puromycin (Gibco). EL4 cells were transferred to RPMI media containing 10\% FBS, 2mM L-glutamine, 100 U/mL penicillin/streptomycin, and 2 $\mu$g/mL puromycin (Gibco). The antibiotic selection continued for one week before transferring cells to the antibiotic-free media.\\

\noindent {\bf TCR-expressing Jurkat 76.7 cell lines.} Nucleotide sequences of TCRa and TCRb chains were generated using stitchr~\cite{Heather2022-nr}. Both chains were modified to incorporate murine constant segments (TRAC*01 and TRBC2*01) to enhance TCR surface expression. For NYESO and EBV-specific TCRs, the TCRa and TCRb chains were connected through a P2A ribosomal skip motif, synthesized as a single bicistronic gene fragment by Genscript, and cloned into the lentiviral pLVX-EF1a-P2A-mCherry-IRES-Puro (Clontech). For MAGE-specific TCR, gene fragments for the TCRa and TCRb chains were separately synthesized by Genscript and cloned into either the lentiviral pLVX-EF1a-P2A-mCherry-IRES-Puro (Clontech) or pLVX-EF1$\alpha$-IRES-G418 (modified in-house from pLVX-EF1$\alpha$-IRES-Puro). The 293T packaging cell line (ATCC CRL-3216) was transfected with TCR-encoding lentiviral vectors, psPAX2 packaging plasmid (Addgene plasmid \#12260), and pMD2.G envelope plasmid (Addgene plasmid \#12259) using Lipofectamine 3000 Transfection Reagent (Invitrogen). The lentivirus-containing supernatant was collected after 24 and 48-hours after transfection, filtered through a 0.45$\mu$m SFCA filter (Thermo Scientific), concentrated using Lenti-X Concentrator (Takara Bio) according to the manufacturer’s protocol, resuspended in DPBS (Gibco), and stored at -80C for future use. To generate Jurkat cells expressing the paired TCR, TCR-null CD8-positive Jurkat 76.7 cells with NFAT-eGFP reporter were transduced with lentivirus. At 48 hours post-transduction, cells were transferred to RPMI media containing 10\% FBS, 2mM L-glutamine, 100 U/mL penicillin/streptomycin, and 1 $\mu$g/mL puromycin (Gibco). For MAGE-specific cell line generation Jurkat cells were co-transduced with lentiviruses encoding TCRa and TCRb chains and cultured with both 1 $\mu$g/mL puromycin and 500 $\mu$g/mL Genectin (Gibco). The antibiotic selection continued for one week. The abTCR expression on the surface was confirmed by mCherry expression and staining with a 1:100 dilution of anti-mouse TCRb APC-Fire750 (clone H57-597, 109246, Biolegend).\\

\noindent {\bf Specificity validation of TCR-expressing Jurkat.} Original cognate peptides and their variants were synthesized as a crude peptide library with by Genscript, and then diluted to a 4mM stock solution in DMSO (Sigma). To examine the specificity of the TCR-expressing Jurkat cell lines, we co-cultured 100,000 Jurkat cells with equal amounts of artificial APCs expressing the corresponding HLA allele in a 96-well round-bottom tissue culture plate. We tested each specificity—MAGE, EBV, NYESO—under 96 different conditions. These included one or two positive control original peptides and an unstimulated “no peptide” control. The cells were cultured in RPMI media supplemented with 10\% FBS, 2mM L-glutamine, 100 U/mL penicillin/streptomycin, and 1 $\mu$g/ml 
each of anti-human CD28 (BD Biosciences, 555725) and CD49d (BD Biosciences, 555501). The co-cultures were pulsed with 1$\mu$M peptide in triplicate. We added DMSO to unstimulated control wells (no peptide) at concentrations equivalent to those in the peptide-stimulated wells. After incubating the cells for 24 hours at 37$^\circ$C and 5\% CO2, we washed them once in DPBS (Gibco) and resuspended them in 150$\mu$l DPBS for flow cytometry analysis using a custom-configured BD Fortessa with FACSDiva software (Beckton Dickenson). We determined GFP percentages using FlowJo version 10.8.1 software (BD Biosciences). Alternatively, we measured the GFP percentage in wells of a 96-well round-bottom plate using Cellcyte X (Echo), using 4x objective in spheroid mode, acquisition in green channel (800 ms exposure 3dB gain). Image analysis was done with CellCyte Studio software with total spheroid area recipe and standard defaults (50 a.u. Contrast Sensitivity, 2 a.u Smoothing, 100 $\mu$m$^2$ filled hole size 100 $\mu$m Min. object size), relative GFP+ spheroid area was used as output. For NY-ESO, we compared the T-cell activity level measured using  flow cytometry with  fluorescence microscopy, and show consistency between the two (Fig.~\ref{fig:flow_vs_microscopy}). Given the consistency between the two  experimental approaches, we relied on fluorescence microscopy only to measure GFP levels induced by different peptides for MAGE and EBV.  \\

\noindent {\bf Selecting the GFP expression threshold for T-cell activity.} Experimental measurements are done across three replicates, and the reported GFP levels in Figs.~4,~\ref{fig:design_scatterplots} for each condition are averaged over the three replicates. We consider a peptide to have activated T-cells  if its induced replicate-averaged GFP level met two criteria: 1) it exceeded $0.5\%$, and 2) it surpassed  the mean plus three standard deviations of the GFP level negative controls (no-peptide). 
According to these criteria, the  GFP activation cutoffs were 0.5\% for the NY-ESO system,  0.64\% for the EBV system, and  0.76\% for the MAGE system.

\section{Experimental activities of the designed peptides}
\label{sec:exp_analysis}

\noindent {\bf Comparing design success rates across the three TCR-pMHC systems.} Among the designed peptides, 50\% in the  NY-ESO system (Fig. 4A), 16\% in the  EBV system (Fig.~4B), and 26\% in the MAGE system induce significant T-cell activities. However, it should be noted that the designed NY-ESO peptides were closer in sequence to their wild-type templates (2-7 amino acid differences) than the peptides designed for the EBV (3-9 differences) or MAGE (3-8 differences) systems. We find that the activity of designs decays with increasing distance from the respective wild-type, with almost no designs beyond 5 amino-acid distance from the wild-types activating their respective T-cells (Fig.~4,~\ref{fig:design_experimental_nyeso},~\ref{fig:design_experimental_ebv},~\ref{fig:design_experimental_mage}). This underscores the difficulty of exploring peptide variants far from the wild-type sequences. If any positive designs do exist at these greater distances, they represent an exponentially smaller fraction of the sequence space, and when in complex with the TCR-MHC, are likely to adopt a different binding mode compared to that of the wild-type.

Minor adjustments to the design protocols, aimed at better exploring the design space, may have also contributed to the observed differences in performance across the systems. For NY-ESO, which achieved the highest design success rate, peptides with highest PAE scores were  chosen for experimental validation, whereas in EBV and MAGE, peptides were sampled uniformly above the set PAE thresholds.  In MAGE, fixing the amino-acid at position 1 to a Glutamic Acid (E) for the one-third of the designs led to an elevated success rate in this subset (Fig.~\ref{fig:design_experimental_mage}D);  E$_1$  is shared between the two wild-type peptides and this design choice allowed us to assess the importance of this amino acid for function. \\

\noindent{\bf The design potential of  different HERMES models.} In panel A of Figures~\ref{fig:design_experimental_nyeso}~(NY-ESO),~\ref{fig:design_experimental_ebv}~(EBV), and~\ref{fig:design_experimental_mage}~(MAGE), we show the fidelity of designs using the four HERMES models: HERMES-{\em fixed} with no noise, HERMES-{\em fixed} with 0.5~\AA~noise, HERMES-{\em relaxed} with no noise, and HERMES-{\em relaxed} with 0.5~\AA~noise. Although PyRosetta relaxations allows HERMES-{\em relaxed} to explore the sequence space more broadly, the resulting designs that are made at greater distances from the wild-types seldom succeed in activating T-cells. This pattern is also observed with using models trained with 0.5~\AA~Gaussian noise.   Indeed, successful designs lie within five amino acid substitutions from the  wild-types, and for these designs, the un-noised HERMES-{\em fixed} model demonstrates the highest overall fidelity. To better explore the peptide energy landscape and to design sequences further from the wild-types, it may be necessary to adopt a more specialized relaxation model that accounts for the intricacies of TCR-pMHC poses, while being consistent with the neural network's model of protein environments.\\

\noindent{\bf Comparing TCRdock and AF3 as design filters.} Using our T-cell activity measurements as the ground truth, we sought to quantify how well the TCRdock and AF3 PAE scores can be used as predictors of T-cell activation and filtering mechanisms for peptide design. For this purpose, we focus primarily on the NY-ESO system, for which we validated a balanced  mix of positive and negative designs, chosen based on their TCRdock PAE scores. Here, we found that TCRdock PAE could moderately distinguish true positive from true negative designs \GM{(AUROCs = 0.91 and 0.88, Fig.~\ref{fig:design_roc_curves}A)}. Notably, using TCRdock PAE resulted in a very low false negative rate (0.03\%) but a considerably higher false positive rate (50\%); see Fig.~\ref{fig:design_scatterplots}A top. AF3 PAE scores performed equivalently to TCRdock PAE in differentiating positives from negatives (AUROCs = 0.89 and 0.94, Fig.~\ref{fig:design_roc_curves}A), though they correlated only moderately with TCRdock’s PAE values predictions (Spearman correlation $\rho=0.5$, Fig.~\ref{fig:design_scatterplots}C). Indeed, some peptides deemed positive by TCRdock were classified as negative by AF3, and vice versa (Fig.~\ref{fig:design_scatterplots}C).

\GM{When restricting the analysis to peptides that were deemed as positive based on their TCRdock and AF3 PAE scores, we did not expect either TCRdock  or AF3 PAE to effectively discriminate between peptides that did or did not elicit a T-cell response in our experimental assay (i.e., true vs. false positives). However, we found that both TCRdock PAE and AF3 PAE scores retained significant predictive power in the NY-ESO system (AUROCs = 0.81 and 0.77, and AUROCs = 0.80 and 0.89, Fig.~\ref{fig:design_roc_curves}A). In the EBV and MAGE systems,  the TCRdock  and AF3 PAE values for  peptides that we classified positive  showed lower predictive power in distinguishing  true from false positives compared to the NY-ESO system (Fig.~\ref{fig:design_roc_curves}B-C). The PAE scores of TCRdock and AF3 differed most markedly in the MAGE system (Spearman correlation $\rho = 0.09$ for MAGE), with TCRdock assigning relatively unfavorable PAE scores to the wild-type peptides, whereas AF3 ranked those same peptides near the top (Fig.~\ref{fig:design_scatterplots}I).}

The PAE values reported by the structure prediction algorithms have been shown to be noisy indicators of protein folding fidelity and design quality~\cite{Roney2022-am}. While larger experimental libraries will be needed to draw definitive conclusions, it appears that PAEs from TCRdock and AF3 have some complementary predictive strengths, and therefore, combining their information could lead to more robust  design decisions.
\GM{Moreover, both of these scores appear to be uncorrelated or only weakly correlated with HERMES scores across systems (Figs.~\ref{fig:design_scatterplots},~\ref{fig:design_tcrdock_nyeso}H,~\ref{fig:design_tcrdock_ebv}H,~\ref{fig:design_tcrdock_mage}H), suggesting that AF3 and TCRdock PAE's  can serve as complementary filters to HERMES during design.\\
}

\GM{
\noindent {\bf  HERMES energy  as design filter.} We further sought to assess whether HERMES energy score itself could serve as a predictor of T-cell activation and as a filtering criterion for  design.
For each of the three systems, we scored all peptides we designed with all four combinations of HERMES protocols, regardless of which protocol was originially used for design.  For each system, we then computed two sets of ROC curves, separately for peptides designed  with HERMES-{\em fixed} and those designed with  HERMES-{\em relaxed}.
We found that scores from the HERMES-{\em relaxed} models  consistently performed poorly in predicting T-cell activation, with only one out of six cases achieving an  AUROC better than random; HERMES-{\em relaxed} 0.00 as scoring criterion on the peptides designed with HERMES-{\em fixed} in  the EBV system showing AUROC = 0.66 (Fig.~\ref{fig:design_roc_curves}B).
In contrast, HERMES-{\em fixed} 0.00   was consistently the best filtering criterion, outperforming even TCRdock and AF3 PAE score in several cases: in the EBV system,   HERMES-{\em fixed} 0.00 scores on peptides designed by HERMES-{\em fixed} reached AUROC~$=0.74$; in the MAGE system, AUROC were 0.65 and 0.89 for   and peptides designed with HERMES-{\em fixed} and HERMES-{\em relaxed}, respectively (Fig.~\ref{fig:design_roc_curves}).
These results highlight two main takeaways: first, that HERMES-{\em fixed} 0.00 is the most reliable model for both scoring and designing peptides to elicit T-cell response; second, selecting high-scoring peptides using this model can improve hit rates, though potentially at the cost of reduced diversity in the peptide pool.
}

\section{T-cell and MHC specificity as the entropy of the associated peptide distribution}
 We characterize the specificity of a TCR-MHC complex and that of the MHC by the diversity of peptides that they can recognize.  For TCR-MHC specificity, we starting with a TCR-pMHC structural template, and use  HERMES-{\em fixed} model to generates a peptide position weight matrix (PWM) that captures the ensemble of peptides recognized by the TCR and presented by the MHC. In Figure 5, we analyzed 105 human and 36 mouse MHC class I TCR-pMHC structures---curated by the TCRdock template database \cite{Bradley2023-rl}---which included only MHC alleles represented in at least two structures. For each MHC allele in this set, we then characterized the distribution of peptides that the MHC can present (independently of the TCR) using PWMs inferred by the MHC Motif Atlas \cite{Tadros2023-ij}. In both cases, we use the entropy of these PWMs, $S= -\sum_{\alpha, i} p_i(\alpha) \log_2p_i(\alpha)$, as a proxy for degeneracy of the target (TCR-MHC or MHC), where $\alpha \in \{1,\dots 20\}$ denotes the amino acid types, and $i \in\{ 1, \dots, \ell\}$ are positions along the  peptide.
 
\section{Predicting MHC-I peptide presentation preferences with HERMES.} HERMES is designed to be broadly applicable to diverse protein-interaction contexts. As such, we attempted to use HERMES-{\em fixed} 0.00 model to recapitulate the ensemble of peptides presented by an MHC-I allele. To do so, we gathered 416 peptide-MHC-I crystal structures from the TCR3D database~\cite{gowthaman_tcr3d_2019} (Table~\ref{table:pmhc_info}). For each structure, we computed a peptide position position weight matrix (PWM) with the HERMES-{\em fixed} 0.00 design protocol, conditioning on the surrounding MHC structure and the masked wild-type peptide as the template. The resulting PWMs were then averaged  across complexes that shared the same MHC allele and peptide length to produce an allele-specific peptide motif (for a given peptide length). 

We compared these HERMES-generated PWMs with the corresponding PWMs from the MHC Motif Atlas~\cite{Tadros2023-ij}, which we consider as ground truth; we used relative entropy (KL-Divergence), $D_{KL}= \sum_{\alpha, i} p^{\text{motif}}_i(\alpha) \log_2 \frac{p^{\text{motif}}_i(\alpha)}{p^{\text{HERMES}}_i(\alpha)}$ as well as the entropy difference between the two PWMs as metrics; here, $p_i^{(\cdot)}(\alpha)$ is the probability of  amino acid $\alpha$ at position $i$ of a peptide, given the morel specified in the superscript. When structural coverage is extensive (roughly more than 30 structures of distinct peptides with a given MHC allele),  HERMES-derived motifs align closely with those from  MHC Motif Atlas, as quantified by the low KL-divergence and comparable entropy values (Fig.~\ref{fig:pmhc_pwm_entropy_comparisons}). Under sparse coverage, HERMES motifs begin to underestimate the peptide diversity  that can be presented by an MHC allele (measured by motif entropy), a limitation we attribute to incomplete sampling of peptide backbone conformations by HERMES-{\em fixed} (Fig.~\ref{fig:pmhc_pwm_entropy_comparisons}C). Overall, we see a clear pattern that the KL-Divergence increases with decreasing number of structural templates for a given MHC allele (Fig.~\ref{fig:pmhc_pwm_entropy_comparisons}B). Importantly, even with only a handful of templates, HERMES still captures substantially more diversity  than a na\"ive baseline PWM generated by simply aligning the template peptides themselves (Fig.~\ref{fig:pmhc_pwm_entropy_comparisons}A). Moreover, in almost all cases, HERMES generated PWMs  are more similar (lower KL-Divergence) to that of the MHC Motif Atlas than the na\"ive baseline PWM are (Fig.~\ref{fig:pmhc_pwm_entropy_comparisons}A).
 
 These results suggest that accurate modeling of MHC presentation with HERMES-{\em fixed} design protocol requires both sufficient  structure templates and peptide diversity. It should be noted that  MHC Motif Atlas's allele‑specific motifs--derived from the large body of peptide binding data curated for individual MHC-I alleles--are very reliable, so we continue to use them as the benchmark for assessing motif specificity. Developing design protocols that better sample peptide backbone conformations remains an important avenue for future work.

\clearpage{}
\newpage{}

\begin{table}
\centering
\begin{tabular}{l l l}
\hline
\textbf{TCR / Ab} & \textbf{(PDBid, WT peptide)} & \textbf{MHC allele} \\
\hline \hline
1G4 TCR & (2BNR~\cite{chen_structural_2005}, SLLMWITQC); (2BNQ~\cite{chen_structural_2005}, SLLMWITQV) & HLA-A*02:01 \\
A6 TCR & (1AO7~\cite{Garboczi1996-fr}, LLFGYPVYV); (1QSF~\cite{ding_four_1999}, LLFGYPVAV) & HLA-A*02:01 \\
\hline
TCR1 &                    (5D2N~\cite{Yang2015-cj}, NLVPMVATV) & HLA-A*02:01 \\
TCR2 &               (AF3 model, NLVPMVATV) & HLA-A*02:01 \\
TCR3 &               (AF3 model, NLVPMVATV) & HLA-A*02:01 \\
TCR4 &               (AF3 model, IMDQVPFSV) & HLA-A*02:01 \\
TCR5 &               (AF3 model, IMDQVPFSV) & HLA-A*02:01 \\
TCR6 &               (AF3 model, IMDQVPFSV) & HLA-A*02:01 \\
TCR7 &               (AF3 model, GRLKALCQR) & HLA-B*27:05 \\
H2-scDb (Fab) &                    (6W51~\cite{Hsiue2021-ag}, HMTEVVRHC) & HLA-A*02:01 \\
\hline
\end{tabular}
\caption{\textbf{Structure templates used for the binding affinity and T-cell activity analyses.} \normalfont{The first two rows contain information related to the analyses on binding affinity data, the rest are related to T-cell activity data. Every structure identified by a 4-letter PDBid was downloaded from the Protein Data Bank. The raw PDB file for 5D2N~\cite{Yang2015-cj} has two copies of the TCR-pMHC complex, so we deleted one before running any analysis. Similarly, the raw PDB file for 6W51~\cite{Hsiue2021-ag} has the Antibody and pMHC coordinates of two different copies within the crystal, so we expanded the crystal using PyMOL's \texttt{symexp} command and kept a single bound Antibody-pMHC complex. The sequence inputs used to fold TCR-pMHC complexes with AlphaFold3~\cite{Abramson2024-jp} can be found in \GM{Dataset~\ref{dataset:af3_sequences}}}.\\\\\\
}
\label{table:affinity_and_activity_info}
\end{table}

\begin{table}[ht]
\centering
\begin{tabular}{l l c c}
\hline
\GM{Model type} & Model & 1G4 TCR & A6 TCR \\
\hline \hline
\multirow{3}{*}{\GM{Substitution Matrix}} & BLOSUM62 & {\bf\underline{0.83}} \scriptsize{(0.00)} & 0.56* \scriptsize{(0.02)} \\
& \GM{Luksza et al. $M$} & \GM{0.44* \scriptsize{(0.12)} }& \GM{0.24 \scriptsize{(0.38)}} \\
& \GM{Luksza et al. $d \cdot M$} & \GM{0.48* \scriptsize{(0.08)}} & \GM{0.29 \scriptsize{(0.27)}} \\
\hline
\multirow{2}{*}{\GM{Structure Predictor}} & TCRdock & 0.62* \scriptsize{(0.02)} & {\bf\underline{0.88}} \scriptsize{(0.00)} \\
 & TCRdock-{\em no template} & 0.25 \scriptsize{(0.39)} & -0.14 \scriptsize{(0.62)} \\
\hline
 \multirow{2}{*}{\GM{Sequence-based}} &\GM{TAPIR} &  \GM{0.29 \scriptsize{(0.32)} }& \GM{0.35 \scriptsize{(0.19)}} \\
 & \GM{NetTCR-2.2} & \GM{-0.25 \scriptsize{(0.40)}} & \GM{-0.27 \scriptsize{(0.36)} }\\
\hline
\multirow{9}{*}{\GM{Structure-based}} & \GM{ESM-IF1} & \GM{0.38* \scriptsize{(0.17)}} & \GM{0.17 \scriptsize{(0.53)}} \\
& ProteinMPNN 0.02 \GM{Pep 1-by-1} & -0.01 \scriptsize{(0.97)} & 0.49 \scriptsize{(0.05)} \\
 & ProteinMPNN 0.20 \GM{Pep 1-by-1} & 0.12 \scriptsize{(0.69)} & 0.29 \scriptsize{(0.27)} \\
& \GM{ProteinMPNN 0.02 Pep}  & \GM{-0.01 \scriptsize{(0.97)}} & \GM{0.49 \scriptsize{(0.06)}} \\
 & \GM{ProteinMPNN 0.20 Pep} & \GM{0.14 \scriptsize{(0.63)} }& \GM{0.29 \scriptsize{(0.27)}} \\
 & \GM{ProteinMPNN 0.02 Pep and TCR} & \GM{0.07 \scriptsize{(0.81)}} & \GM{0.54 \scriptsize{(0.03)}} \\
 & \GM{ProteinMPNN 0.20 Pep and TCR} & \GM{-0.08 \scriptsize{(0.79)} }& \GM{0.26 \scriptsize{(0.33)}} \\
\hline
 \multirow{6}{*}{\GM{Structure-based}} & HERMES-{\em fixed} 0.00 & 0.29 \scriptsize{(0.31)} & 0.63* \scriptsize{(0.01)} \\
 & HERMES-{\em fixed} 0.50 & 0.19 \scriptsize{(0.51)} & -0.28 \scriptsize{(0.29)} \\
 & HERMES-{\em relaxed} 0.00 & 0.72* \scriptsize{(0.00)} & 0.56* \scriptsize{(0.03)} \\
 & HERMES-{\em relaxed} 0.50 & 0.65* \scriptsize{(0.01)} & 0.06 \scriptsize{(0.83)} \\
 & \GM{HERMES-{\em relaxed}-min\_energy 0.00} & \GM{0.58* \scriptsize{(0.03)}} & \GM{0.71* \scriptsize{(0.00)}} \\
 & \GM{HERMES-{\em relaxed}-min\_energy 0.50} & \GM{0.52* \scriptsize{(0.06)}} & \GM{0.51 \scriptsize{(0.04)}} \\
\hline
\end{tabular}
\caption{{\bf Benchmarking of models for predicting TCR-pMHC binding affinities.}
\normalfont{The table lists the Spearman correlation~$\rho \,{\rm (p-value)}$ of models' predictions against experimentally measured binding affinities for (i) 1G4 TCR in complex with HLA-A*02:01 binding to the  variants of the NY-ESO-1 peptide, and (ii) the A6 TCR in complex with HLA-A*02:01  binding to the variants of the Tax peptide. \GM{The HERMES model names specify the scoring scheme (fixed, relaxed, and relaxed with peptide energy computed from the relaxed conformation with the minimum Rosetta energy),  and the noise amplitude (in \AA) injected to the coordinates of the structure data during training. Similarly, the ProteinMPNN model names capture scoring scheme (1-by-1 (eq.~\ref{eq:peplogp_proteinmpnn}), the whole peptide (eq.~\ref{eq:peplogp_proteinmpnn_pep}), and  the peptide and TCR interface together (eq.~\ref{eq:peplogp_proteinmpnn_tcr_and_pep})), and the noise amplitude injected to the training data. We show in \textbf{bold} the best model for each dataset, and any model whose performance is not significantly different  from the best model (p-value$>0.05$ for the difference in the Fisher's z-transformed Spearman correlations) is indicated with  an asterisk (*). }
}}
\label{table:spearmanr_affinity}
\end{table}

\begin{table}[ht]
\centering
\resizebox{1.0\textwidth}{!}{
\begin{tabular}{ l l c c c c c c c c c }
\hline
\GM{Model type} & Model & H2-scDb (Fab) & TCR1 & TCR2 & TCR3 & TCR4 & TCR5 & TCR6 & TCR7 \\
 \hline \hline
\multirow{3}{*}{\GM{Sub. Matrix}} & BLOSUM62 & -0.05 \scriptsize{(0.52)} & -0.07 \scriptsize{(0.47)} & 0.14 \scriptsize{(0.20)} & 0.38 \scriptsize{(0.00)} & 0.33* \scriptsize{(0.01)} & 0.42* \scriptsize{(0.00)} & 0.55* \scriptsize{(0.00)} & -0.01 \scriptsize{(0.89)} \\
& \GM{Luksza et al. $M$} & \GM{0.19 \scriptsize{(0.02)}} & \GM{$^\dagger$0.25 \scriptsize{(0.01)}} & \GM{$^\dagger$0.39 \scriptsize{(0.00)}} & \GM{$^\dagger$0.53* \scriptsize{(0.00)}} & \GM{$^\dagger$0.25* \scriptsize{(0.03)}} & \GM{$^\dagger$0.27 \scriptsize{(0.00)}} & \GM{$^\dagger$0.36* \scriptsize{(0.00)}} & \GM{$^\dagger$0.01 \scriptsize{(0.94)}} \\
& \GM{Luksza et al. $d \cdot M$} & \GM{0.46* \scriptsize{(0.00)}} & \GM{$^\dagger$0.50* \scriptsize{(0.00)}} & \GM{$^\dagger$0.66* \scriptsize{(0.00)}} & \GM{$^\dagger$0.40 \scriptsize{(0.00)}} & \GM{$^\dagger${\bf \underline{0.38}} \scriptsize{(0.00)}} & \GM{$^\dagger$ {\bf \underline{0.55}} \scriptsize{(0.00)}} & \GM{$^\dagger${0.42}* \scriptsize{(0.00)}} & \GM{$^\dagger$0.27* \scriptsize{(0.00)}} \\
\hline
\multirow{2}{*}{\GM{Struc. Pred.}} & TCRdock & N/A & 0.02 \scriptsize{(0.83)} & 0.13 \scriptsize{(0.24)} & 0.21 \scriptsize{(0.03)} & 0.18 \scriptsize{(0.15)} & -0.02 \scriptsize{(0.83)} & 0.19 \scriptsize{(0.04)} & 0.06 \scriptsize{(0.50)} \\
& TCRdock-{\em no template} & N/A & -0.03 \scriptsize{(0.77)} & 0.44 \scriptsize{(0.00)} & 0.11 \scriptsize{(0.27)} & -0.22 \scriptsize{(0.08)} & 0.28 \scriptsize{(0.00)} & 0.14 \scriptsize{(0.14)} & 0.06 \scriptsize{(0.50)} \\
\hline
\multirow{4}{*}{\GM{Seq.-based}} & \GM{TAPIR} & \GM{N/A} & \GM{0.36 \scriptsize{(0.00)}} & \GM{0.36 \scriptsize{(0.00)}} & \GM{-0.00 \scriptsize{(0.97)}} & \GM{0.12 \scriptsize{(0.32)}} & \GM{0.39* \scriptsize{(0.00)}} & \GM{0.08 \scriptsize{(0.42)} }& \GM{-0.01 \scriptsize{(0.94)}} \\
& \GM{NetTCR-2.2} & \GM{N/A} & \GM{-0.05 \scriptsize{(0.59)}} & \GM{0.20 \scriptsize{(0.06)}} & \GM{0.36 \scriptsize{(0.00)}} & \GM{-0.21 \scriptsize{(0.09)}} & \GM{0.23 \scriptsize{(0.01)}} & \GM{0.20 \scriptsize{(0.03)}} & \GM{-0.22 \scriptsize{(0.02)}} \\
& TULIP & N/A & 0.12 \scriptsize{(0.24)} & 0.18 \scriptsize{(0.09)} & -0.12 \scriptsize{(0.20)} & 0.04 \scriptsize{(0.73)} & 0.24 \scriptsize{(0.00)} & 0.29 \scriptsize{(0.00)} & N/A \\
& TULIP* (from~\cite{Meynard-Piganeau2024-po}) & N/A & 0.47* \scriptsize{(0.00)} & 0.44 \scriptsize{(0.00)} & 0.14 \scriptsize{(0.07)} & 0.23* \scriptsize{(0.00)} & 0.20 \scriptsize{(0.01)} & 0.09 \scriptsize{(0.25)} & N/A \\
\hline
\multirow{9}{*}{\GM{Struc.-based}} & \GM{ESM-IF1} & \GM{0.12 \scriptsize{(0.13)}} & \GM{0.43* \scriptsize{(0.00)}} & \GM{0.41 \scriptsize{(0.00)}} & \GM{0.19 \scriptsize{(0.05)}} & \GM{-0.15 \scriptsize{(0.22)}} & \GM{0.28 \scriptsize{(0.00)}} & \GM{0.25 \scriptsize{(0.01)}} & \GM{0.08 \scriptsize{(0.40)}} \\
& ProteinMPNN 0.02 \GM{1-by-1} & 0.53* \scriptsize{(0.00)} & 0.46* \scriptsize{(0.00)} & 0.53 \scriptsize{(0.00)} & 0.49* \scriptsize{(0.00)} & -0.23 \scriptsize{(0.06)} & 0.45* \scriptsize{(0.00)} & 0.42* \scriptsize{(0.00)} & {\bf \underline{0.31}} \scriptsize{(0.00)} \\
& ProteinMPNN 0.20 \GM{1-by-1} & 0.03 \scriptsize{(0.65)} & 0.27 \scriptsize{(0.01)} & 0.46 \scriptsize{(0.00)} & 0.40 \scriptsize{(0.00)} & -0.23 \scriptsize{(0.06)} & 0.26 \scriptsize{(0.00)} & {\bf \underline{0.49}} \scriptsize{(0.00)} & 0.02 \scriptsize{(0.79)} \\
& \GM{ProteinMPNN 0.02 Pep} & \GM{{\bf\underline{0.55}} \scriptsize{(0.00)}} & \GM{0.46* \scriptsize{(0.00)}} & \GM{0.52 \scriptsize{(0.00)}} & \GM{0.49* \scriptsize{(0.00)}} & \GM{-0.28 \scriptsize{(0.02)}} & \GM{0.45* \scriptsize{(0.00)}} & \GM{0.41* \scriptsize{(0.00)}} & \GM{0.31 \scriptsize{(0.00)}} \\
& \GM{ProteinMPNN 0.20 Pep} & \GM{0.04 \scriptsize{(0.64)}} & \GM{0.27 \scriptsize{(0.01)}} & \GM{0.44 \scriptsize{(0.00)}} & \GM{0.40 \scriptsize{(0.00)}} & \GM{-0.23 \scriptsize{(0.05)}} & \GM{0.25 \scriptsize{(0.00)}} & \GM{0.48* \scriptsize{(0.00)}} & \GM{0.02 \scriptsize{(0.83)}} \\
& \GM{ProteinMPNN 0.02 Pep and TCR} & \GM{0.50* \scriptsize{(0.00)}} & \GM{0.44* \scriptsize{(0.00)}} & \GM{0.53 \scriptsize{(0.00)}} & \GM{0.34 \scriptsize{(0.00)}} & \GM{-0.24 \scriptsize{(0.05)}} & \GM{0.51* \scriptsize{(0.00)}} & \GM{0.46* \scriptsize{(0.00)}} & \GM{0.29* \scriptsize{(0.00)}} \\
& \GM{ProteinMPNN 0.20 Pep and TCR} & \GM{-0.02 \scriptsize{(0.84)}} & \GM{0.20 \scriptsize{(0.04)}} & \GM{0.47 \scriptsize{(0.00)}} & \GM{0.30 \scriptsize{(0.00)}} & \GM{-0.23 \scriptsize{(0.06)}} & \GM{0.32 \scriptsize{(0.00)}} & \GM{0.48* \scriptsize{(0.00)}} & \GM{-0.02 \scriptsize{(0.81)}} \\
\hline
\multirow{6}{*}{\GM{Struc.-based} }& HERMES-{\em fixed} 0.00 & 0.46* \scriptsize{(0.00)} & {\bf \underline{0.56}} \scriptsize{(0.00)} & {\bf \underline{0.71}} \scriptsize{(0.00)} & {\bf \underline{0.59}} \scriptsize{(0.00)} & -0.06 \scriptsize{(0.60)} & 0.37 \scriptsize{(0.00)} & 0.42* \scriptsize{(0.00)} & -0.01 \scriptsize{(0.95)} \\
& HERMES-{\em fixed} 0.50 & 0.26 \scriptsize{(0.00)} & 0.42* \scriptsize{(0.00)} & 0.61* \scriptsize{(0.00)} & 0.54* \scriptsize{(0.00)} & 0.03 \scriptsize{(0.83)} & 0.26 \scriptsize{(0.00)} & 0.31 \scriptsize{(0.00)} & 0.02 \scriptsize{(0.82)} \\
& HERMES-{\em relaxed} 0.00 & 0.29 \scriptsize{(0.00)} & 0.31 \scriptsize{(0.00)} & 0.57 \scriptsize{(0.00)} & 0.55* \scriptsize{(0.00)} & -0.20 \scriptsize{(0.11)} & 0.37 \scriptsize{(0.00)} & 0.42* \scriptsize{(0.00)} & 0.14 \scriptsize{(0.15)} \\
& HERMES-{\em relaxed} 0.50 & 0.16 \scriptsize{(0.04)} & 0.37 \scriptsize{(0.00)} & 0.50 \scriptsize{(0.00)} & 0.45* \scriptsize{(0.00)} & -0.16 \scriptsize{(0.20)} & 0.23 \scriptsize{(0.00)} & 0.06 \scriptsize{(0.49)} & 0.08 \scriptsize{(0.37)} \\
& \GM{HERMES-{\em relaxed}-min\_energy 0.00} & \GM{0.18 \scriptsize{(0.02)}} & \GM{0.32 \scriptsize{(0.00)}} & \GM{0.56* \scriptsize{(0.00)}} & \GM{0.50* \scriptsize{(0.00)}} & \GM{-0.10 \scriptsize{(0.39)}} & \GM{0.37 \scriptsize{(0.00)}} & \GM{0.45* \scriptsize{(0.00)}} & \GM{0.04 \scriptsize{(0.65)}} \\
& \GM{HERMES-{\em relaxed}-min\_energy 0.50} & \GM{0.13 \scriptsize{(0.09)} }& \GM{0.39 \scriptsize{(0.00)}} & \GM{0.51 \scriptsize{(0.00)}} & \GM{0.46* \scriptsize{(0.00)}} & \GM{-0.12 \scriptsize{(0.34)}} & \GM{0.15 \scriptsize{(0.07)}} & \GM{0.17 \scriptsize{(0.07)}} & \GM{0.08 \scriptsize{(0.38)}} \\
\hline
\end{tabular}}
\caption{{\bf Benchmarking of models for predicting peptide-induced T-cell activities.}
\normalfont{The table lists the Spearman correlation~$\rho \,{\rm (p-value)}$ of models' predictions against experimentally measured T-cell activities.  The correlations for TCR1-7 are reported based on the EC$_{50}$ values obtained using the procedure outlined in Section~\ref{sec:aff_act} and Eq.~\ref{eq:EC50}, consistent with  correlations reported in Fig.~3. In addition, we present the Spearman correlations between TULIP predictions and the fitted EC$_{50}$ values as reported in ref.~\cite{Luksza2022-gx}, which are the same correlations reported in the TULIP paper~\cite{Meynard-Piganeau2024-po} (TULIP* row). The TULIP* correlations also include datapoints for which T-cell responses were  below the detection limit, and which we exclude from our analysis (gray points in Fig.~3). \GM{We note that the substitution matrix - with ($C$) and without ($C/d$) the position-based scaling $d$ - by Luksza et al.~\cite{Luksza2022-gx} was constructed precisely using experimental activity data from TCR1-7; we note predictions made on these measurements with a dagger$^\dagger$ to indicate that they are not representative of the matrix' ability to predict T-cell activity in new settings. 
The HERMES model names specify the scoring scheme (fixed, relaxed, and relaxed with peptide energy computed from the relaxed conformation with the minimum Rosetta energy),  and the noise amplitude (in \AA) injected to the coordinates of the structure data during training. Similarly, the ProteinMPNN model names capture scoring scheme (1-by-1 (eq.~\ref{eq:peplogp_proteinmpnn}), the whole peptide (eq.~\ref{eq:peplogp_proteinmpnn_pep}), and  the peptide and TCR interface together (eq.~\ref{eq:peplogp_proteinmpnn_tcr_and_pep})), and the noise amplitude injected to the training data. 
We show in \textbf{bold} the best model for each dataset, and any model whose performance is not significantly different  from the best model (p-value$>0.05$ for the difference in the Fisher's z-transformed Spearman correlations) is indicated with  an asterisk (*). }
}}
\label{table:spearmanr_activity}
\end{table}

\begin{table}[ht]
\centering
\AN{\begin{tabular}{l c c c c c c c c}
 & \multicolumn{4}{c}{with templates} & \multicolumn{4}{c}{without templates} \\
 \cline{2-9}
\textbf{TCR / Ab} & RMSD & PAE & ipTM & pTM & RMSD & PAE & ipTM & pTM \\
\hline \hline
1G4 TCR & 0.72 & 1.06 & 0.94 & 0.94 & 0.82 & 1.08 & 0.93 & 0.94 \\
1G4 TCR w/ 9V on peptide & 0.72 & 1.07 & 0.94 & 0.94 & 0.82 & 1.08 & 0.93 & 0.94 \\
A6 TCR & 0.87 & 0.91 & 0.95 & 0.95 & 0.83 & 0.91 & 0.95 & 0.95 \\
A6 TCR w/ 8A on peptide & 0.67 & 1.00 & 0.94 & 0.94 & 0.73 & 1.03 & 0.94 & 0.95 \\
\hline
TCR1 & 0.92 & 1.52 & 0.88 & 0.91 & 1.12 & 2.46 & 0.82 & 0.86 \\
TCR2 & N/A & 1.63 & 0.86 & 0.89 & N/A & 1.86 & 0.84 & 0.87 \\
TCR3 & N/A & 0.84 & 0.95 & 0.95 & N/A & 0.86 & 0.95 & 0.95 \\
TCR4 & N/A & 2.07 & 0.80 & 0.83 & N/A & 2.01 & 0.84 & 0.86 \\
TCR5 & N/A & 1.17 & 0.91 & 0.92 & N/A & 1.19 & 0.91 & 0.93 \\
TCR6 & N/A & 1.14 & 0.92 & 0.94 & N/A & 1.16 & 0.91 & 0.93 \\
TCR7 & N/A & 1.37 & 0.84 & 0.87 & N/A & 1.42 & 0.89 & 0.91 \\
H2-scDb (Fab) & 13.89 & 19.44 & 0.48 & 0.54 & 11.57 & 19.2 & 0.47 & 0.55 \\
\hline
eng. TCR w/ MageA3 peptide & 0.62 & 1.03 & 0.92 & 0.94 & 0.69 & 1.03 & 0.92 & 0.94 \\
eng. TCR w/ Titin peptide & 0.64 & 1.05 & 0.92 & 0.94 & 0.65 & 1.03 & 0.92 & 0.94 \\
TK3 TCR w/ HPVG peptide & 1.05 & 1.11 & 0.92 & 0.94 & 1.05 & 1.05 & 0.93 & 0.94 \\
TK3 TCR w/ HPVG EQ5 peptide & 1.01 & 1.07 & 0.92 & 0.94 & 1.06 & 1.07 & 0.92 & 0.94 \\
\hline
\end{tabular}}
\caption{\GM{\textbf{AlphaFold3 folding metrics for  structures used in our analyses.}
\normalfont{
Root-mean-square deviation (RMSD) is computed between the $\alpha$-C's of AF3 predicted structures and  the corresponding experimentally determined structure. The reported PAE values are averaged across the interfaces of TCR ($\alpha$ and $\beta$ chains) and pMHC. \AN{ipTM and pTM values are reported  from the AF3's confidence output.} For template-guided predictions, we used the default template cutoff in the AlphaFold Server (September 9th 2021). If an AF3 prediction incorrectly docked the TCR on the opposite side of the pMHC, the run was discarded  and repeated. In the case of  the H2-scDb bi-specific Fab-pMHC complex,  although AF3  placed the Fab on the correct side of  the pMHC, the docking pose was consistently inaccurate, resulting in high RMSD and  PAE values. This was the only system for which AF3 failed to produce a realistic structure model. Additional details on experimental structures, peptide, TCR, and MHC sequences used in AF3 runs can be found in Tables~\ref{table:affinity_and_activity_info},~\ref{table:design_info}, Dataset~\ref{dataset:af3_sequences}. Example AF3-generated structures are shown in Figure~\ref{fig:af3_folded_structures_example}.}}}
\label{table:af3_folding_metrics}
\end{table}

\begin{table}
\centering
\begin{tabular}{l l l l}
\hline
\textbf{System} & \textbf{TCR} & \textbf{(PDBid, WT peptide)} & \textbf{MHC allele} \\
\hline \hline
NY-ESO & 1G4 TCR & (2BNR~\cite{chen_structural_2005}, SLLMWITQC); (2BNQ~\cite{chen_structural_2005}, SLLMWITQV) & HLA-A*02:01 \\
EBV & TK3 TCR & (3MV7*~\cite{gras_allelic_2010}, HPVGEADYFEY); (4PRP*~\cite{Liu2014-cz}, HPVGQADYFEY)	 & HLA-B*35:01 \\
MAGE & eng. TCR & (5BRZ*~\cite{Raman2016-ew}, EVDPIGHLY); (5BS0*~\cite{Raman2016-ew}, ESDPIVAQY) & HLA-A*01:01 \\
\hline
\end{tabular}
\caption{\textbf{Structure templates used for  peptide design.} \normalfont{For the PDB entries marked with an asterisk (*), we used the curated structures from the TCRdock database~\cite{Bradley2023-rl} instead of the PDB entries.} 
}
\label{table:design_info}
\end{table}

\begin{table}[ht]
\centering
\renewcommand\arraystretch{1.2} 
\resizebox{1.0\textwidth}{!}{
\begin{tabular}{l p{16cm}}
\hline
\textbf{MHC allele} & \textbf{(PDBid, WT peptide)} \\
\hline \hline
HLA-A*01:01 & (5BS0, ESDPIVAQY); (5BRZ, EVDPIGHLY) \\
HLA-A*02:01 & (4FTV, LLFGYPVYV); (5D2L, NLVPMVATV); (2F53, SLLMWITQC); (5TEZ, GILGFVFTL); (6VQO, HMTEVVRHC); (3HG1, ELAGIGILTV); (2P5E, SLLMWITQC); (7N6E, YLQPRTFLL); (6RP9, SLLMWITQV); (5C08, RQWGPDPAAV); (3H9S, MLWGYLQYV); (5C0B, RQFGPDFPTI); (6EQA, AAGIGILTV); (5EUO, GILGFVFTL); (3D39, LLFGPVYV); (6VRN, HMTEVVRHC); (5YXU, KLVALGINAV); (5EU6, YLEPGPVTV); (1BD2, LLFGYPVYV); (1OGA, GILGFVFTL); (2VLK, GILGFVFTL); (5MEN, ILAKFLHWL); (3D3V, LLFGPVYV); (1AO7, LLFGYPVYV); (5ISZ, GILGFVFTL); (5JHD, GILGFVFTL); (6DKP, ELAGIGILTV); (1QSE, LLFGYPRYV); (3QDM, ELAGIGILTV); (5JZI, KLVALGINAV); (2P5W, SLLMWITQC); (5C0C, RQFGPDWIVA); (7N1E, RLQSLQTYV); (4QOK, EAAGIGILTV); (5E6I, GILGFVFTL); (2VLR, GILGFVFTL); (6RSY, RMFPNAPYL); (3QEQ, AAGIGILTV); (3O4L, GLCTLVAML); (5C0A, MVWGPDPLYV); (5C07, YQFGPDFPIA); (5D2N, NLVPMVATV); (2F54, SLLMWITQC); (2VLJ, GILGFVFTL); (5NHT, ELAGIGILTV); (6TRO, GVYDGREHTV); (7N1F, YLQPRTFLL); (3GSN, NLVPMVATV); (6D78, AAGIGILTV); (3QDG, ELAGIGILTV); (5NQK, ELAGIGILTV); (5HHO, GILEFVFTL); (2PYE, SLLMWITQC); (5HHM, GILGLVFTL); (2GJ6, LLFGKPVYV); (5NMF, SLYNTIATL); (3UTS, ALWGPDPAAA); (5E9D, ELAGIGILTV); (6RPA, SLLMWITQV); (1QSF, LLFGYPVAV); (4EUP, ALGIGILTV); (5YXN, KLVALGINAV); (2BNQ, SLLMWITQV); (6RPB, SLLMWITQV); (5HYJ, AQWGPDPAAA); (3QFJ, LLFGFPVYV); (3UTT, ALWGPDPAAA); (3QDJ, AAGIGILTV); (4L3E, ELAGIGILTV); (5C09, YLGGPDFPTI); (6VRM, HMTEVVRHC); (1QRN, LLFGYAVYV); (2BNR, SLLMWITQC); (3PWP, LGYGFVNYI); (6AMU, MMWDRGLGMM); (6AM5, SMLGIGIVPV); (4MNQ, ILAKFLHWL); (5NME, SLYNTVATL); (5NMG, SLFNTIAVL) \\
HLA-A*11:01 & (5WKH, GTSGSPIINR); (5WKF, GTSGSPIVNR) \\
HLA-A*24:02 & (3VXR, RYPLTFGWCF); (3VXM, RFPLTFGWCF); (3VXS, RYPLTLGWCF) \\
HLA-B*07:02 & (6VMX, RPPIFIRRL); (6AVF, APRGPHGGAASGL); (6AVG, APRGPHGGAASGL) \\
HLA-B*08:01 & (1MI5, FLRGRAYGL); (3FFC, FLRGRAYGL); (4QRP, HSKKKCDEL); (3SJV, FLRGRAYGL) \\
HLA-B*27:05 & (4G8G, KRWIILGLNK); (4G9F, KRWIIMGLNK) \\
HLA-B*35:01 & (6BJ2, IPLTEEAEL); (4PRP, HPVGQADYFEY); (3MV7, HPVGEADYFEY); (3MV9, HPVGEADYFEY); (3MV8, HPVGEADYFEY) \\
HLA-B*35:08 & (4PRH, HPVGDADYFEY); (4PRI, HPVGEADYFEY) \\
HLA-B*44:05 & (3DXA, EENLLDFVRF); (3KPR, EEYLKAWTF); (3KPS, EEYLQAFTY) \\
H2Db & (3PQY, SSLENFRAYV); (5WLG, SQLLNAKYL); (5TJE, KAVYNFATM); (7JWI, ASNENMETM); (5SWS, ASNENMETM); (5TIL, KAPYNFATM); (7JWJ, ASNENMETM); (6G9Q, KAPYDYAPI); (5M02, KAPFNFATM); (5M00, KAVANFATM); (5M01, KAPANFATM) \\
H2Kb & (3PQY, SSLENFRAYV); (5WLG, SQLLNAKYL); (5TJE, KAVYNFATM); (7JWI, ASNENMETM); (5SWS, ASNENMETM); (5TIL, KAPYNFATM); (7JWJ, ASNENMETM); (6G9Q, KAPYDYAPI); (5M02, KAPFNFATM); (5M00, KAVANFATM); (5M01, KAPANFATM) \\
H2Ld & (2OI9, QLSPFPFDL); (4NHU, GGGAPWNPAMMI); (4MVB, QPAEGGFQL); (4MXQ, SPAPRPLDL); (2E7L, QLSPFPFDL); (4N5E, VPYMAEFGM); (6L9L, SPSYAYHQF); (3E3Q, QLSPFPFDL); (3E2H, QLSPFPFDL); (3TPU, FLSPFWFDI); (3TFK, QLSDVPMDL); (4MS8, SPAEAGFFL); (4N0C, MPAGRPWDL); (3TJH, SPLDSLWWI) \\
\hline
\end{tabular}}
\caption{\textbf{Structure templates to compute TCR-MHC specificities.} \normalfont{The table list the structures used to compute TCR-MHC specificities in Fig.~5 for different MHC class I alleles in humans and mice. Structures were collected from the TCRdock curated database~\cite{Bradley2023-rl}.}}
\label{table:specificity_info}
\end{table}

\begin{table}[ht]
\renewcommand\arraystretch{1.2} 
\resizebox{1.0\textwidth}{!}{
\GM{
\begin{tabular}{c c p{15cm}}
\hline
\textbf{MHC allele} & \textbf{Peptide Length} & \textbf{PDBid}                                                                                                                                                                                                                                                                                                                                                                                                                                                                                                                                                                                                                                       
\\
\hline \hline
HLA-A*01:01    & 9              & 3BO8; 4NQV; 4NQX                                                                                                                                                                                                                                                                                                                                                                                                                                                                                                                                                                                                                               \\
HLA-A*01:01    & 10             & 6AT9; 6MPP                                                                                                                                                                                                                                                                                                                                                                                                                                                                                                                                                                                                                                     \\
HLA-A*02:01    & 9              & 1B0G; 1DUZ; 1EEY; 1EEZ; 1HHG; 1HHI; 1HHJ; 1HHK; 1I1F; 1I1Y; 1I7R; 1I7T; 1I7U; 1JHT; 1QEW; 1QR1; 1S9W; 1S9X; 1S9Y; 1TVB; 2C7U; 2GTW; 2GTZ; 2GUO; 2V2W; 2V2X; 2VLL; 2X4O; 2X4S; 2X4U; 3BGM; 3FT2; 3FT3; 3FT4; 3GSO; 3GSQ; 3GSR; 3GSU; 3GSV; 3GSW; 3GSX; 3H7B; 3H9H; 3HPJ; 3I6G; 3IXA; 3KLA; 3MR9; 3MRB; 3MRC; 3MRD; 3MRE; 3MRF; 3MRG; 3MRH; 3MRI; 3MRJ; 3MRK; 3MRL; 3MYJ; 3PWJ; 3PWL; 3PWN; 3QFD; 3TO2; 3V5D; 3V5H; 3V5K; 4I4W; 4K7F; 4L29; 5E00; 5EU3; 5HHN; 5HHP; 5HHQ; 5MEO; 5MEP; 5N6B; 5WSH; 6SS7; 6SS8; 6SS9; 6SSA; 6VR1; 6VR5; 6Z9W; 7EU2; 7LG2; 7LG3; 7MJ6; 7MJ9; 7MKB; 7N1A; 7N1B; 7RTD; 7SA2; 7SIS; 7U1R; 7U21; 7UM2; 7ZUC; 8FU4; 8RBV \\
HLA-A*02:01    & 10             & 1HHH; 1I4F; 1JF1; 2CLR; 2GT9; 3BH8; 3GIV; 3I6K; 3MGO; 3MGT; 3MRM; 3MRN; 3MRO; 3MRP; 3MRQ; 3MRR; 3UTQ; 4GKN; 4GKS; 4JFO; 4JFP; 4JFQ; 5C0D; 5C0G; 5C0I; 5FDW; 5N1Y; 6AMT; 6G3J; 6G3K; 6TRN; 7M8S; 7Q98; 7UR1                                                                                                                                                                                                                                                                                                                                                                                                                                     \\
HLA-A*02:01    & 11             & 5D9S; 5EOT; 7MJ7                                                                                                                                                                                                                                                                                                                                                                                                                                                                                                                                                                                                                               \\
HLA-A*02:06    & 9              & 6UJO; 6UJQ                                                                                                                                                                                                                                                                                                                                                                                                                                                                                                                                                                                                                                     \\
HLA-A*03:01    & 9              & 3RL1; 7L1B; 7L1C; 7MLE; 7UC5                                                                                                                                                                                                                                                                                                                                                                                                                                                                                                                                                                                                                   \\
HLA-A*03:271   & 10             & 8JHV; 8JHW                                                                                                                                                                                                                                                                                                                                                                                                                                                                                                                                                                                                                                     \\
HLA-A*11:01    & 9              & 1Q94; 7S8R; 8I5E                                                                                                                                                                                                                                                                                                                                                                                                                                                                                                                                                                                                                               \\
HLA-A*11:01    & 10             & 1QVO; 5WJN; 7OW3; 7S8Q; 7S8S; 8RBU; 8RH6; 8RHQ                                                                                                                                                                                                                                                                                                                                                                                                                                                                                                                                                                                                 \\
HLA-A*11:01    & 11             & 4MJ5; 4MJ6                                                                                                                                                                                                                                                                                                                                                                                                                                                                                                                                                                                                                                     \\
HLA-A*24:02    & 8              & 4F7T; 4WU5; 5HGA; 5HGB                                                                                                                                                                                                                                                                                                                                                                                                                                                                                                                                                                                                                         \\
HLA-A*24:02    & 9              & 2BCK; 3I6L; 6XQA; 7F4W; 7JYV; 7JYW; 7MJA; 7NMD                                                                                                                                                                                                                                                                                                                                                                                                                                                                                                                                                                                                 \\
HLA-A*24:02    & 10             & 3NFN; 3VXN; 3VXO; 3VXP; 3WL9; 3WLB; 4F7M; 5HGD; 5HGH; 7JYU                                                                                                                                                                                                                                                                                                                                                                                                                                                                                                                                                                                     \\
HLA-A*26:01    & 9              & 8XES; 8XFZ; 8XKC                                                                                                                                                                                                                                                                                                                                                                                                                                                                                                                                                                                                                               \\
HLA-A*68:01    & 10             & 1TMC; 6EI2                                                                                                                                                                                                                                                                                                                                                                                                                                                                                                                                                                                                                                     \\
HLA-B*07:02    & 9              & 4U1K; 5EO0; 5EO1; 5WMN; 7LFZ; 7LG0; 7LGD; 7RZD; 7RZJ; 7S7D; 7S7E; 7S7F; 7S8A; 7S8E; 7S8F                                                                                                                                                                                                                                                                                                                                                                                                                                                                                                                                                       \\
HLA-B*08:01    & 8              & 8E13; 8E8I                                                                                                                                                                                                                                                                                                                                                                                                                                                                                                                                                                                                                                     \\
HLA-B*08:01    & 9              & 3SKM; 3SKO; 4QRQ; 4QRT; 5WMQ; 5WMR; 7NUI                                                                                                                                                                                                                                                                                                                                                                                                                                                                                                                                                                                                       \\
HLA-B*15:01    & 9              & 5TXS; 6UZS; 6VB3; 7XF3; 8ELG; 8ELH                                                                                                                                                                                                                                                                                                                                                                                                                                                                                                                                                                                                             \\
HLA-B*15:02    & 9              & 6UZM; 6UZN; 6UZO; 6VB0; 6VB7; 6VIU                                                                                                                                                                                                                                                                                                                                                                                                                                                                                                                                                                                                             \\
HLA-B*18:01    & 8              & 4XXC; 8RNG; 8ROO; 8ROP                                                                                                                                                                                                                                                                                                                                                                                                                                                                                                                                                                                                                         \\
HLA-B*18:01    & 9              & 6MT3; 8RNH                                                                                                                                                                                                                                                                                                                                                                                                                                                                                                                                                                                                                                     \\
HLA-B*27:05    & 9              & 1JGE; 1UXS; 1W0V; 2A83; 2BSR; 2BST; 3B6S; 3BP4; 3DTX; 3LV3; 5IB1; 5IB2; 6PYW; 6Y2A; 7CIS                                                                                                                                                                                                                                                                                                                                                                                                                                                                                                                                                       \\
HLA-B*27:05    & 10             & 2BSS; 7CIQ; 7CIR; 7DYN                                                                                                                                                                                                                                                                                                                                                                                                                                                                                                                                                                                                                         \\
HLA-B*27:09    & 9              & 1K5N; 1W0W; 3B3I; 3BP7; 3CZF; 3HCV; 6Y27; 6Y2B                                                                                                                                                                                                                                                                                                                                                                                                                                                                                                                                                                                                 \\
HLA-B*35:01    & 9              & 1A9B; 1A9E; 1CG9; 3LKN; 3LKO; 3LKP; 3LKQ; 3LKR; 3LKS; 7SIG; 8EMF; 8EMG; 8EMI; 8EMJ                                                                                                                                                                                                                                                                                                                                                                                                                                                                                                                                                             \\
HLA-B*35:08    & 13             & 3VFM; 3VFR; 3VFT; 3VFU; 3VFV                                                                                                                                                                                                                                                                                                                                                                                                                                                                                                                                                                                                                   \\
HLA-B*37:01    & 9              & 6MT4; 6MT5; 6MT6                                                                                                                                                                                                                                                                                                                                                                                                                                                                                                                                                                                                                               \\
HLA-B*44:03    & 9              & 1SYS; 3KPO                                                                                                                                                                                                                                                                                                                                                                                                                                                                                                                                                                                                                                     \\
HLA-B*44:05    & 9              & 1SYV; 7TUC                                                                                                                                                                                                                                                                                                                                                                                                                                                                                                                                                                                                                                     \\
HLA-B*53:01    & 9              & 7R7V; 7R7W                                                                                                                                                                                                                                                                                                                                                                                                                                                                                                                                                                                                                                     \\
HLA-B*57:01    & 9              & 6D2T; 7R7X                                                                                                                                                                                                                                                                                                                                                                                                                                                                                                                                                                                                                                     \\
HLA-B*57:01    & 10             & 3VRI; 3VRJ; 6D29; 6V2O; 8F7M                                                                                                                                                                                                                                                                                                                                                                                                                                                                                                                                                                                                                   \\
HLA-B*57:03    & 9              & 2BVP; 6V2Q                                                                                                                                                                                                                                                                                                                                                                                                                                                                                                                                                                                                                                     \\
HLA-B*58:01    & 8              & 7X1B; 7X1C                                                                                                                                                                                                                                                                                                                                                                                                                                                                                                                                                                                                                                     \\
HLA-B*58:01    & 9              & 5VWH; 7WZZ; 7X00                                                                                                                                                                                                                                                                                                                                                                                                                                                                                                                                                                                                                               \\
HLA-B*81:01    & 9              & 4U1I; 4U1L; 4U1S                                                                                                                                                                                                                                                                                                                                                                                                                                                                                                                                                                                                                               \\
H2-Db      & 9              & 1BZ9; 1FFN; 1FFO; 1FFP; 1FG2; 1HOC; 1INQ; 1JPG; 1JUF; 1LD9; 1N5A; 1S7U; 1S7V; 1S7W; 1S7X; 1ZHB; 2F74; 2ZOL; 3BUY; 3CC5; 3CCH; 3CH1; 3CPL; 3FTG; 3QUK; 3QUL; 3TBS; 3TBT; 3TBV; 3TBW; 3TBY; 3WS3; 3WS6; 4HUU; 4HUV; 4HUW; 4HUX; 4HV8; 4IHO; 4L8B; 4L8C; 4L8D; 4NSK; 5E8N; 5E8O; 5E8P; 5JWD; 5MZM; 5WLI; 6G9R; 6L9M; 6L9N; 7N9J; 7P0A; 7P0T                                                                                                                                                                                                                                                                                                       \\
H2-Db      & 10             & 1N3N; 1WBX; 1WBY; 1YN6; 1YN7; 3L3H; 5JWE; 6H6D; 6H6H; 7N5Q                                                                                                                                                                                                                                                                                                                                                                                                                                                                                                                                                                                     \\
H2-Db      & 11             & 1JPF; 6X00                                                                                                                                                                                                                                                                                                                                                                                                                                                                                                                                                                                                                                     \\
H2-Dd      & 9              & 5KD7; 5T7G; 8D5E; 8D5F; 8FHL; 8FHU                                                                                                                                                                                                                                                                                                                                                                                                                                                                                                                                                                                                             \\
H2-Dd      & 10             & 1BII; 3E6H; 3ECB; 5KD4; 5WEU; 6NPR                                                                                                                                                                                                                                                                                                                                                                                                                                                                                                                                                                                                             \\
H2-Kb      & 8              & 1FZJ; 1G7Q; 1KBG; 1KJ3; 1KPU; 1LEG; 1LEK; 1LK2; 1NAN; 1OSZ; 1RK0; 1RK1; 1T0M; 1VAC; 2FO4; 2MHA; 2VAA; 2ZSV; 2ZSW; 3P4M; 3P4N; 3P4O; 3P9L; 3P9M; 3PAB; 3ROL; 3ROO; 4HS3; 4PV8; 4PV9; 6JQ2; 6JQ3                                                                                                                                                                                                                                                                                                                                                                                                                                                 \\
H2-Kb      & 9              & 1FZK; 1G7P; 1KPV; 1S7R; 1VAD; 1WBZ; 2VAB; 4PGB; 4PGD; 7WCY; 8P43 \\
H2-Kd      & 9              & 2FWO; 4WDI; 4Z76; 5TRZ; 5TS1; 8D5J; 8D5K \\
\hline
\end{tabular}}}
\caption{\GM{\textbf{Structure templates to compute PWMs of peptide presentation on MHC.}
The table lists the structures used to compute peptide PWMs in Fig.~\ref{fig:pmhc_pwm_entropy_comparisons} for different MHC class I alleles in humans and mice. Structures were collected from the Protein Data Bank.
}}
\label{table:pmhc_info}
\end{table}

\begin{figure}[ht]
\centering
	\includegraphics[width=0.63\textwidth]{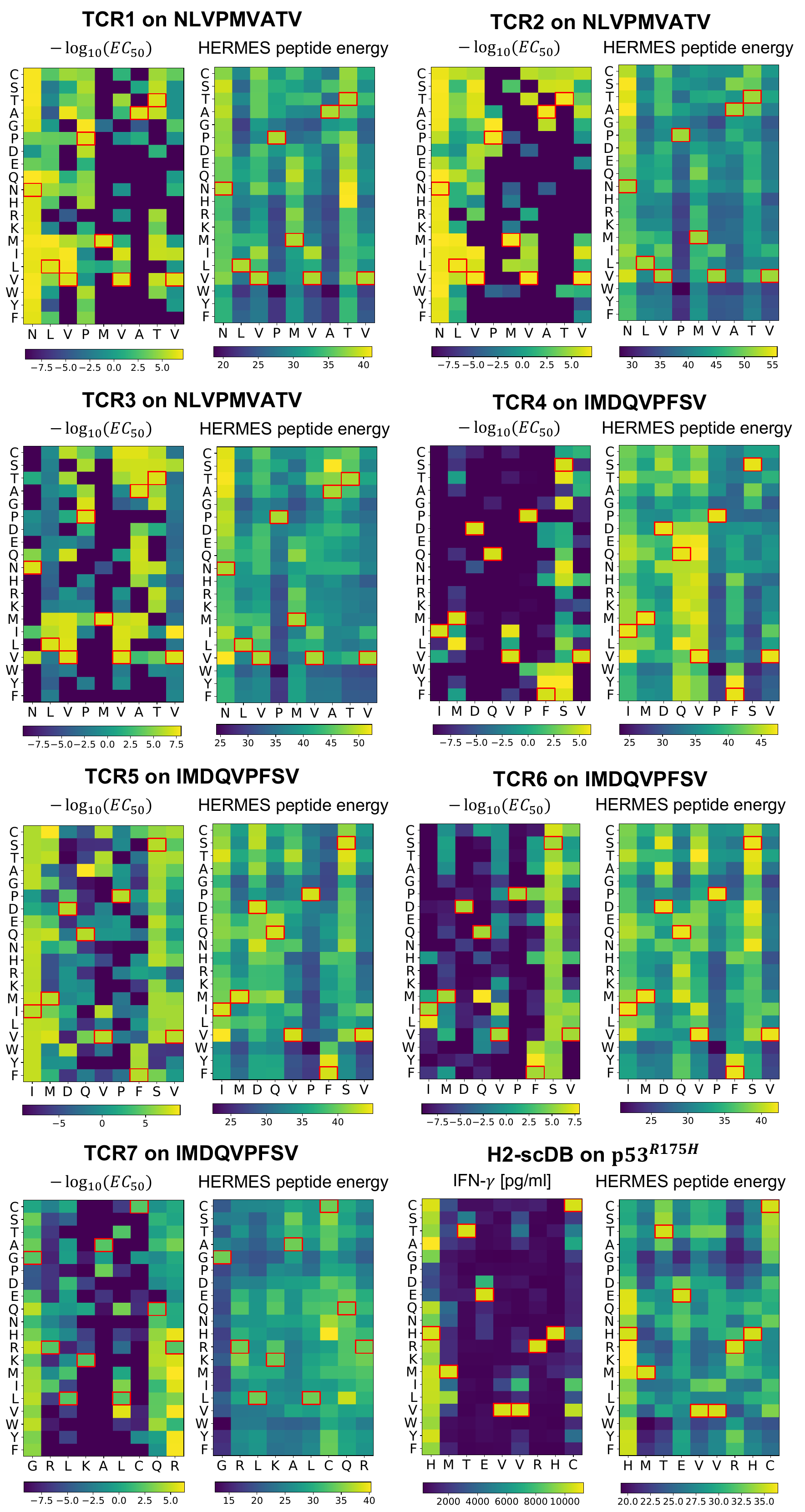}
\caption{{\bf Mutational effect on T-cell activities}. The heatmaps show the experimental measurements and the predictions from the HERMES-{\em fixed} model for the effect of  all single point mutations on peptides interacting with  TCR1-7~\cite{Luksza2022-gx}, and the designed bi-specific antibody Fab in the H2-scDB system~\cite{Hsiue2021-ag} (similar to the data presented in scatter plots in Fig.~3). In the heatmaps, each column corresponds to a position in the peptide and is labeled with its wild-type amino acid (indicated by a red square on the heatmaps), and the rows show the amino acid substitutions. }
\label{fig:activity_heatmaps}
\end{figure}

\begin{figure}[ht]
\centering
	\includegraphics[width=\textwidth]{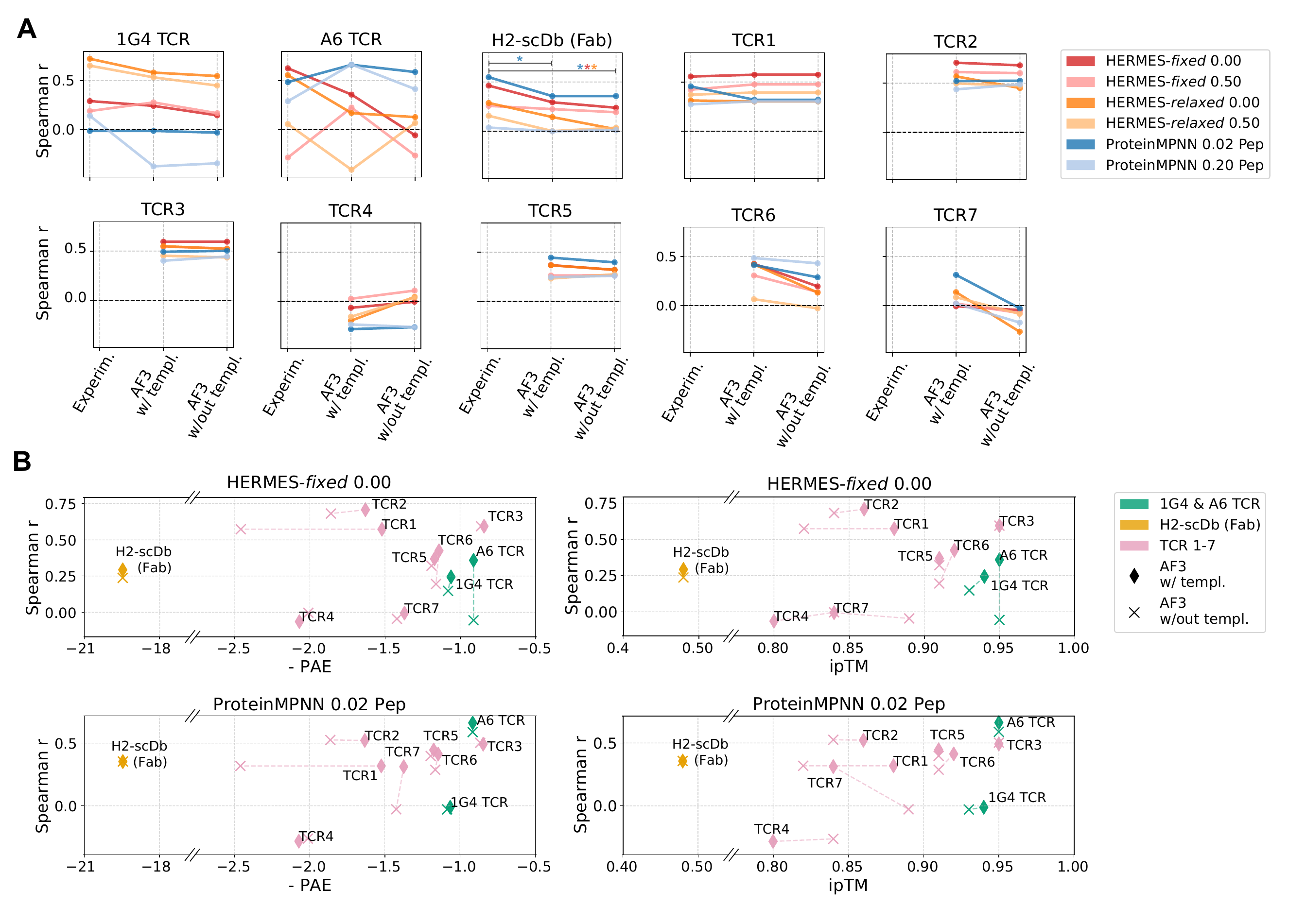}
\caption{\GM{\textbf{
 HERMES and ProteinMPNN prediction accuracies for AlphaFold3-predicted   structures.} {\bf (A)} For all systems analyzed, we used the AlphaFold Server to model the structures of complexes with the wild-type peptides, both with and without  template guidance; see Table~\ref{table:af3_folding_metrics} for details on the confidence of these AF3 models. For each system, the Spearman correlation ($r$) between   experimental   affinity or activity data and predictions from  HERMES and ProteinMPNN models (colors) is shown for three structure inputs: experimental structures (when available), AF3 model with template, and AF3 model without template.
Statistically significant drop in accuracy between models using experimental vs. AF3 structures--defined as cases with  p-value$<0.05$ for  difference in the Fisher's $z$-transformed Spearman correlations--are indicated with an asterisk (*) in the same color as that of the model (from the legend). Such drops are only observed for the H2-scDb Fab-pMHC system, where AF3 failed to generate a realistic structure. \AN{{\bf (B)} Spearman correlation ($r$) between experimental affinity/activity data and predictions from the overall best HERMES-{\em fixed} (top) and ProteinMPNN (bottom) models from (A), plotted against AF3 confidence metrics: negative predicted alignment error ($-$PAE; left) and interchain predicted TM-score (ipTM; right). Results are shown for AF3 structures generated with and without template guidance. The confidence metrics are reported in Table~\ref{table:af3_folding_metrics}.}
 The HERMES model names specify the scoring scheme (fixed, relaxed),  and the noise amplitude (in \AA) injected to the coordinates of the structure data during training. Similarly, the ProteinMPNN model names capture scoring scheme, i.e., the whole peptide (eq.~\ref{eq:peplogp_proteinmpnn_pep}), and the noise amplitude injected to the training data. 
}}
\label{fig:plots_af3_struc_ablation}
\end{figure}

\begin{figure}[ht]
\centering
	\includegraphics[width=0.8\textwidth]{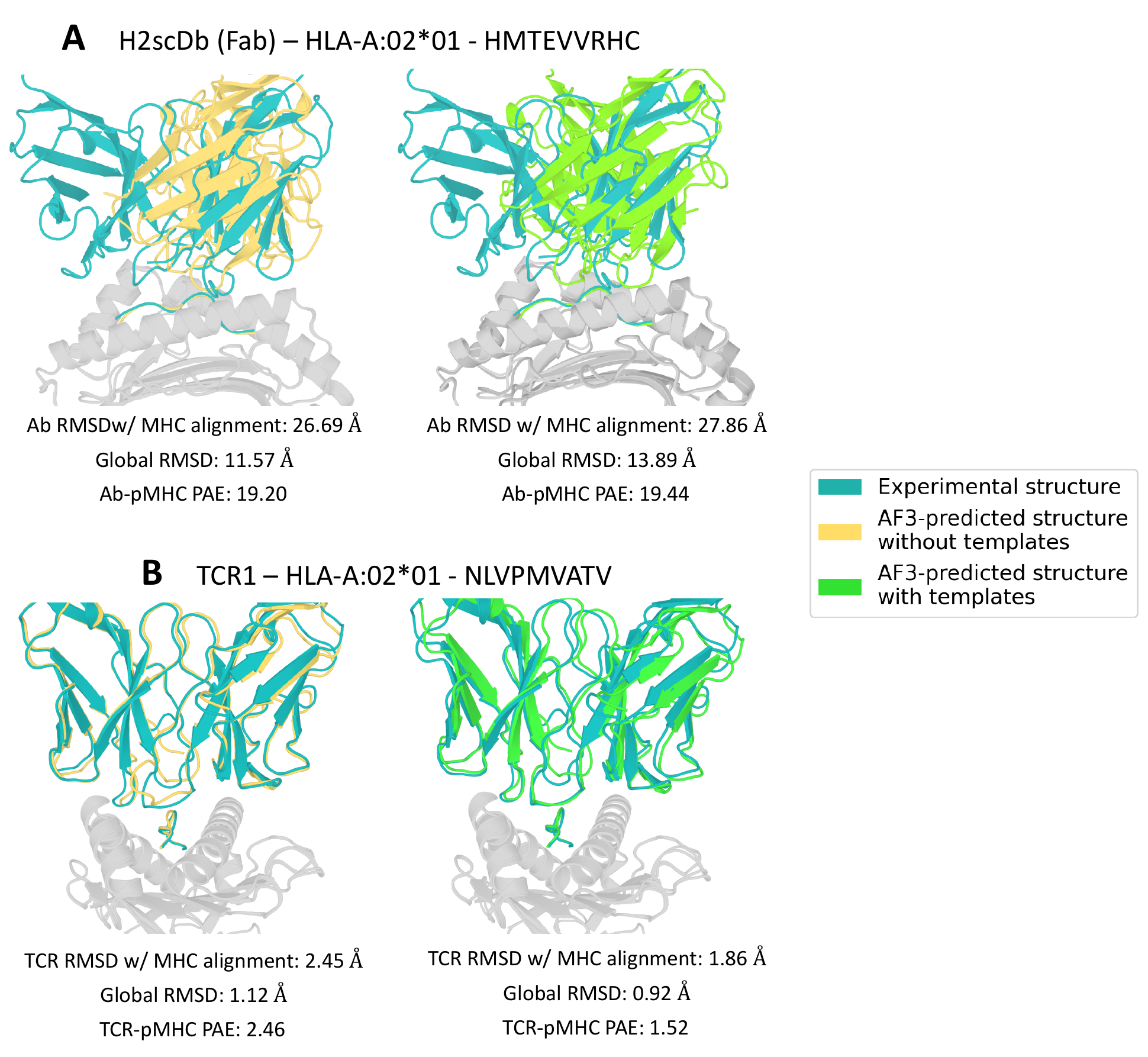}
\caption{\GM{\textbf{Comparison of experimentally determined structures with AlphaFold3 predicted structures.}
The experimentally determined structure of {\bf (A)} H2-scDb Fab-pMHC complex, and {\bf (B)} TCR1-pMHC complex are 
shown against the AF3 predicted structures generated without (left) and with (right) template guidance. All structures were aligned on the MHC using PyMOL to enable consistent visualiziation of  TCR/Fab docking geometries. AF3 failed to produce a realistic structure model for the H2-scDb system, as evidenced by the high RMSD and  interface  PAE values reported in Table~\ref{table:af3_folding_metrics}.
}}
\label{fig:af3_folded_structures_example}
\end{figure}

\begin{figure}[ht]
\centering
	\includegraphics[width=0.95\textwidth]{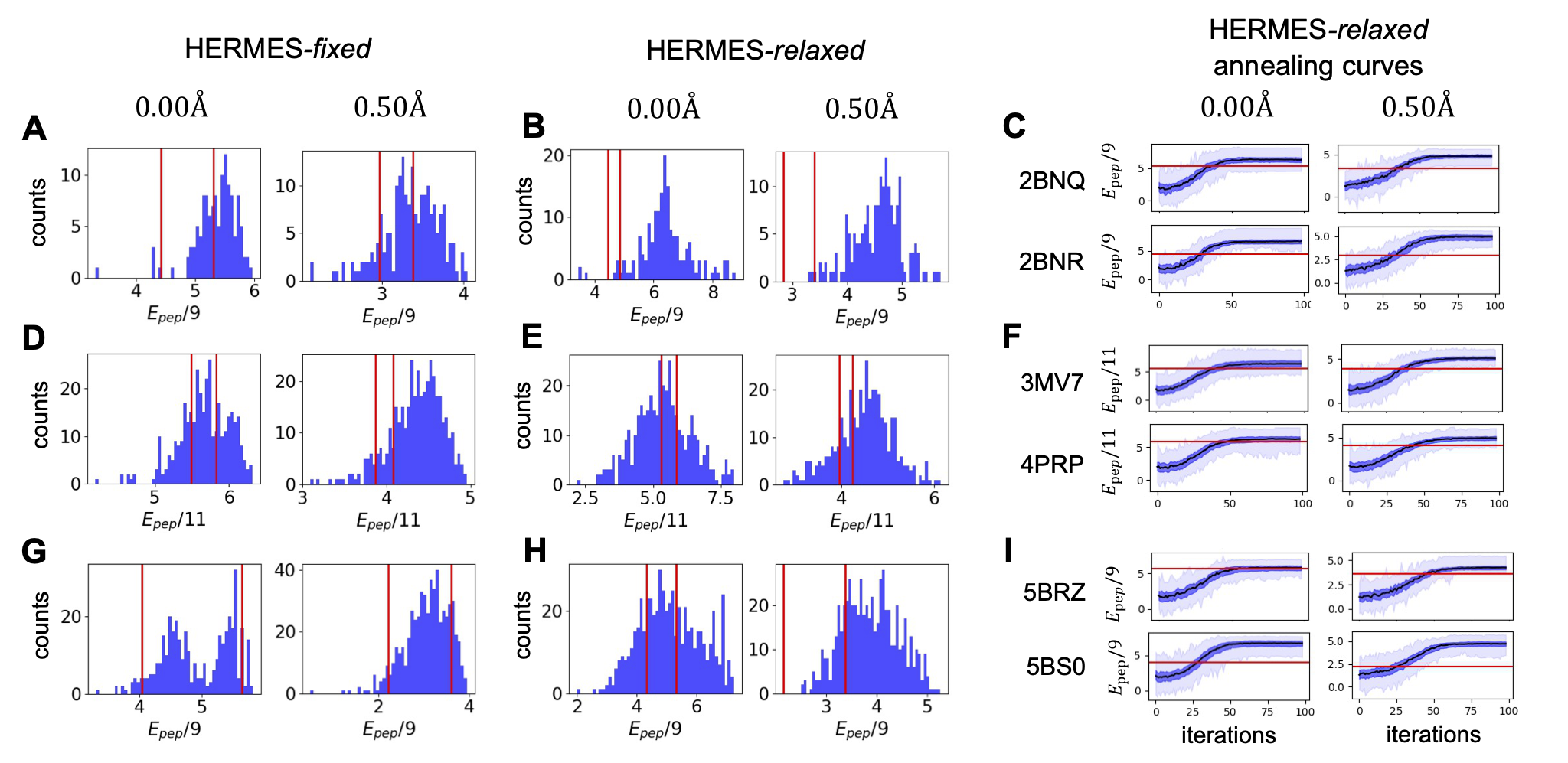}
\caption{ \GM{{\bf Peptide design energy profiles across systems and HERMES models.}
For the NY-ESO system, panels {\bf (A)} and {\bf (B)} show the  distribution of peptide energies (eq.~\ref{eq:pepEn}) divided by the  length, for peptides designed with HERMES-{\em fixed}  and HERMES-{\em relaxed} protocols, respectively.
Designs were pooled  across the two structure templates used (2BNQ and 2BNR), with the peptide energies computed using their respective template structures. In all panels, the red lines indicates the energies of the wild-type peptides in their corresponding structures. {\bf(C)} The scaled energy of the designed peptides are shown as a function of the MCMC iterations during  the HERMES-{\em relaxed} protocol, using the two different  structural templates for the NY-ESO system: 2BNQ (top) and 2BNR (bottom). The red line indicates the energy of the wild-type peptide in its respective structure. In panels (A-C), designs are grouped by the noise amplitude applied during   HERMES training (no noise: 0.00\AA, and moderate noise:  0.50\AA), as indicated above each panel. The HERMES-{\em relaxed} protocol consistently produces peptides with more favorable (higher) peptide energies than that of the wild-type, indicating that the protocol can successfully optimize within the peptide energy landscapes; models trained with noise are more successful in finding peptides with higher energy values.
Panels {\bf (D-F)}  and {\bf (G-I)} present analogous design statistics for the EBV and the MAGE systems, respectively.}}
\label{fig:joint_design_distributions}
\end{figure}

\begin{figure}[ht]
\centering
	\includegraphics[width=0.7\textwidth]{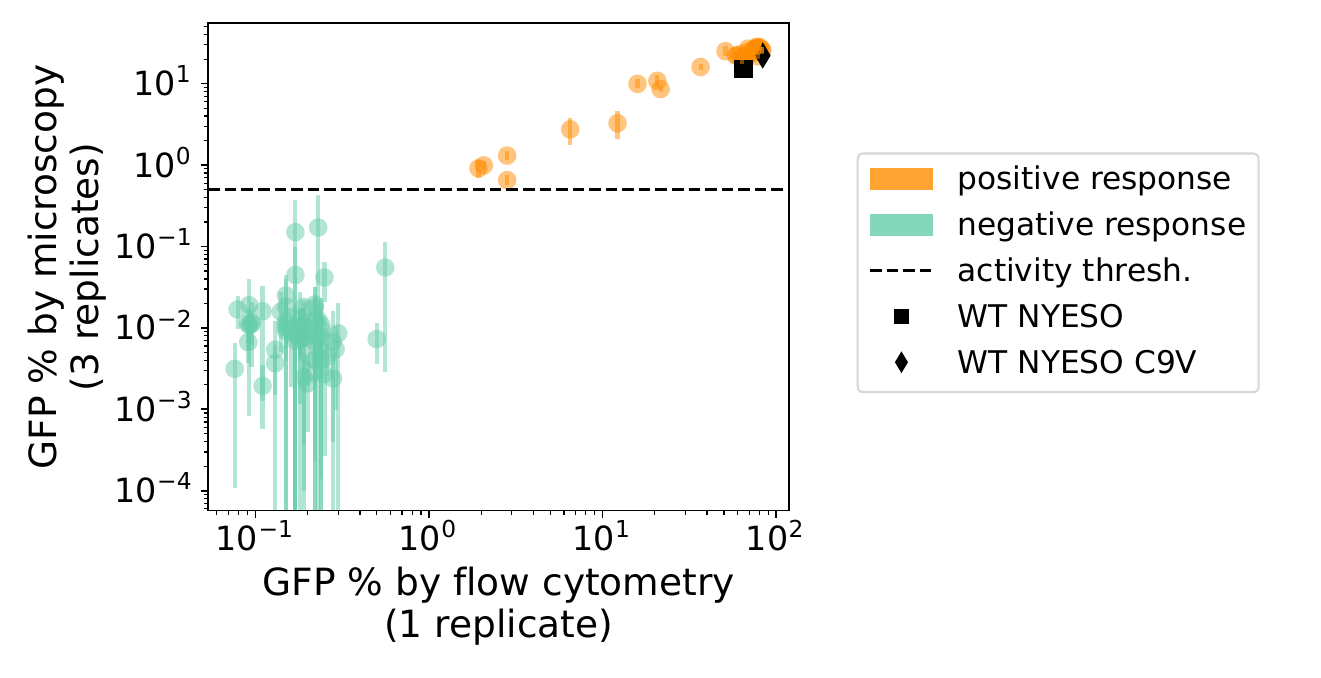}
\caption{\textbf{Measurements of GFP levels by  flow cytometry and fluorescence microscopy.} For the NY-ESO system, we performed measurements of GFP levels from peptide-induced  T-cell activities,  using both flow cytometry and  fluorescence microscopy. Fluorescence microscopy measurements were performed in  three independent replicates;  markers denote the mean GFP level,  and error bars show the standard deviation across the three replicates. Flow cytometry was done only on a single replicate.
The two measurement methods demonstrate good agreement, with a clear separation between designs that induced significant T-cell activation (orange) and those that did not (green). The black markers indicate the activities induced by the wild-type peptides.
}
\label{fig:flow_vs_microscopy}
\end{figure}

\begin{figure}[ht]
\centering
	\includegraphics[width=0.90\textwidth]{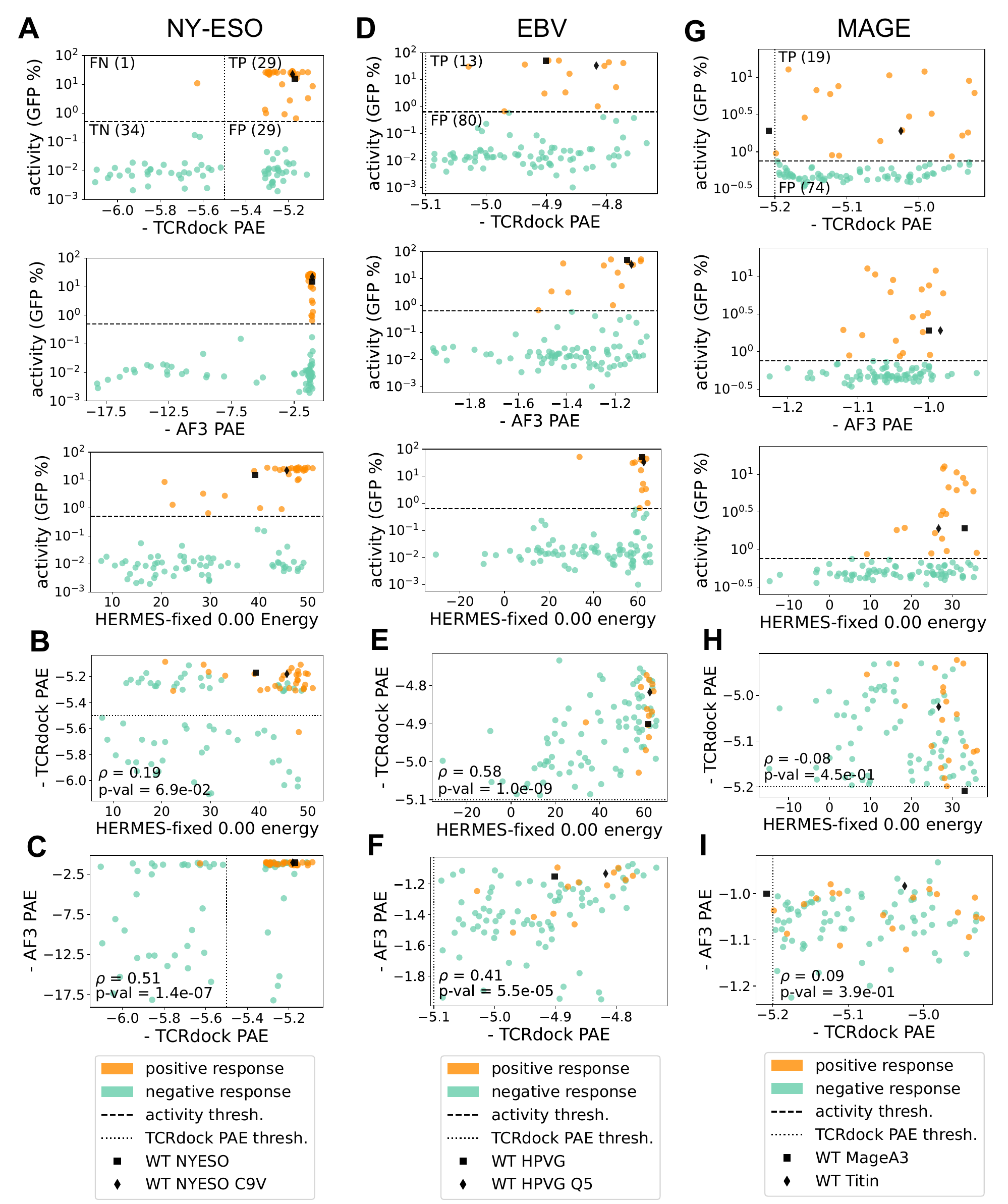}
\caption{\GM{{\bf TCRdock and AlphaFold3 PAEs for HERMES-designed peptides}. {\bf(A)} For the NY-ESO system, the induced T-cell activity by the designed peptides is shown against the computed TCRdock negative-PAE (first row), the AlphaFold3 negative-PAE (second row), and the energy score predicted with HERMES-{\em fixed} 0.00 (non noise during training) (third row); higher negative-PAE's indicate better designs. For the designed peptides in the NY-ESO system {\bf(B)} shows the relationship between negative TCRdock PAE values  and the  peptide energies from the HERMES-{\em fixed} 0.00. model, and {\bf (C)} shows the negative AlphaFold3 PAE versus TCRdock PAE values.  Peptides that induced significant T-cell responses are shown in orange, and non-responders are shown in green. The vertical dashed line indicate the filtering threshold for TCRdock PAE, with points to the right representing the negative design set. Panels {\bf (D-F)} and {\bf (G-I)} show analogous statistics for the EBV and the MAGE system, respectively. No negative designs (i.e., designs below the respective TCRdock negative-PAE thresholds) were experimentally  validated for these systems. We observe that the HERMES scores and TCRdock PAE values are uncorrelated or only weakly correlated across systems, and so are TCRdock and AF3 PAE, suggesting that AF3 and TCRdock PAE's  can serve as complementary filters to HERMES during design.
}}
\label{fig:design_scatterplots}
\end{figure}

\begin{figure}[ht]
\centering
	\includegraphics[width=0.99\textwidth]{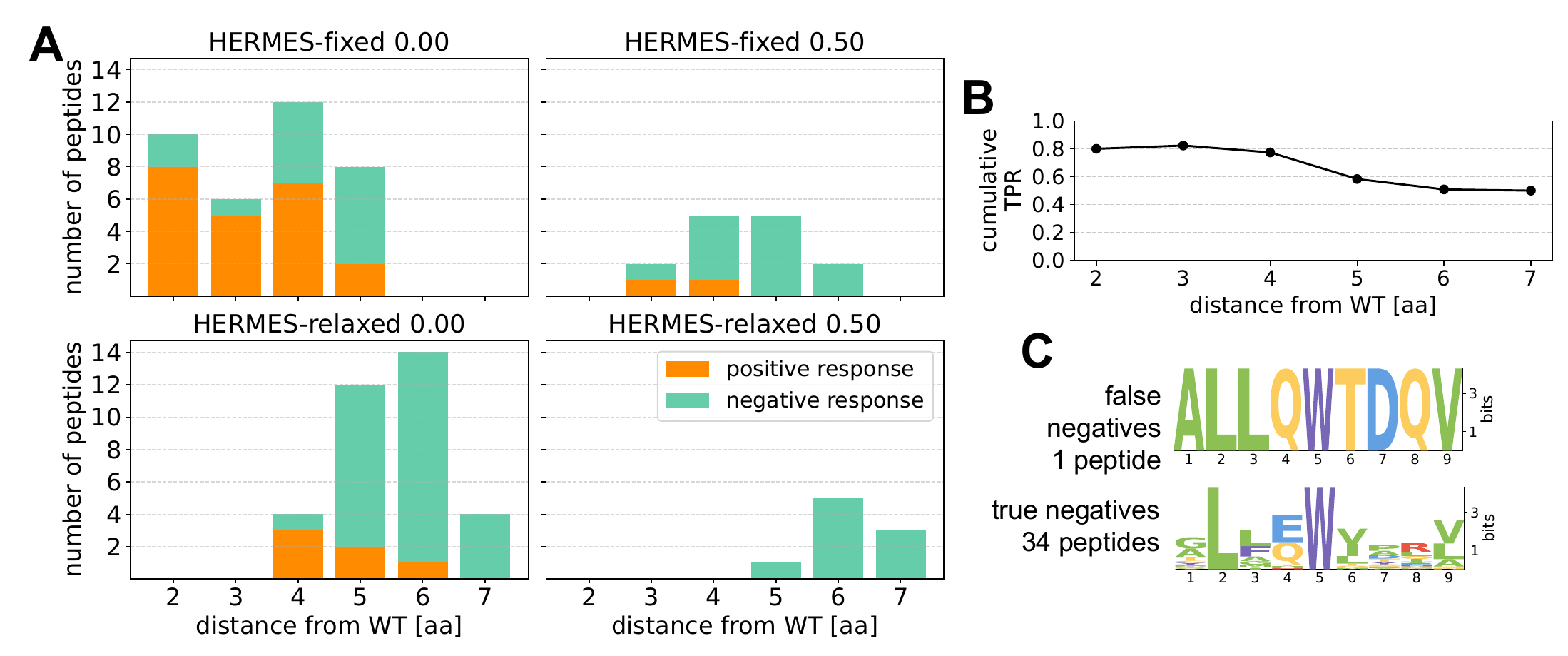}
\caption{\textbf{Experimental T-cell activation by HERMES-designed peptides in the NY-ESO system}. {\bf (A)} The number of peptides designed by different HERMES models (panels) that induced GFP expression beyond the T-cell activation threshold (positives; shown in orange) versus those that did not (negatives; shown in green) is shown. The designed peptides are grouped according to their minimum Hamming distance from any of the wild-type sequences used as design templates (Table~\ref{table:design_info}). \GM{The HERMES model names specify the scoring scheme (fixed, relaxed),  and the noise amplitude (in \AA) injected to the coordinates of the structure data during training. }
{\bf (B)} The cumulative True Positive Rate (TPR) for designed peptides is shown as a function of their Hamming distance from any of the wild-types. {\bf (C)} Logo plots show the sequence composition of the False Negative and the True Negative designs. }
\label{fig:design_experimental_nyeso}
\end{figure}

\begin{figure}[ht]
\centering
	\includegraphics[width=0.99\textwidth]{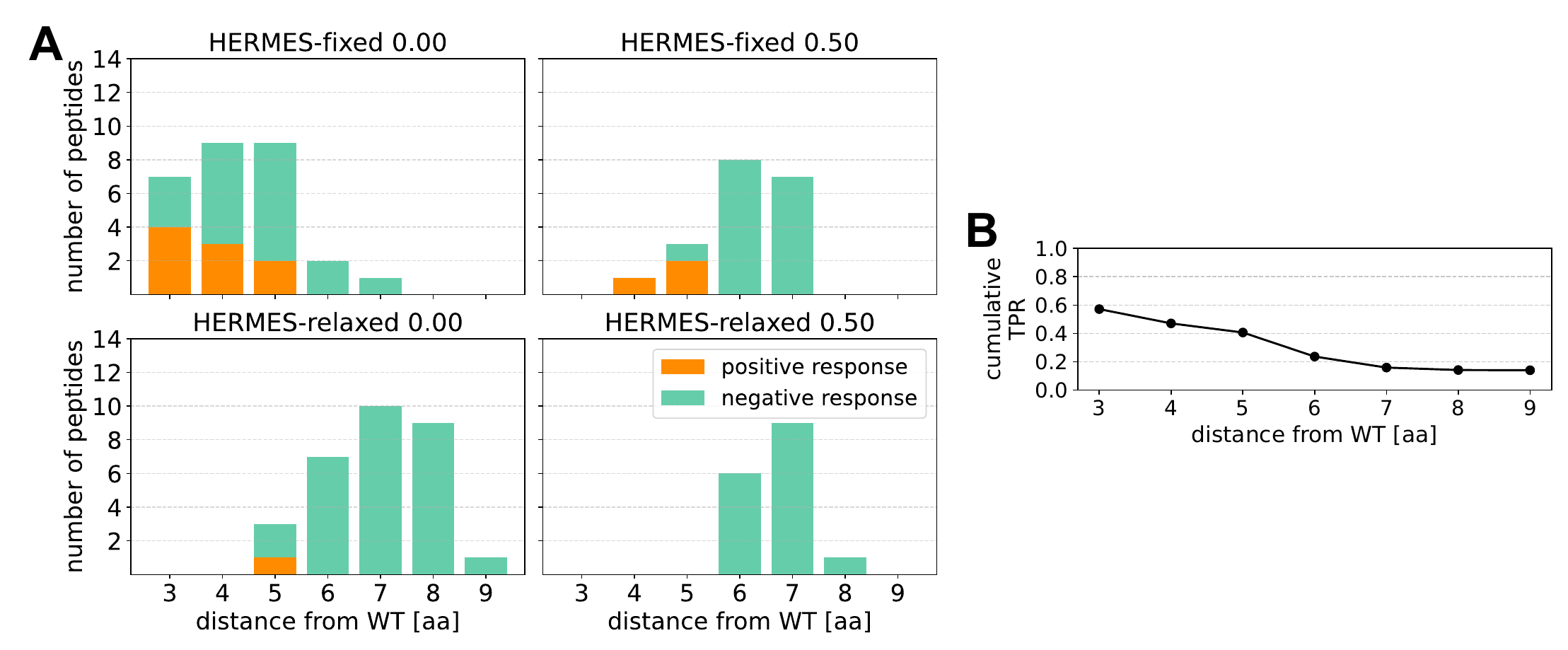}
\caption{\textbf{Experimental T-cell activation by HERMES-designed peptides in the EBV system.} {\bf (A-B)} Similar to (A-B) of Fig.~\ref{fig:design_experimental_nyeso} but for the EBV system.
}
\label{fig:design_experimental_ebv}
\end{figure}

\begin{figure}[ht]
\centering
	\includegraphics[width=0.99\textwidth]{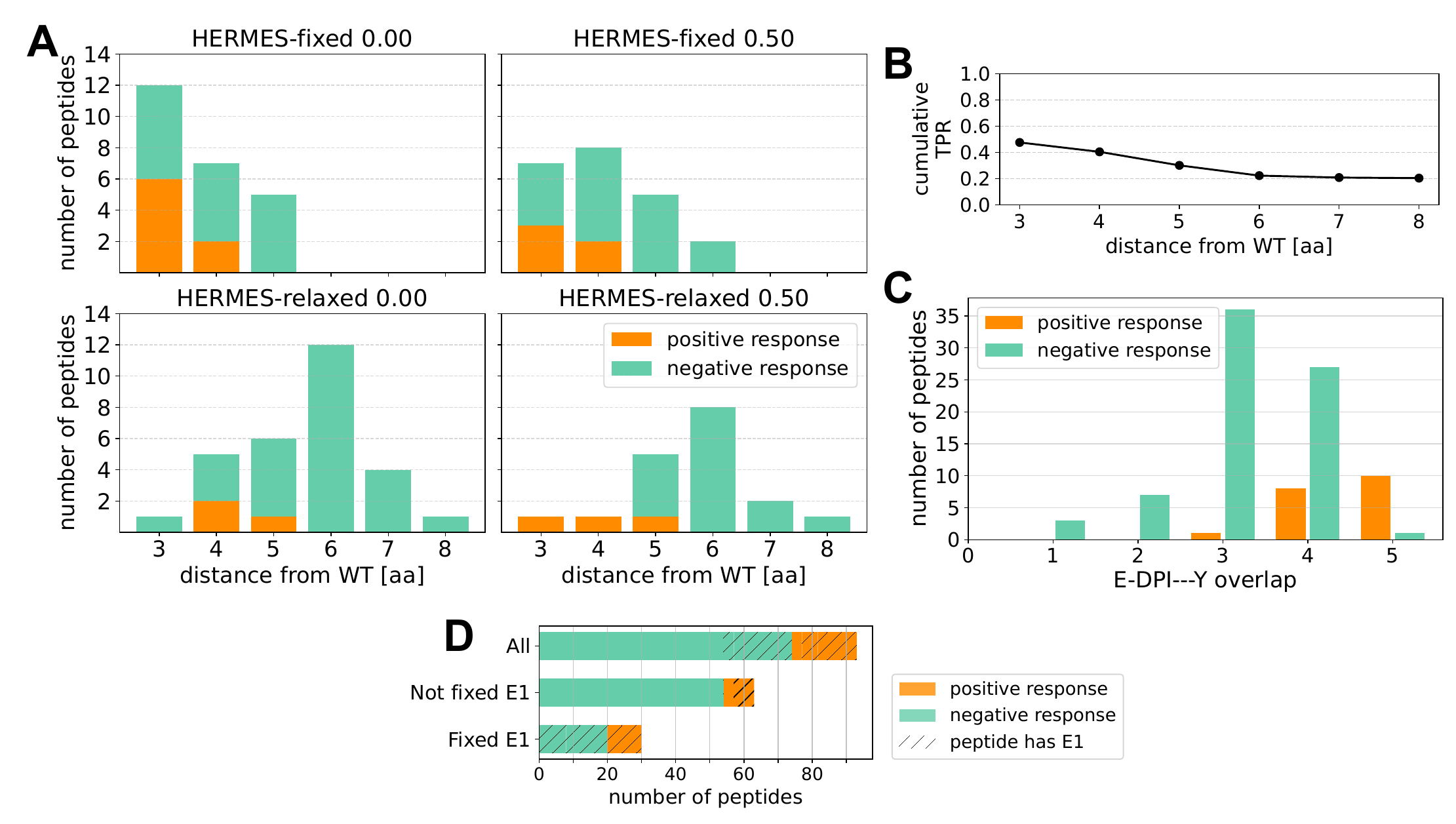}
\caption{\textbf{Experimental T-cell activation by HERMES-designed peptides in the MAGE system.}
{\bf (A-B)} Similar to (A-B) of Fig.~\ref{fig:design_experimental_nyeso} but for the MAGE system.
\textbf{(C)} The number of  designs that induced GFP expression beyond the T-cell activation threshold (positive response;  in orange) versus those that did not (negative response;  in green) 
 is shown as a function of the designs' overlap with the  E-DPI-\,-\,-Y motif;  this motif has  been shown to enhance T-cell activation in this system~\cite{Cameron2013-vw}.  The overlap is   the number of  amino acids that a design shares with the E-DPI-\,-\,-Y motif (maximum of 5). The fraction of positive designs increases with greater similarity to the motif. \textbf{(D)} Shown are the numbers of True Positives (orange) and False Positives (green) among designs generated under two protocols: one that enforced glutamic acid at position one (E$_1$)---labeled ``fixed E$_1$"----and one that did not enforce E$_1$---labeled ``not fixed E$_1$".  For each protocol, the dashed segments indicate the number of designs that contain E$_1$. The ``All" bar presents the aggregate data from both protocols.
}
\label{fig:design_experimental_mage}
\end{figure}

\begin{figure}[ht]
\centering
	\includegraphics[width=1.0\textwidth]{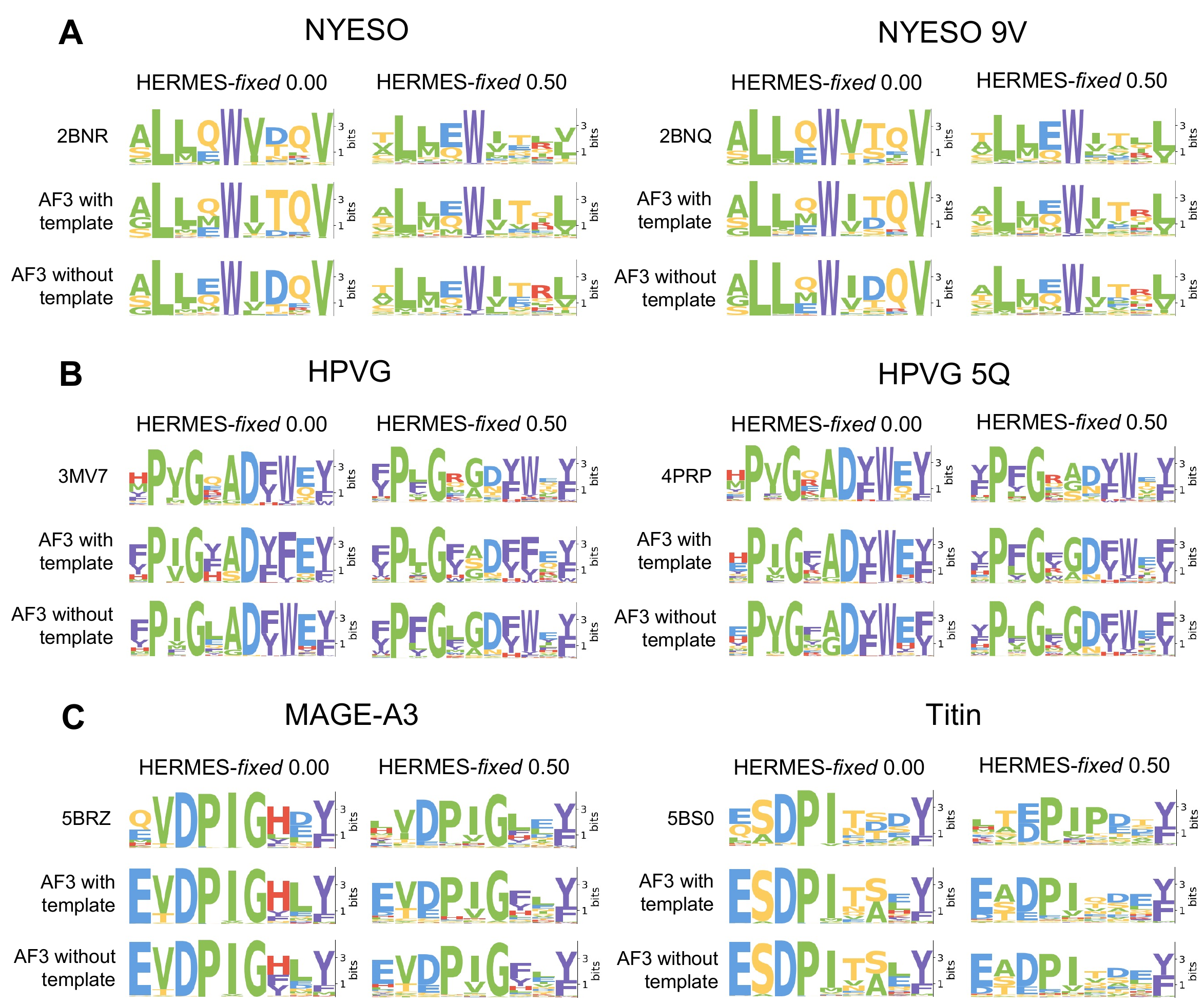}
\caption{\GM{{\bf Composition of designed peptides using experimental versus AlphaFold3 structures as design templates.} For the {\bf(A)} NY-ESO, {\bf(B)} EBV, and {\bf (C)} MAGE, each panel displays the position weight matrices (PWMs) for peptides designed using the HERMES-{\em fixed} protocol. Designs were generated  using three types of structural templates as indicated in each panel:  the experimentally determined  wild-type  structure (first row), and the AlphaFold3-predicted models of that structure generated with (middle) and without (bottom) template guidance. PWMs are shown separately for the HERMES-{\em fixed} models trained with different noise amplitudes  (as indicated above the panels). For each system, results are provided for the two   wild-type peptides used as input structural templates (shown in the left and right columns of each panel).    }
}
\label{fig:joint_logoplots_for_design}
\end{figure}

\begin{figure}[ht]
\centering
	\includegraphics[width=0.85\textwidth]{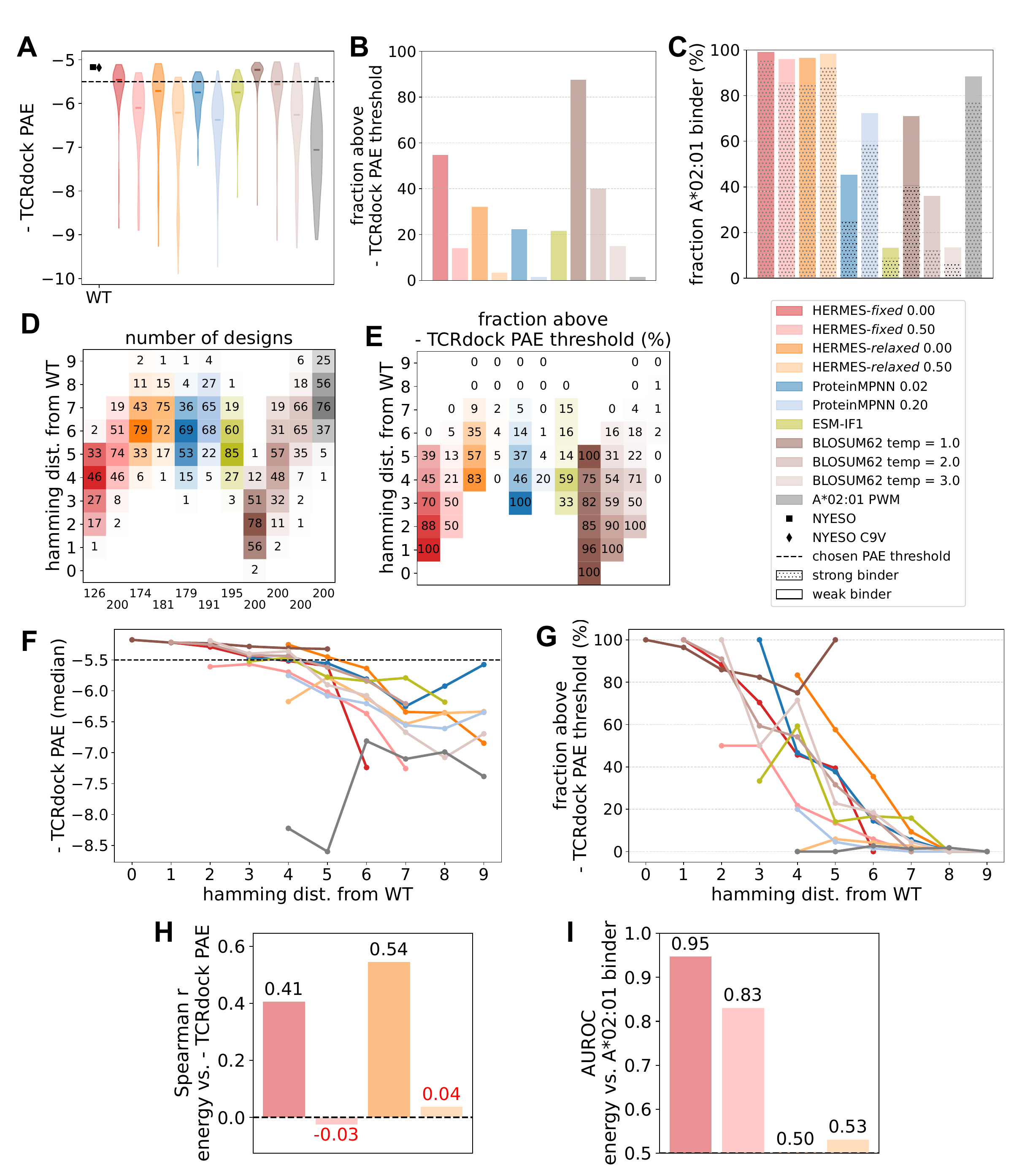}
\caption{\GM{\textbf{Benchmarking of models for peptide design in the NY-ESO system.}}
\textbf{(A)} The distribution of the TCRdock negative-PAE's for  peptides designed by different computational protocols (colors) are shown; in each case the score distribution for 100 independently designed peptides is shown. The horizontal dashed line indicates the PAE threshold chosen to classify positive designs.
\textbf{(B)} The fractions of designed peptides  that pass the TCRdock PAE threshold is shown for each method.
\textbf{(C)} The fraction of designs that are predicted by NetMHCPan~\cite{reynisson_netmhcpan-41_2020} to be presented by the template's MHC allele are shown; dotted regions correspond to strong binders (EL\_rank $\leq 0.5$), while the bar indicates all weak and strong binders  (EL\_Rank $\leq 2.0$). HERMES designs are most consistently predicted to be presented by the MHC. 
\textbf{(D} The number of designs,  and {\bf (E)} the fraction of designs that pass the TCRdock PAE threshold are shown for each model (colors) at different Hamming distances from the wild-types; The numbers at the bottom of each column in (D) indicate the total number of unique designs. 
{\bf (F)} The median of TCRdock negative-PAE across designs, and  {\bf (G)} the fraction of designs that pass the TCRdock PAE  are shown  for each model (colors) as a function of the  Hamming distance from the wild-types.
\GM{{\bf (H)} Spearman correlation between HERMES-predicted peptide energy and TCRdock PAE is shown, with data  pooled across design models. In  each case, HERMES energies are computed using the same protocol applied during design. Correlations that are not statistically significant (p-value $> 0.05$) are highlighted in red.
{\bf (I)} Shows the AUROC for classifying whether a designed peptide is presented by the template's MHC allele (i.e., predicted to be a strong binder according to  NetMHCPan~\cite{reynisson_netmhcpan-41_2020}), using HERMES peptide energy as the classifier. As in (H), for each peptide, HERMES energies are computed via the same protocol used for their deign.  
\GM{The HERMES model names specify the design scheme (fixed, relaxed),  and the noise amplitude (in \AA) injected to the coordinates of the structure data during training. The ProteinMPNN design uses baseline scoring scheme with peptide 1-by-1 (eq.~\ref{eq:peplogp_proteinmpnn}), with the training noise amplitude specified in the model name.}
}
}
\label{fig:design_tcrdock_nyeso}
\end{figure}

\begin{figure}[ht]
\centering
	\includegraphics[width=0.9\textwidth]{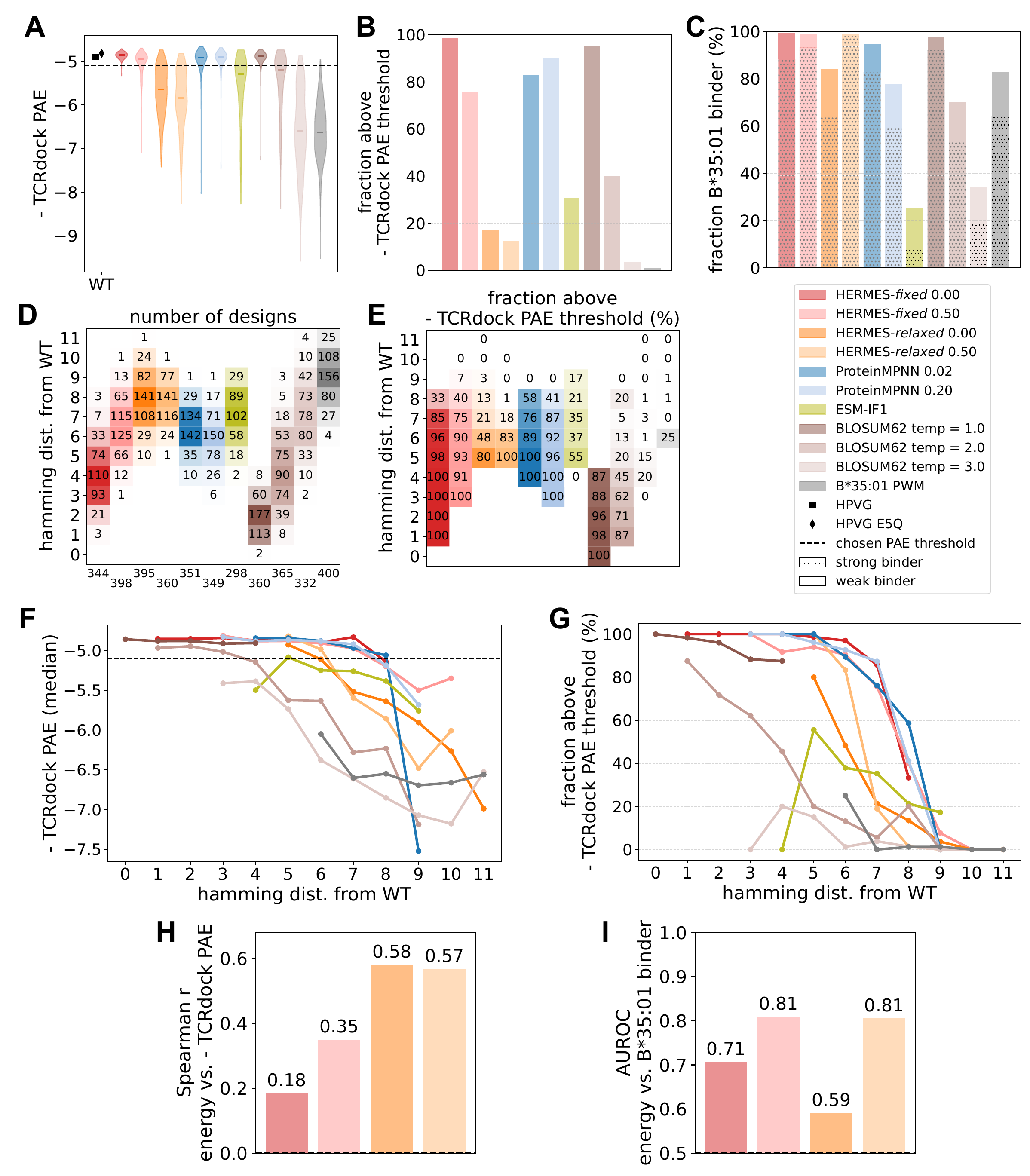}
\caption{\GM{\textbf{Benchmarking of models for peptide design in the EBV system.}} Similar to Fig.~\ref{fig:design_tcrdock_nyeso} but for the EBV system.}
\label{fig:design_tcrdock_ebv}
\end{figure}

\begin{figure}[ht]
\centering
	\includegraphics[width=0.9\textwidth]{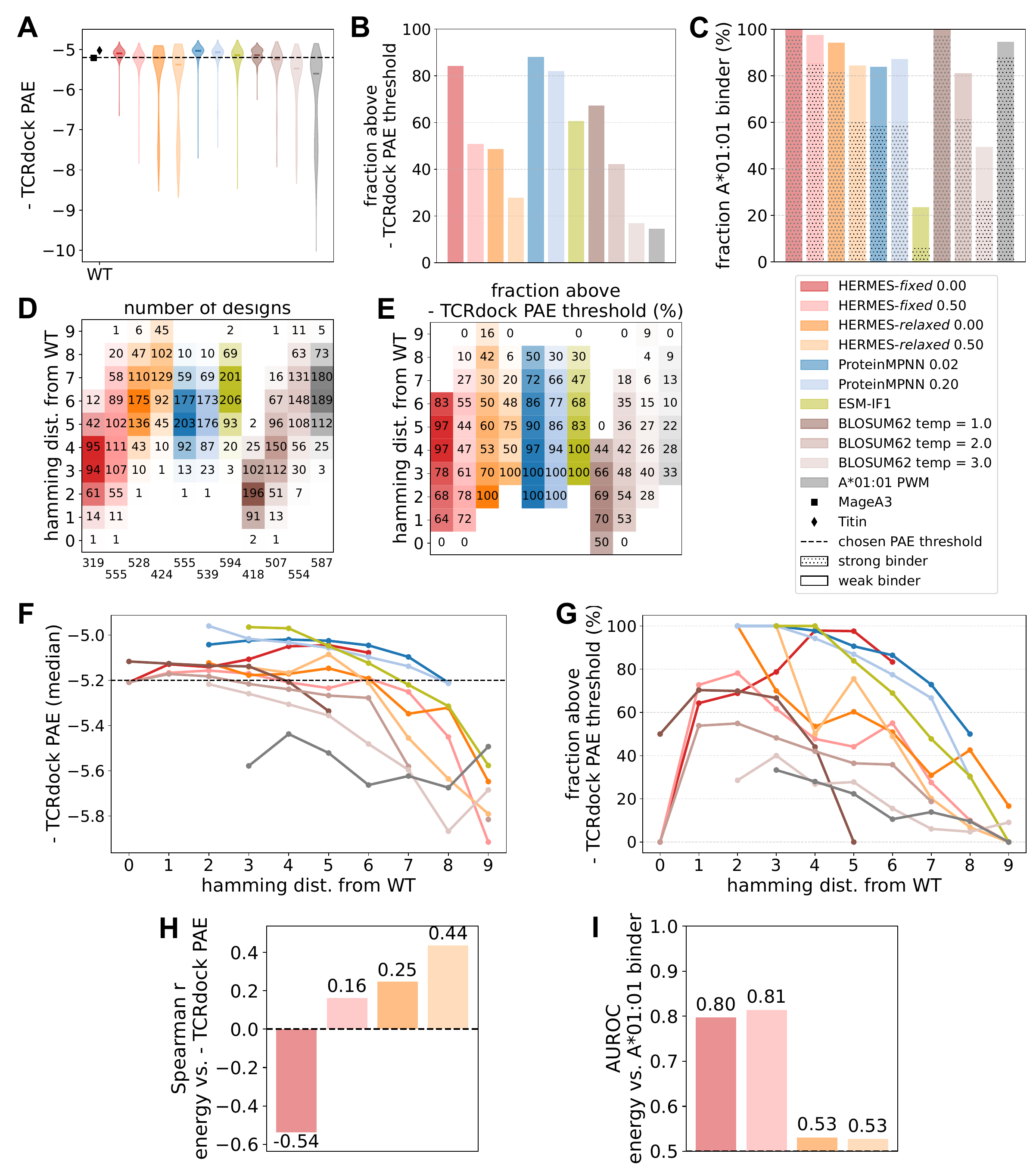}
\caption{\GM{\textbf{Benchmarking of models for peptide design in the MAGE system.}} Similar to Fig.~\ref{fig:design_tcrdock_nyeso} but for the MAGE system.
}
\label{fig:design_tcrdock_mage}
\end{figure}

\begin{figure}[ht]
\centering
	\includegraphics[width=0.8\textwidth]{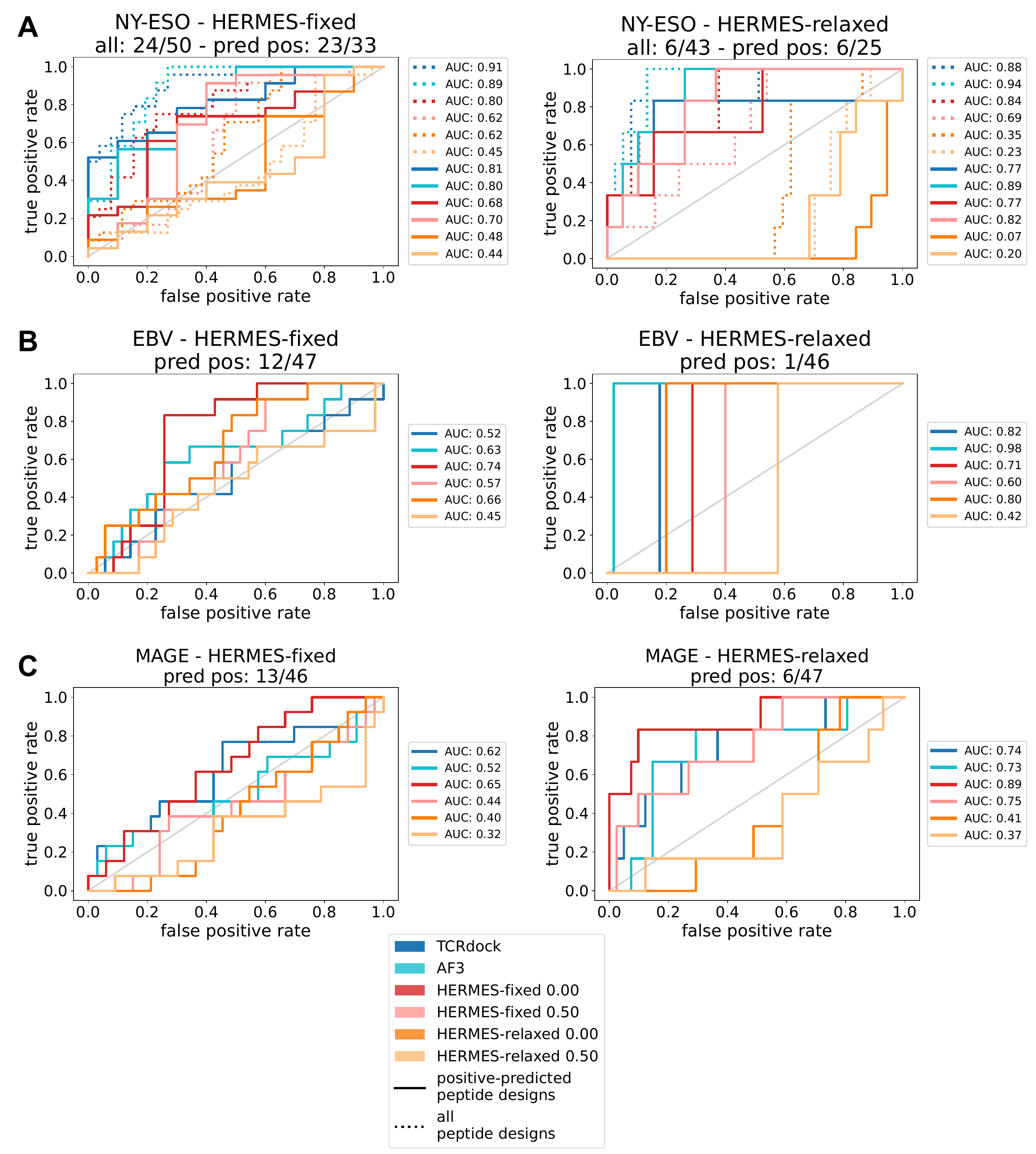}
\caption{\GM{{\bf Discrimination ability of different scoring schemes as design filters.}
The ROC curves show  true positive versus false positive rates  when using scores from different models--HERMES, AlphaFold3, and TCRdock (color coded as  in the legend)--as  scoring criteria for peptide design. Results are shown separately for designs generated  using the HERMES-{\em fixed} (left) and the HERMES-{\em relaxed} (right) protocols,  for the {\bf (A)} NY-ESO, {\bf (B)} EBV, and {\bf (C)} MAGE systems. Experimentally measured  peptide-induced T-cell activity (GFP\%) is used as the ground truth for evaluating the predictive performance in each case. {For the NY-ESO system only {(A)} we show ROC curves computed on both peptides that were predicted to activate T-cells (positives) based on their TCRdock PAE scores (full lines), and the mixture of positive and negative peptides (dotted lines). For EBV and MAGE  {(B-C)}, the scoring criterion is only tested for  the peptides that we deemed to be activating (positive) based on their TCRdock PAE scores (full lines), as these were the ones that we tested experimentally. } The HERMES model names specify the design scheme (fixed, relaxed),  and the noise amplitude (in \AA) injected to the coordinates of the structure data during training.
}}
\label{fig:design_roc_curves}
\end{figure}

\begin{figure}[ht]
\centering
	\includegraphics[width=1.0\textwidth]{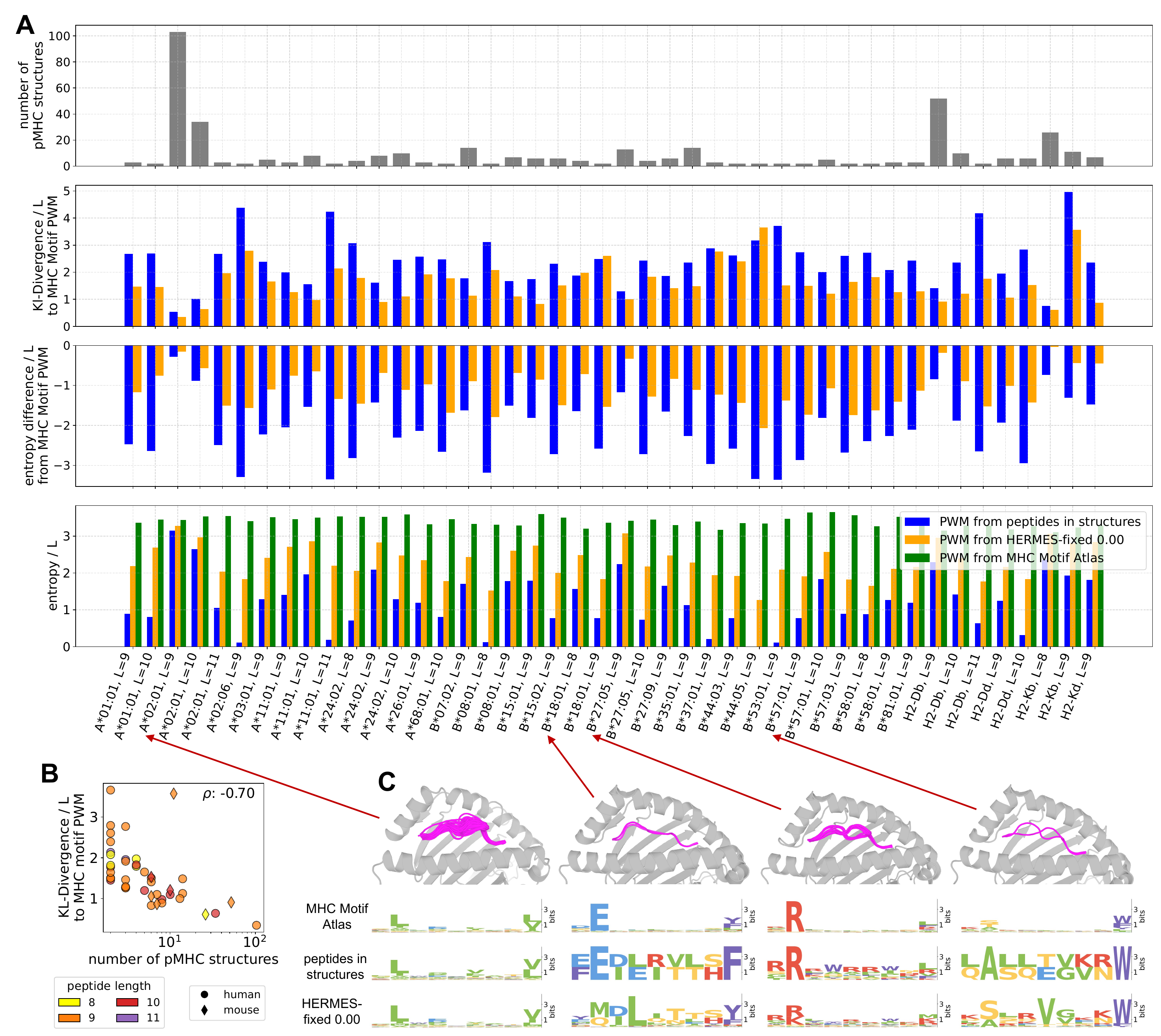}
\caption{\GM{\textbf{Predicting MHC-I peptide presentation preferences with HERMES.}
We evaluated HERMES's ability to recover MHC class I peptide presentation preferences by analyzing pMHC structures from the TCR3d database~\cite{gowthaman_tcr3d_2019}, grouped by MHC allele and peptide length (L), retaining only those combinations with at least two structures. As ground truth, we used position weight matrices (PWMs) from the MHC Motif Atlas~\cite{Tadros2023-ij}, derived from large datasets of peptide sequences known to be presented by each allele.
HERMES-{\em fixed} was used to design peptides for each structure in a given allele-length group, and the resulting PWMs were averaged across structures to generate a motif reflecting the peptide preferences for that MHC I allele (for a given length).
As the baseline, we used PWMs generated directly from the peptides found in the structural dataset. PWM quality was evaluated using the Kullback-Leibler divergence, and the entropy difference from the ground-truth PWM.  {\bf (A)} Shown are the summaries of data and  performances across all MHC I allele-length combinations: number of available structures (first row), KL-divergence between the HERMES-derived PWMs (or the baseline) to the ground truth  (second row), the entropy difference to the ground truth (third row), and the entropy of the PWMs based on the three approaches (fourth row). {\bf (B)} Shown is the KL-divergence of HERMES-derived PWMs versus the number of available pMHC structures for across different allele-length classes. {\bf (C)} Examples of pMHC structures and the corresponding PWMs are shown for four allele-length combinations, spanning a range of  structure counts. In each case, structures are aligned along the MHC chain to show the diversity of peptide poses captured by the ensemble of available structures, used as design templates. }}
\label{fig:pmhc_pwm_entropy_comparisons}
\end{figure}

\clearpage{}
\newpage{}

\renewcommand{\figurename}{Dataset}
\setcounter{figure}{0}

\begin{figure}[h!]
{}
\caption{Amino-acid sequences used as input to AF3 within the AlphaFold Server to fold the TCR-pMHC and Fab-pMHC systems considered in this study.
\label{dataset:af3_sequences}}
\end{figure}

\begin{figure}[h!]
{}
\caption{Measured T-cell activity levels and metadata for the peptide designs reported in this study. The excel file contains three sheets, one for each system considered in this study: {\em NYESO-designs}, {\em EBV-designs} and {\em MAGE-designs}. Three replicates of activity levels (GFP \%), measured by fluorescence microscopy are reported under columns {\em r1}, {\em r2}, and {\em r3}; their mean values are reported under {\em r\_mean} and the binary response label is under {\em resp}. For the NYESO system only, we also report the measured GFP levels through flow cytometry (one replicate). Notable metadata includes: peptide sequence, HERMES model and protocol used for the design, TCRdock PAE score, AF3 PAE score, hamming distance from the wild-types. Activity levels induced by the wild-type peptides as well as under unstimulated (negative control) condition are  provided.}
\end{figure}

\clearpage{}
\newpage{}